\documentclass[12pt,english]{article}
\usepackage{fontenc}
\usepackage[latin9]{inputenc}
\usepackage{geometry}
\usepackage{rotating}
\usepackage{amsmath}
\usepackage{amssymb}
\usepackage{graphicx}
\usepackage{setspace}
\usepackage[authoryear]{natbib}
\usepackage{amsfonts}
\usepackage{dsfont}
\usepackage{booktabs}
\usepackage{lscape}
\usepackage{babel}
\usepackage{hyperref}
\usepackage{eurosym}
\hypersetup{
	colorlinks=true,        
	allcolors=blue,
	pdftitle={Linear, matrix normal state space models},
	pdfauthor={Creal, Medeiros, Sarlo},
	pdfstartview={FitV},
	pdfpagemode={UseNone},
	pdfnewwindow=true,      
}
\DeclareRobustCommand{\officialeuro}{%
	\ifmmode\expandafter\text\fi
	{\fontencoding{U}\fontfamily{eurosym}\selectfont e}}

\renewcommand{\thefootnote}{\fnsymbol{footnote}}

\makeatletter

\RequirePackage{comment}
\RequirePackage{graphicx}

\renewcommand{\epsilon}{\varepsilon}

\newcounter{remark}
\makeatother
\newcounter{fig}
\newcommand{\Lp}{\left(}
\newcommand{\Rp}{\right)}
\newcommand{\Lb}{\left[}
\newcommand{\Rb}{\right]}

\newcommand{\Lv}{\left|}
\newcommand{\Rv}{\right|}

\newcommand{\V}{\mathds{V}}

\newcommand{\eye}{\mathbf{I}}

\newcommand{\ps}{\mathbf{s}}

\newcommand{\pA}{\mathbf{A}}
\newcommand{\pB}{\mathbf{B}}
\newcommand{\pC}{\mathbf{C}}
\newcommand{\pD}{\mathbf{D}}
\newcommand{\pE}{\mathbf{E}}
\newcommand{\pF}{\mathbf{F}}
\newcommand{\pFcal}{\pmb{\mathcal{F}}}
\newcommand{\pJ}{\mathbf{J}}

\newcommand{\pG}{\mathbf{G}}
\newcommand{\pH}{\mathbf{H}}
\newcommand{\pL}{\mathbf{L}}
\newcommand{\pK}{\mathbf{K}}
\newcommand{\pM}{\mathbf{M}}
\newcommand{\pN}{\mathbf{N}}

\newcommand{\pP}{\mathbf{P}}
\newcommand{\pQ}{\mathbf{Q}}
\newcommand{\pR}{\mathbf{R}}
\newcommand{\pS}{\mathbf{S}}
\newcommand{\pT}{\mathbf{T}}
\newcommand{\pU}{\mathbf{U}}
\newcommand{\pV}{\mathbf{V}}
\newcommand{\pW}{\mathbf{W}}
\newcommand{\pX}{\mathbf{X}}
\newcommand{\pY}{\mathbf{Y}}
\newcommand{\pZ}{\mathbf{Z}}

\newcommand{\pzero}{\pmb{0}}

\newcommand{\pbeta}{\pmb{\beta}}

\newcommand{\ptheta}{\pmb{\theta}}

\newcommand{\pPhi}{\pmb{\Phi}}
\newcommand{\pPsi}{\pmb{\Psi}}
\newcommand{\pSigma}{\pmb{\Sigma}}
\newcommand{\pLambda}{\pmb{\Lambda}}
\newcommand{\pUpsilon}{\pmb{\Upsilon}}
\newcommand{\pGamma}{\pmb{\Gamma}}
\newcommand{\pDelta}{\pmb{\Delta}}
\newcommand{\pPi}{\pmb{\Pi}}
\newcommand{\pOmega}{\pmb{\Omega}}
\newcommand{\pTheta}{\pmb{\Theta}}

\newtheorem{proposition}{Proposition}

\newtheorem{lemma}{Lemma}

\makeatletter
\@addtoreset{table}{section}
\makeatother

\begin{document}
	\thispagestyle{empty}
	\enlargethispage{1in}
	\newlength{\oldparindent}
	\oldparindent=\parindent
	\parindent=0.0in
	
	\vspace*{0.1in}

	\noindent{\Large\bf Conditionally linear, matrix normal state space models}

	\vspace{0.05in}
	\noindent{%
		Drew D. Creal\footnote{
			Department of Economics, University of Illinois Urbana-Champaign;
			dcreal@illinois.edu.},
		Marcelo C. Medeiros,\footnote{
			Department of Economics, University of Illinois Urbana-Champaign;
			marcelom@illinois.edu.}
		and
		Rodrigo Sarlo,\footnote{
			Department of Electrical Engineering, Pontifical Catholic University of Rio de Janeiro;
			rodrigosarlofilho@gmail.com.}
	}

	\vspace{0.05in}
	\noindent{This version: \today}
	
	{
		{\bf Abstract}
		
		\medskip
		We develop a class of linear state space models for matrix-valued time series data where the state is a latent matrix normal process. We derive matrix versions of the Kalman filter, log-likelihood, and smoother enabling estimation of the latent state matrix as well as the model's parameters. To conduct Bayesian inference, we provide algorithms that draw from the joint posterior distribution of the latent state matrices conditional on the observed data and parameters. We apply these methods to a large panel of U.S. macroeconomic time series across the 50 U.S.\ states. The proposed framework accommodates mixed-frequency data, heteroskedasticity, and outliers within a unified matrix-valued structure. Empirically, we find that a small number of latent factors captures the joint dynamics across states and variables, providing a parsimonious and scalable approach to modeling high-dimensional macroeconomic systems.
		\medskip
		
		
		
		\medskip
		
	}
	
	
	\setcounter{footnote}{0}
	\renewcommand{\thefootnote}{\arabic{footnote}}
	\noindent \thispagestyle{empty}\addtocounter{page}{-1}\newpage{}%
	\setcounter{equation}{0}
	
	\section{Introduction}
	
	Conditionally linear Gaussian state space models and the Kalman filter are central tools in modern time series analysis. They enable estimation and forecasting in a wide range of empirically important models, including vector autoregressions, structural time series models, and dynamic factor models; see, e.g., \cite{CappeMoulinesRyden(05)}, \cite{DurbinKoopman(12)}, and \cite{ShumwayStoffer(25)}. 
	
	At the same time, matrix-valued time series data have become increasingly common in economics, engineering, and the biological sciences; see \cite{Tsay(24)} for a recent survey. In many applications, the data naturally take the form of an $m \times n$ matrix $\pY_t$, where rows correspond to variables and columns correspond to related cross-sectional units. For example, in macroeconomics, one may observe multiple economic series across U.S. states, where each column corresponds to a state and each row corresponds to a variable such as employment or income. 
	
	Despite the growing importance of matrix-valued data, the matrix analogue of conditionally linear Gaussian state space models and their associated estimation methods remain only partially developed in the literature. Matrix-variate dynamic linear models originate with \cite{QuintanaWest(87)}, are treated within the general Bayesian forecasting framework of \cite{WestHarrison(97)}, and are further developed by \cite{CarvalhoWest(07)} and \cite{WangWest(09)}. \cite{CarvalhoWest(07)} derive the posterior mean and covariance of the smoothed state matrix, but only for a model in which every column shares the same known regression vector and evolution matrix -- there is no estimated loading matrix, and no row/column dimension mismatch for a companion-form representation to resolve. None of these papers derive the matrix-variate log-likelihood, a simulation smoother for drawing the state matrices jointly from their posterior distribution, or a precision sampler. Existing work therefore does not provide a unified framework for likelihood evaluation, state smoothing, simulation smoothing, and posterior inference for the more general class of matrix-valued state space models we consider.

	Our main contribution is twofold. First, we introduce a companion-form representation for matrix normal processes that substantially generalizes existing matrix state space formulations (e.g., \cite{WangWest(09)}), accommodating a broad class of models, including bilinear dynamic factor models. Second, building on this representation, we derive the matrix Kalman filter, log-likelihood, and smoothing distributions. We also develop simulation smoothing and precision sampling methods for drawing the latent state matrix jointly from its posterior distribution. These extend the simulation smoother of \cite{DurbinKoopman(02)} and the precision-based sampler of \cite{ChanJeliazkov(09)} to the matrix setting. By preserving the matrix structure of the data throughout, the framework achieves substantial dimension reduction relative to vectorized representations while remaining fully likelihood-based. Vectorizing the observation and transition equations recovers an ordinary vector state space model, so the filtering and smoothing distributions we derive coincide exactly with those of that vector model. Our contribution is computing them without ever forming or factoring the resulting large Kronecker covariance, instead working directly with much smaller matrices tied to the row and column dimensions of the data and the state.
	
	The framework readily accommodates additional features commonly used in empirical applications, such as time-varying parameters, heteroskedasticity, and mixture models. In particular, we extend the methods of \cite{GerlachCarterKohn(00)} and \cite{DoucetAndrieu(01)} to draw discrete latent states without conditioning on the continuous state matrix, enabling efficient Bayesian inference in matrix-valued state space models.

	\subsection{Relation to the Literature}
	
	\cite{CarvalhoWest(07)} introduce a matrix-normal dynamic linear model with graphically structured cross-sectional covariance, deriving both a forward filter and a backward smoother for the posterior moments of the state matrix. \cite{WangWest(09)} extend this to genuine matrix observations with a matrix state space representation and filtering algorithms. In both cases, however, every column shares the same known regression vector and evolution matrix, so there is no estimated loading matrix and no row/column dimension mismatch to resolve, and their smoother gives only the smoothed mean and covariance, not a joint draw of the state path. We generalize this to a companion-form representation that accommodates estimated, series-specific loading matrices and a general row/column dimension mismatch, and add likelihood evaluation, smoothing, simulation smoothing, and precision-based posterior simulation.
	
	\cite{ChoukrounWeissBarItzhackOshman(06)} derive a more general matrix Kalman filter for a sum of bilinear terms with fully unrestricted, non-separable covariance matrices, obtained by vectorizing the system and applying the standard vector Kalman filter as a minimum-variance estimator. Their approach abandons the Kronecker-separable, matrix normal structure our framework relies on.
	
	A growing literature develops matrix autoregressive models for matrix-valued time series (\cite{ChenXiaoYang(21)}, \cite{WangLiuChen2019}), typically estimated by moment-based or least-squares methods, with a large-scale Bayesian version in \cite{ChanQi(25)}. A related strand imposes low-rank, two-way factor structure on matrix data (\cite{YuanGaoHeHuangGuo(23)}, \cite{QinWangZhuShia(25)}). None of this casts the problem as a state space model; doing so is what gives us likelihood-based filtering, smoothing, and joint posterior simulation of the latent state.
	
	Finally, our framework also builds on the classic Bayesian dynamic factor model literature (e.g., \cite{AguilarWest(00)}, \cite{LopesWest2004}), which treats the data as vector-valued. We extend this to matrix-valued data by keeping the matrix structure explicit throughout: the matrix normal distribution with separable covariance is what buys the dimension reduction and computational gains over a vectorized factor model.
	
	\section{Matrix normal state space models} \label{model}
	
	\subsection{Model}
	
	Let $\pY_{t}$ denote an $m \times n$ matrix time series observed for $t=1,\ldots,T$. We study models that can be fit into the state space representation
	\begin{eqnarray}
		\pY_{t} & = & \pD_{t} + \pZ_{t}\pA_{t}\pW_{t}^{\top} + \pE_{1t}, \qquad \pE_{1t} \sim \text{MN}\Lp \pzero,\pH_{t},\pU_{\pY,t}\Rp, \label{obsrvation equation} \\
		\pA_{t+1} &=& \pC_{t} + \pT_{t}\pA_{t} + \pR_{t}\pE_{2t}, \qquad \pE_{2t} \sim \text{MN}\Lp \pzero,\pQ_{t},\pU\Rp, \label{transition equation} \\
		\pA_{1} & \sim & \text{MN}\Lp \pA_{1|0},\pP_{1|0},\pU\Rp. \label{initial condition}
	\end{eqnarray}
	The state matrix $\pA_{t}$ is $s \times r$ and is potentially unobserved or latent. The model requires that $n \geq r$. The matrices of shocks $\pE_{1t} $ and $\pE_{2t}$ have dimensions $m \times n$ and $q \times r$, respectively. Both have matrix normal (MN) distributions with mean zero and covariance matrices $\V\Lb \text{vec}\Lp\pE_{1t} \Rp\Rb \ = \ \pU_{\pY,t} \otimes \pH_{t}$ and $\V\Lb \text{vec}\Lp\pE_{2t} \Rp\Rb \ = \ \pU \otimes \pQ_{t}$. 
	The system matrices are functions of a $k \times 1$ vector of parameters $\ptheta$ that need to be estimated. We take $\ptheta$ as known in Sections \ref{model}-\ref{MKFS} and discuss their estimation below.

	The state space representation (\ref{obsrvation equation})-(\ref{initial condition}) is written in companion form enabling it to encompass a wide range of models, which we illustrate through several examples.
	
	\textbf{Example \#1:} Consider a matrix version of the local level model
	\begin{eqnarray*}
		\pY_{t} &=& \pA_{t} + \pE_{1t}, \qquad \pE_{1t} \sim \text{MN}\Lp \pzero,\pSigma,\pU\Rp,\\
		\pA_{t+1} &=& \pA_{t} + \pE_{2t},  \qquad \pE_{2t} \sim \text{MN}\Lp \pzero,\pOmega,\pU\Rp,
	\end{eqnarray*}
	which can be placed in state space form (\ref{obsrvation equation})-(\ref{transition equation}) by defining $\pD_{t} = \pzero$, $\pZ_{t} = \eye$, $\pW_{t} = \eye$, $\pH_{t} = \pSigma$, $\pC_{t} = \pzero$, $\pT_{t} = \eye$, $\pR_{t} = \eye$, and $\pQ_{t} = \pOmega$. This model can be extended to include trends, seasonals, and cycles as in the literature on structural time series models; see e.g. \cite{DurbinKoopman(12)}.
	
	\textbf{Example \#2:} Consider a matrix bi-linear dynamic factor model
	\begin{eqnarray}
		\pY_{t} &=& \pLambda \pFcal_{t}\pW^{\top} + \pE_{1t}, \qquad \pE_{1t} \sim \text{MN}\Lp \pzero,\pSigma_{t},\pU_{\pY}\Rp, \label{Example 2a} \\
		\pFcal_{t+1} &=& \pPhi_{1}\pFcal_{t} + \pPhi_{2}\pFcal_{t-1} + \pE_{2t},  \qquad \pE_{2t} \sim \text{MN}\Lp \pzero,\pOmega_{t},\pU\Rp. \label{Example 2c}
	\end{eqnarray}
	The left factor loadings $\pLambda$ control the common factors across series while the right factor loadings $\pW$ control the common factors across spatial units. This model can be placed in state space form by defining the state and its  matrices as
	\[  \pA_{t} \ = \ \Lp\begin{matrix}
		\pFcal_{t} \\
		\pFcal_{t-1}
	\end{matrix}\Rp \ \ \pT_{t} \ = \ \Lp\begin{matrix}
		\pPhi_{1} &  \pPhi_{2} \\
		\eye & \pzero
	\end{matrix}\Rp \ \ \pC_{t} \ = \ \Lp\begin{matrix}
		\pzero \\
		\pzero
	\end{matrix}\Rp  \ \ \pR_{t} \ = \ \Lp\begin{matrix}
		\eye  \\
		\pzero 
	\end{matrix}\Rp  \]
	and
	$\pD_{t} = \pzero$, $\pZ_{t} = \Lp \pLambda \  \pzero\Rp $, $\pW_{t} = \pW$, $\pH_{t} = \pSigma_{t}$, and $\pQ_{t} = \pOmega_{t}$. We estimate a version of this model in our empirical application.
	
	\textbf{Example \#3:} Consider the matrix autoregressive moving average process model
	\begin{eqnarray*}
		\pY_{t} &=& \pPhi_{1}\pY_{t-1}\pUpsilon_{1}^{\top} + \ldots +  \pPhi_{\overline{p}}\pY_{t-\overline{p}}\pUpsilon_{\overline{p}}^{\top} + \pE_{t} + \pTheta_{1}\pE_{t-1}\pGamma_{1}^{\top} + \ldots  + \pTheta_{\overline{q}}\pE_{t-\overline{q}}\pGamma_{\overline{q}}^{\top}, 
	\end{eqnarray*}
	with $\pE_{t} \sim \text{MN}\Lp \pzero,\pSigma,\pU\Rp$.
	Under the restrictions that $\pGamma_{\ell} = \eye$ for all lags $\ell$, this model can be fit into the companion form (\ref{obsrvation equation})-(\ref{transition equation}). For example, if $\overline{p} = 2$ and $\overline{q} = 2$, we define $\pD_{t} \ = \ \pPhi_{1}\pY_{t-1}\pUpsilon_{1}^{\top} + \pPhi_{2}\pY_{t-2}\pUpsilon_{2}^{\top}$ 
	and
	\[  \pA_{t} \ = \ \Lp\begin{matrix}
		\pE_{t} \\
		\pE_{t-1} \\
		\pE_{t-2}
	\end{matrix}\Rp \ \ \pT_{t} \ = \ \Lp\begin{matrix}
		\pzero &  \pzero &  \pzero \\
		\eye & \pzero & \pzero \\
		\pzero & \eye & \pzero
	\end{matrix}\Rp \ \ \pC_{t} \ = \ \Lp\begin{matrix}
		\pzero \\
		\pzero \\
		\pzero
	\end{matrix}\Rp \ \ \pR_{t} \ = \ \Lp\begin{matrix}
		\eye \\
		\pzero \\
		\pzero
	\end{matrix}\Rp.  \]
	The remaining matrices are $\pZ_{t} \ = \ \Lp\begin{matrix}
		\eye & \pTheta_{1} & \pTheta_{2}
	\end{matrix}\Rp$, $\pH_{t} = \pzero$, $\pW_{t} = \eye$, and $\pQ_{t} = \pSigma$. If the dynamics of $\pY_{t}$ follow a general MARMA$(\overline{p},\overline{q})$, the observations $\pY_{t}$ cannot be placed in the state matrix unless the additional restrictions that $\pUpsilon_{j} = \eye$ are imposed for all lags $j$. Right multiplication destroys the matrix normal structure. The transition equation propagates the state by left multiplication only, so every lag must share the same right scale matrix, ruling out lag-specific $\pGamma_{\ell}$ or $\pUpsilon_{j}$.
	
	An important special case of the model is when the system matrices of the transition equation (\ref{transition equation}) are time-invariant and the transition matrix $\pT$ has all eigenvalues less than one in modulus. Then, the stochastic process $ p\Lp \pA_{t+1}|\pA_{t},\ptheta\Rp$ has a stationary distribution that is matrix normal $p\Lp \pA_{t}|\ptheta\Rp \ = \ \text{MN}\Lp \overline{\pA},\overline{\pP},\pU\Rp $ with mean matrix $\overline{\pA} \ = \ \Lp \eye_{s} - \pT\Rp^{-1}\pC$ and left scale matrix $\text{vec}\Lp\overline{\pP}\Rp \ = \ \Lp \eye_{s^{2}} - \pT\otimes\pT\Rp^{-1}\text{vec}\Lp \pR\pQ\pR^{\top}\Rp$.
	The stationary distribution is often chosen for the initial condition of the model. 
	
	In order for the filtering and smoothing distributions to be known in closed form, the measurement and transition equations must have a matrix normal distribution with a common right scale matrix $\pU$ at each date, for two reasons. First, a sum of matrix normals is itself matrix normal only when one pair of scale matrices is proportional. The Kronecker factorization $\pP\otimes\pU$ is identified only up to such a rescaling. Therefore, we hold $\pU$ fixed and shared across dates to keep the joint distribution $p\Lp \pA_{1},\ldots,\pA_{T}|\ptheta\Rp$ matrix normal.
	
	Secondly, the proof of the matrix Kalman filter and related algorithms relies on a key property of the matrix normal distribution, which is Theorem 2.3.12 in \cite{GuptaNagar(00)}. We re-state it using our notation.
	
	\begin{lemma}\label{Theorem 1} Let $\pX$ be an $m \times n$ matrix with distribution $\pX \ \sim \ \text{MN}\Lp \pM,\pP,\pU\Rp$, where $\pP$ is $m\times m$ and $\pU$ is $n\times n$. Consider a partition of the matrices as
		\[ \pX \ = \ \Lp \begin{matrix}
			\pX_{1} \\
			\pX_{2}
		\end{matrix}\Rp \qquad \pM \ = \ \Lp\begin{matrix}
			\pM_{1} \\
			\pM_{2}
		\end{matrix}\Rp \qquad \pP \ = \ \Lp\begin{matrix}
			\pP_{11} & \pP_{12} \\
			\pP_{21} & \pP_{22}
		\end{matrix}\Rp \]
		such that $\pX_{1}$ is $m_{1} \times n$ and $\pX_{2}$ is $m_{2} \times n$ with $m = m_{1} + m_{2}$. 
		Then, the conditional distribution $p\Lp\pX_{2}|\pX_{1}\Rp$ is matrix normal $\text{MN}\Lp \pM_{2|1},\pP_{2|1},\pU\Rp$ with
		\[\pM_{2|1} \ = \ \pM_{2} + \pP_{21}\pP_{11}^{-1}\Lp \pX_{1} - \pM_{1}\Rp,  \qquad 
		\pP_{2|1} \ = \ \pP_{22} - \pP_{21}\pP_{11}^{-1}\pP_{12}. \]
	\end{lemma}
	\vskip -0.2cm
	The Kalman filter and related algorithms are a recursive application of this lemma.

	\subsection{Collapsing the observation equation when $n > r$}
	
	When the column dimension of $\pY_t$ exceeds that of the state matrix $\pA_t$, i.e.\ $n > r$, the measurement and transition equations in (\ref{obsrvation equation})-(\ref{transition equation})  do not share a common right scale matrix. The measurement disturbances have right scale $\pU_{\pY,t}$ while the state disturbances have right scale 
	$\pU$. This prevents direct application of Lemma \ref{Theorem 1}.
	
	Our solution is to right multiply the original observation equation (\ref{obsrvation equation}) by a matrix $\pJ_{t} = \Lp \pJ_{t}^{*,\top} \ \pJ_{t}^{+,\top} \Rp^{\top}$ that is a function of $\pW_{t}$ and $\pU_{\pY,t}$.  The transformation $\pJ_{t}$ and the scale matrices $\pU_{\pY,t}$ and $\pU$ must be jointly specified
	to satisfy the conditions
	\begin{eqnarray}
		\pJ_{t}^{*}\pW_{t} &=& \eye_{r}, \label{condition 1} \\
		\pJ_{t}^{+}\pW_{t} &=& \pzero, \label{condition 1b} \\
		\pJ_t^* \pU_{\pY,t} \pJ_t^{*,\top} &=& \pU, \label{condition 2} \\
		\pJ_t^* \pU_{\pY,t} \pJ_t^{+,\top} &=& \pzero. \label{condition 3}
	\end{eqnarray}
	and $\pJ_{t}$ must be full rank. Condition (\ref{condition 1b}) ensures that the transformed data $\pY_{t}^{+} \ \equiv \ \pY_{t}\pJ_{t}^{+,\top}$ does not depend on the state $\pA_t$. Condition (\ref{condition 2}) fixes the right scale matrix of $\pY_{t}^{*} \ \equiv \ \pY_{t}\pJ_{t}^{*,\top}$ to equal $\pU$. Condition (\ref{condition 3}) ensures that $\pY_t^*$ and $\pY_t^+$ are independent.
	Under these conditions, the transformations 
	create a new set of observation equations 
	\begin{eqnarray}
		\pY_{t}^{*} & = & \pD_{t}^{*} + \pZ_{t}\pA_{t} + \pE_{1t}^{*}, \qquad \pE_{1t}^{*} \sim \text{MN}\Lp \pzero,\pH_{t},\pU\Rp, \label{obs equation 1} \\
		\pY_{t}^{+} & = & \pD_{t}^{+} + \pE_{1t}^{+},   , \qquad \pE_{1t}^{+} \sim \text{MN}\Lp \pzero,\pH_{t},\pPsi_t\Rp, \label{obs equation 2}
	\end{eqnarray}
	where $\pD_{t}^{*} = \pD_{t}\pJ_{t}^{*,\top}$, $\pD_{t}^{+} = \pD_{t}\pJ_{t}^{+,\top}$, and $\pPsi_t = \pJ_t^+ \pU_{\pY,t} \pJ_t^{+,\top}$ is a $(n - r) \times (n-r)$  positive definite matrix. The matrix $\pPsi_t$ will typically contain unknown parameters. By construction, $\pY_t^+$ contains the information about $\pPsi_t$.
	
	In order to satisfy conditions (\ref{condition 1})-(\ref{condition 3}), a researcher must first decide how to model $\pU_{\pY,t}$ and the identifying restrictions they want to impose on the model. Once these are chosen, the matrix $\pJ_t^*$ is pinned down uniquely, as the generalized least squares projection onto the column space of $\pW_t$,
	\begin{eqnarray}
		\pJ_t^* &=& \Lp \pW_t^\top \pU_{\pY,t}^{-1} \pW_t \Rp^{-1} \pW_t^\top \pU_{\pY,t}^{-1}. \label{Jstar GLS}
	\end{eqnarray}
	$\pJ_t^+$ can be any $(n-r)\times n$ matrix whose rows are orthogonal to the columns of $\pW_t$, so that $\pY_t^+\equiv\pY_t\pJ_t^{+,\top}$ captures the part of $\pY_t$ that the state matrix does not explain. Unlike $\pJ_t^*$, the matrix $\pJ_t^+$ is not pinned down by the conditions. Any basis for this leftover space works, and different choices only relabel the coordinates of $\pY_t^+$, leaving $\pJ_t^*$ and $\pPsi_t$ untouched. After discussing identification, we give concrete examples of how to parameterize $\pU_{\pY,t}$.
	
	The idea of collapsing the measurement equation in a linear state space model into a lower dimensional representation was introduced by \cite{JungbackerKoopman(15)}. In their setting, the observation vector is collapsed (by left multiplication) to reduce the computational burden.
	In our setting, collapsing the column dimension of the state space model is required in order to apply Lemma \ref{Theorem 1}. We show in the online appendix that their left collapse can be applied simultaneously with our column collapse, further reducing the dimension of the filter.
	
	\subsection{Identification and choice of parameterization}
	
	The model (\ref{obsrvation equation})--(\ref{transition equation}) is invariant to rotations $\pA_t \to \pR_{\ell} \pA_t \pR_{r}^{-1}$, $\pZ_t \to \pZ_t \pR_{\ell}^{-1}$, and $\pW_t \to \pW_t \pR_{r}^{\top}$ for any nonsingular $\pR_{\ell},\pR_{r}$, leaving the likelihood unchanged:
\begin{eqnarray*}
	\Lp\pZ_t\pR_\ell^{-1}\Rp\Lp\pR_\ell\pA_t\pR_r^{-1}\Rp\Lp\pW_t\pR_r^\top\Rp^\top &=& \pZ_t\pA_t\pW_t^\top.
\end{eqnarray*}
The transition equation and initial condition are invariant under the same reparameterization provided $\pT_t \to \pR_\ell\pT_t\pR_\ell^{-1}$, $\pC_t \to \pR_\ell\pC_t\pR_r^{-1}$, $\pR_t \to \pR_\ell\pR_t$, $\pU \to \Lp\pR_r^{-1}\Rp^\top\pU\pR_r^{-1}$, $\pP_{1|0}\to\pR_\ell\pP_{1|0}\pR_\ell^\top$, and $\pA_{1|0}\to\pR_\ell\pA_{1|0}\pR_r^{-1}$. Both the conditional mean $\pT_t\pA_t+\pC_t$ and the transition equation's covariance matrix $\pU\otimes\pR_t\pQ_t\pR_t^\top$ transform consistently under this substitution. Identification therefore requires restrictions on both the row and column spaces of the state and on the scale matrices. A convenient choice imposes orthonormality, $\pZ_t^\top \pZ_t = \eye$ and $\pW_t^\top \pW_t = \eye$, and restricts $\pOmega$ and $\pU$ to be diagonal with ordered eigenvalues. These restrictions eliminate ordering and scaling indeterminacies, though not all rotational indeterminacy. 

Two further normalizations are standard here, as in any factor model identified up to rotation. The sign of each factor is fixed by convention rather than by the model. Flipping the sign of a column of $\pZ_t$ (or $\pW_t$) together with the corresponding row (or column) of $\pA_t$ leaves the orthonormality restrictions and ordered scale matrices unchanged. A researcher must designate one observed series as an anchor for each factor and flip the sign of the entire column whenever that anchor's loading comes out negative. A second normalization applies only if two eigenvalues of $\pOmega$ or $\pU$ are exactly tied, in which case any rotation within that pair leaves the restrictions unchanged. This does not bind once the corresponding eigenvalues are well separated, which is easy to check directly.
	
An alternative identifying assumption is to impose a triangular structure on the factor loadings by requiring that $\pZ_t$ or $\pW_t$ are each lower triangular with positive diagonal elements as in \cite{GewekeZhou(96)}, while allowing the scale matrices in the transition equation to be unrestricted. Although this achieves identification, the resulting factors depend on which variables are ordered first in $\pY_t$. Relabeling the rows changes which loadings are restricted to zero, an arbitrary dependence absent a natural ordering among the series. 
	
Finally, the Kronecker structure implies a scalar identification problem. For any scalar $c > 0$, rescaling $\pU \to c\pU$, $\pQ_t \to c^{-1}\pQ_t$, $\pH_t \to c^{-1}\pH_t$, and $\pPsi_t \to c\pPsi_t$ leaves $\pU\otimes\pQ_t$, $\pU\otimes\pH_t$, and $\pPsi_t\otimes\pH_t$ unchanged. Since $\pH_t$ is shared between $\pY_t^*$ and $\pY_t^+$, this is a single redundancy. Fixing the scale of a single parameter in any of the scale matrices solves the problem. For example, consider the matrix $\pU$, we could set $U_{1,1}=1$, $\Lv \pU\Rv = 1$, or $\text{tr}\Lp \pU\Rp/r =1$, which then pins down $\pQ_t$, $\pH_t$, and $\pPsi_t$ as well. 
	
	\paragraph{Example 1i: Generalized least squares parameterization.}

	Here, a researcher specifies the scale matrix $\pU_{\pY,t}$ as a set of estimable parameters, and $\pJ_t^*$ is simply (\ref{Jstar GLS}), following \cite{JungbackerKoopman(15)}. To complete the transformation we need a complement $\pJ_t^+$ satisfying $\pJ_t^*\pU_{\pY,t}\pJ_t^{+,\top}=\pzero$ with $\Lp \pJ_t^{*,\top}, \pJ_t^{+,\top} \Rp$ full rank. Any such choice works, and we fix the scale by imposing $\left| \pJ_t^{+} \pU_{\pY,t} \pJ_t^{+,\top} \right| = 1$. The state's own scale matrix $\pU$ is not estimated separately. It is implied by $\pW_t$ and $\pU_{\pY,t}$ through $\pU = \Lp\pW_t^\top \pU_{\pY,t}^{-1} \pW_t\Rp^{-1}$, as is $\pPsi_{t} = \pJ_t^{+} \pU_{\pY,t} \pJ_t^{+,\top}$. Identification requires restricting $\pW_t^\top \pU_{\pY,t}^{-1} \pW_t$, typically to be diagonal with ordered entries, which removes rotational indeterminacy in the column space.

	\paragraph{Example 2i: Orthogonal parameterization.}

	A simpler alternative is to restrict $\pW_t$ itself to have orthonormal columns, $\pW_t^\top \pW_t = \eye_r$, at the cost of imposing structure on $\pU_{\pY,t}$. Rather than estimating $\pU_{\pY,t}$ directly, a researcher estimates $\pU$ and $\pPsi_t$ separately, and $\pU_{\pY,t}$ is built from them. Let $\pW_{t,\perp}$ be an orthonormal basis for the complement of $\pW_t$. Setting $\pJ_t^{*,\top} = \pW_t$ and $\pJ_t^{+,\top} = \pW_{t,\perp}$ satisfies conditions (\ref{condition 1})--(\ref{condition 3}) whenever the right scale matrix takes the form
	\begin{eqnarray}
		\pU_{\pY,t} &=& \pW_t \pU \pW_t^\top + \pW_{t,\perp} \pPsi_t \pW_{t,\perp}^\top, \label{U param}
	\end{eqnarray}
	with both $\pU$ and $\pW_{t,\perp} \pPsi_t \pW_{t,\perp}^\top$ estimated freely. Identification is achieved by imposing that $\pU$ is diagonal with ordered entries, eliminating rotational indeterminacy.

	Example 2i isn't a different rule for building $\pJ_t$. It is (\ref{Jstar GLS}) evaluated at a value of $\pU_{\pY,t}$ that happens to be block diagonal in the $\Lp\pW_t,\pW_{t,\perp}\Rp$ basis.

	\section{Algorithms} \label{MKFS}
	
 Throughout this section, we assume $\pJ_t$ is full rank and satisfies conditions (\ref{condition 1})-(\ref{condition 3}). All proofs are contained in the online appendix.

	\subsection{Filtering and one-step ahead prediction}
	

	\begin{proposition}
		Consider the state space model (\ref{obsrvation equation})-(\ref{initial condition}). For each date $t=1,\ldots,T$, the filtering and one-step ahead predictive distributions are matrix normal
		\[ p\Lp \pA_{t}|\pY_{1:t},\ptheta\Rp \ = \ \text{MN}\Lp \pA_{t|t},\pP_{t|t},\pU\Rp \qquad  	p\Lp \pA_{t+1}|\pY_{1:t},\ptheta\Rp \ = \ \text{MN}\Lp \pA_{t+1|t},\pP_{t+1|t},\pU\Rp\]
		with parameter matrices that can be calculated recursively forwards in time as
		\begin{eqnarray}
			\pV_{t} &=& \pY_{t}\pJ_{t}^{*,\top} - \pD_{t}\pJ_{t}^{*,\top} - \pZ_{t}\pA_{t|t-1}, \label{KF: V} \\
			\pF_{t} &=&  \pZ_{t}\pP_{t|t-1} \pZ_{t}^{\top} + \pH_{t}, \\
			\pK_{t} &=& \pP_{t|t-1}\pZ_{t}^{\top}\pF_{t}^{-1}, \label{Kalman gain} \\
			\pA_{t|t} & = &  \pA_{t|t-1} + \pK_{t}\pV_{t},  \\
			\pP_{t|t} & = & \pP_{t|t-1} -  \pK_{t} \pZ_{t}\pP_{t|t-1}, \\
			\pA_{t+1|t} &=& \pT_{t}\pA_{t|t} + \pC_{t}, \\
			\pP_{t+1|t} &=& \pT_{t}\pP_{t|t}\pT_{t}^{\top} + \pR_{t}\pQ_{t}\pR_{t}^{\top}. \label{Pred} 
		\end{eqnarray}
	\end{proposition}
	\vskip -0.5cm
	These recursions are the same as the standard Kalman filter but where $\pV_{t}, \pA_{t|t}$ and $\pA_{t|t-1}$ are matrix valued instead of vectors. 
	The matrix $\pU$ does not enter the recursions but it does enter the likelihood function, see sub-section \ref{loglikelihood}.
	
	In many empirical studies, the Kalman filter is primarily used to evaluate the likelihood or for simulation smoothing. In this case, the prediction form of the Kalman filter is more computationally efficient because it avoids calculating the filtered values. It redefines the matrix $\pK_{t} \ = \ \pT_{t}\pP_{t|t-1}\pZ_{t}^{\top}\pF_{t}^{-1}$ and replaces (\ref{Kalman gain})-(\ref{Pred}) with
	\[ \pA_{t+1|t} \ = \ \pT_{t}\pA_{t|t-1} + \pC_{t} + \pK_{t}\pV_{t}, \qquad 
	\pP_{t+1|t} \ = \ \pT_{t}\pP_{t|t-1}\pL_{t}^{\top} + \pR_{t}\pQ_{t}\pR_{t}^{\top}, \]
	where $\pL_{t} = \pT_{t} - \pK_{t}\pZ_{t}$. 
	
	\subsection{Computational savings: Kronecker product assumption}
	
		Time series models that fit into the state space representation  (\ref{obsrvation equation})-(\ref{initial condition}) are admittedly restricted because they assume a (conditional) Kronecker structure on their covariance matrix. This structure yields substantial computational savings. A naive vectorized Kalman filter must invert an $mn\times mn$ matrix at every date, which is $O\Lp(mn)^{3}\Rp$. A sufficiently sophisticated vector implementation could in principle recover some of this cost via Woodbury-type updates, but existing Bayesian matrix-factor implementations do not attempt this and instead vectorize directly (\cite{ChanZhang(24)}, \cite{BarigozziTrapin(25)}). Because the recursions $\pA_{t|t},\pP_{t|t}$ of Proposition 1 depend only on $\pZ_t,\pH_t,\pF_t$ and never $\pU$, the matrix Kalman filter instead inverts only the $m\times m$ matrix $\pF_t$ which is $O\Lp m^{3}\Rp$ and independent of $n$ and $r$. The matrix $\pU$ enters only the log-likelihood, requiring a single, cheap $O\Lp r^{3}\Rp$ inversion. In our application ($m=12$, $n=50$, $r=11$), this is $1{,}728$ versus $2.16\times10^{8}$ operations per date. Using the approach of \cite{JungbackerKoopman(15)} to collapse the row dimension as well (see the online appendix) reduces the cost further to $O\Lp s^{3}\Rp$ when $\pZ_t$ has rank $s<m$. The same principle - exploiting a Kronecker-separable covariance matrix to avoid a large joint matrix inversion -- underlies the natural-conjugate priors used to make large Bayesian VARs computationally tractable; see \cite{CarrieroClarkMarcellino(16)}, \cite{CarrieroClarkMarcellino(19)}, and \cite{Chan(20)}.
		
	\subsection{Log-likelihood function} \label{loglikelihood}
	
	We now derive the log-likelihood for the model, when $\pJ_{t}$ rotates the data from $\pY_{t}$ to $\Lb \pY_{t}^{*} \ \pY_{t}^{+}\Rb$ leaving observation equations (\ref{obs equation 1})-(\ref{obs equation 2}).

	\begin{proposition} \label{proposition loglikelihood}
	The log-likelihood of the state space model (\ref{obsrvation equation})-(\ref{initial condition}) is
	\begin{eqnarray*}
		\log p\Lp \pY_{1:T}|\ptheta\Rp &=& \log p\Lp \pY_{1:T}^{*}|\ptheta\Rp + \log p\Lp \pY_{1:T}^{+}|\ptheta\Rp + m\sum_{t=1}^{T}\log\Lv \pJ_{t}\Rv,
	\end{eqnarray*}
	where $\Lv \pJ_{t}\Rv$ is the Jacobian of the transformation from $\pY_{t}$ to $\Lp \pY_{t}^{*,\top},\pY_{t}^{+,\top}\Rp^{\top}$. The log-likelihood contribution for the transformed data $\pY_{t}^{*}$ is
	\begin{eqnarray*}
		\log p\Lp \pY_{1:T}^* \mid \ptheta\Rp
		&=& -\frac{Tmr}{2} \log(2\pi)
		- \frac{r}{2} \sum_{t=1}^T \log |\pF_t|
		- \frac{mT}{2} \log |\pU| \\
		& &
		- \frac{1}{2} \sum_{t=1}^T \mathrm{tr}
		\left(
		\pF_t^{-1} \pV_t \pU^{-1} \pV_t^{\top}
		\right),
	\end{eqnarray*}
	with $\pV_t$ and $\pF_t$ obtained from the matrix Kalman filter applied to $\pY_t^*$, and the log-likelihood contribution of $\pY_{t}^+$ is
	\begin{eqnarray*}
		\log p\Lp \pY_{1:T}^+ \mid \ptheta\Rp
		&=& -\frac{Tm(n-r)}{2} \log(2\pi)
		- \frac{n-r}{2} \sum_{t=1}^T \log \Lv \pH_t\Rv
		- \frac{m}{2}\sum_{t=1}^{T}\log |\pPsi_{t}|\\
		& & - \frac{1}{2} \sum_{t=1}^T \mathrm{tr}
		\left(
		\pH_t^{-1} \Lb\pY_t^+ - \pD_{t}^{+}\Rb\pPsi_{t}^{-1}\Lb\pY_t^{+}-\pD_{t}^{+}\Rb^{\top}
		\right).
	\end{eqnarray*}
	 The decomposition is also invariant to the choice of basis for the complement $\pJ_t^+$. The log-likelihood does not depend on which valid $\pJ_t^+$ is used.
	\end{proposition}
	Although the log-likelihood is invariant, $\pPsi_t = \pJ_t^+\pU_{\pY,t}\pJ_t^{+,\top}$ itself is not. Replacing $\pJ_t^+$ with $\pB\pJ_t^+$ for any nonsingular $\pB$ changes $\pPsi_t$ to $\pB\pPsi_t\pB^\top$, with the resulting change in the likelihood contribution of $\pY_t^+$ exactly offset by the change in the Jacobian term (see the online appendix).
	When $n=r$, no collapsing is needed, $\pW_t=\eye_n$, $\pJ_t=\eye_n$, and $\pY_t^+$ is an empty matrix. Both the $\pY_t^+$ term and the Jacobian term vanish, and the log-likelihood reduces to the $\pY_t^*$ expression above with $r$ replaced by $n$.
	The Jacobian $\Lv \pJ_t \Rv$ depends on the choice of transformation $\pJ_t$. 
	In the GLS parameterization (Example 1i), the transformation depends on $\pW_t$ and $\pU_{\pY,t}$. Under the normalization $\left| \pJ_t^{+} \pU_{\pY,t} \pJ_t^{+,\top} \right| = 1$, the Jacobian is given by
	$
	\Lv \pJ_t \Rv = |\pU_{\pY,t}|^{-1/2} |\pU|^{1/2},
	$ where $\pU = (\pW_t^\top \pU_{\pY,t}^{-1} \pW_t)^{-1}$. In the orthogonal parameterization (Example 2i), $\pJ_t$ is orthonormal so that $\Lv \pJ_t \Rv = 1$.

	\subsection{Smoothing}

	The marginal smoothing distribution $p\Lp \pA_{t}|\pY_{1:T},\ptheta\Rp$ provides information about the state matrix $\pA_{t}$ conditional on all of the data $\pY_{1:T}$. It can be calculated recursively backwards through time after running the Kalman prediction recursions forwards and storing the matrices $\pV_{t}, \pZ^{\top}\pF_{t}^{-1}, \pL_{t}, \pA_{t|t-1}$ and $\pP_{t|t-1}$ for each date $t$. 
	
	\begin{proposition}
		Consider the state space model (\ref{obsrvation equation})-(\ref{initial condition}). For $t=T,\ldots,1$, the marginal smoothing distribution is matrix normal $p\Lp \pA_{t}|\pY_{1:T},\ptheta\Rp \ = \ \text{MN}\Lp \pA_{t|T},\pP_{t|T},\pU\Rp$.
		The mean matrix $\pA_{t|T} $ and scale matrix $\pP_{t|T}$ can be calculated recursively backwards 
		\begin{eqnarray}
			\pG_{t-1} \ = \  \pZ_{t}^{\top}\pF_{t}^{-1}\pV_{t} + \pL_{t}^{\top}\pG_{t}, & & 	\pA_{t|T} \ = \ \pA_{t|t-1} + \pP_{t|t-1}\pG_{t-1}, \label{state smooth 1} \\
			\pN_{t-1} \ = \ \pZ_{t}^{\top}\pF_{t}^{-1}\pZ_{t} + \pL_{t}^{\top}\pN_{t}\pL_{t}, & & \pP_{t|T} \ = \ \pP_{t|t-1} - \pP_{t|t-1}\pN_{t-1}\pP_{t|t-1}, \label{state smooth 2}
		\end{eqnarray}
		for $t=T,\ldots,1$ with initial values $\pG_{T} = \pzero_{s \times r}$ and $\pN_{T} = \pzero_{s\times s}$. 
	\end{proposition}
	\vskip -0.5cm
	There are several different forms of the Kalman smoother in the literature. This is a matrix version of the method developed by \cite{deJong(89)}. 

	\subsection{Drawing from the joint smoothing distribution} \label{simulation smoothing}
	
	\subsubsection{Simulation smoothing}
	
	Simulation smoothing algorithms are methods for drawing from the joint posterior distribution $p\Lp \pA_{1:T}|\pY_{1:T},\ptheta\Rp$. Simulation smoothing algorithms for linear, Gaussian state space models were originally developed by \cite{CarterKohn(94)}, \cite{FruhwirthSchnatter(94)}, \cite{deJongShephard(95)}, and \cite{DurbinKoopman(02)}. 
	The following algorithm is a matrix generalization of \cite{DurbinKoopman(02)}.
	
	\begin{proposition}
		Consider the state space model (\ref{obsrvation equation})-(\ref{initial condition}). A random draw $\pA_{1}^{d}, \ldots,\pA_{T}^{d}$ from $p\Lp \pA_{1:T}|\pY_{1:T},\ptheta\Rp$
		can be obtained using the following algorithm
		\begin{enumerate}
			\item For $t=1,\ldots,T$, simulate a new state matrix $\pA_{t}^{\dagger}$ and data series $\pY_{t}^{*\dagger}$ from the model (\ref{transition equation})-(\ref{initial condition}) and (\ref{obs equation 1}). When running this step, all mean terms are set to zero, i.e. $\pD_{t}^{*} = \pzero$ and $\pC_{t} = \pzero$, including the initial state $\pA_{1}^{\dagger} \sim \text{MN}\Lp \pzero,\pP_{1|0},\pU\Rp$.
			
			\item Construct the artificial data series $\widehat{\pY}_{t}^{*} = \pY_{t}^{*} - \pY_{t}^{*\dagger}$. Run the matrix Kalman filter and smoother on the data series $\widehat{\pY}_{t}^{*}$, storing the smoothed estimates $\widehat{\pA}_{t|T}$.
			
			\item For $t=1,\ldots,T$, calculate $\pA_{t}^{d} = \widehat{\pA}_{t|T} + \pA_{t}^{\dagger}$. 
		\end{enumerate}
	\end{proposition}
	In step 2, only equations (\ref{state smooth 1}) need calculated during the backwards pass of the Kalman smoother. This algorithm works well for most models and has the benefit that minimal changes are needed when going from one model to another.
	
	\subsubsection{Precision sampler}
	
	\cite{ChanJeliazkov(09)} proposed an algorithm for drawing from the joint smoothing distribution in a linear, Gaussian state space model that does not use the Kalman filter. It can be faster for some models because it takes the Cholesky decomposition of the posterior precision matrix which is a lower-triangular, banded matrix instead of the covariance matrix. 
	
	To implement the precision sampler, we need to re-define the state space model for the transformed data $\pY_{t}^{*} = \pY_{t}\pJ_{t}^{*,\top}$ without using the companion form. In this subsection, we use tilde's above the matrices to differentiate the notation. Let $\widetilde{\pA}_{t}$ be a $\widetilde{s} \times r$ latent state with autoregressive dynamics of order $p$ given by
	\begin{eqnarray*}
		\pY_{t}^{*} &=& \widetilde{\pD}_{t}^{*} + \widetilde{\pZ}_{t}\widetilde{\pA}_{t} + \widetilde{\pE}_{1t}, \qquad \widetilde{\pE}_{1t} \sim \text{MN}\Lp \pzero,\pH_{t},\pU\Rp, \\
		\widetilde{\pA}_{t} &=& \widetilde{\pC}_{t} + \widetilde{\pT}_{1}\widetilde{\pA}_{t-1} + \ldots +  \widetilde{\pT}_{p}\widetilde{\pA}_{t-p} + \widetilde{\pE}_{2t}, \qquad \widetilde{\pE}_{2t} \sim \text{MN}\Lp \pzero,\widetilde{\pQ}_{t},\pU\Rp, 
	\end{eqnarray*}
	with initial condition $\widetilde{\pA}_{1} \sim \text{MN}\Lp \widetilde{\pA}_{1|0},\widetilde{\pP}_{1|0},\pU\Rp$. The $r \times r$ matrix $\pU$ and the $m \times m$ matrix $\pH_{t}$ are the same as before. The dimension of all other system matrices are adjusted. The transition matrices $\widetilde{\pT}_{j}$ are also assumed to be constant over time.
	
	Next, we stack the observation equations together
	and define the following matrices
	\[ \underset{Tm \times r}{\widetilde{\pY}^{*}} \ = \ \Lb \begin{matrix}
		\pY_{1}^{*} \\
		\vdots \\
		\pY_{T}^{*}
	\end{matrix}\Rb \qquad \underset{Tm \times Tr}{\widetilde{\pZ}} \ = \ \Lb \begin{matrix}
		\widetilde{\pZ}_{1}  & \ldots & \pzero  \\
		\vdots & \ddots & \vdots \\
		\pzero & \ldots &\widetilde{\pZ}_{T} 
	\end{matrix}\Rb \qquad \underset{Tm \times Tm}{\widetilde{\pH}} \ = \ \Lb \begin{matrix}
		\pH_{1} & \ldots & \pzero  \\
		\vdots & \ddots & \vdots \\
		\pzero & \ldots & \pH_{T} \\
	\end{matrix}\Rb  \]
	with $\widetilde{\pD}^{*} = \Lp \widetilde{\pD}_{1}^{*\top},\widetilde{\pD}_{2}^{*\top},\ldots,\widetilde{\pD}_{T}^{*\top}\Rp^{\top}$.
	We also stack the state matrices into a $T \tilde{s} \times r$ matrix $\widetilde{\pA} = \Lp \widetilde{\pA}_{1}^{\top},\widetilde{\pA}_{2}^{\top},\ldots,\widetilde{\pA}_{T}^{\top}\Rp^{\top}$ and define
	\[ \underset{T\tilde{s} \times T\tilde{s}}{\widetilde{\pT}} \ =\  \Lb \begin{matrix}
		\eye & \pzero  & \ldots & \ldots & \ldots & \pzero  \\
		-\widetilde{\pT}_{1}  & \eye & \pzero & \ldots & \ldots & \vdots  \\
		-\widetilde{\pT}_{2}  & -\widetilde{\pT}_{1} & \eye  & \ldots & \ldots & \vdots  \\
		\vdots &  \vdots &  \vdots & & \ddots & \pzero \\
		\pzero & \ldots  & -\widetilde{\pT}_{p} & \ldots & -\widetilde{\pT}_{1} & \eye \\
	\end{matrix}\Rb \ \  
	\underset{T\tilde{s} \times T\tilde{s}}{\widetilde{\pQ}} \ = \ \Lb \begin{matrix}
		\widetilde{\pP}_{1|0} & \pzero & \ldots & \pzero  \\
		\pzero & \widetilde{\pQ}_{2} & \ldots & \pzero  \\
		\vdots & \vdots & \ddots & \vdots \\
		\pzero & \pzero & \ldots & \widetilde{\pQ}_{T} \\
	\end{matrix}\Rb  \]
	with $\widetilde{\pC} = \Lp \widetilde{\pA}_{1|0}^{\top},\widetilde{\pC}_{2}^{\top},\ldots,\widetilde{\pC}_{T}^{\top}\Rp^{\top}$. Moving average terms can be incorporated by making $\widetilde{\pQ}$ a banded matrix, though the bandwidth grows with the MA order and erodes the sparsity advantage over the simulation smoother.
	
	Using this notation, the model can be written in stacked form as
	\begin{eqnarray*}
		\widetilde{\pY}^{*} &=& \widetilde{\pD}^{*} + \widetilde{\pZ}\widetilde{\pA} + \widetilde{\pE}_{1}^{*}, \qquad \widetilde{\pE}_{1}^{*} \sim \text{MN}\Lp \pzero,\widetilde{\pH},\pU\Rp, \\
		\widetilde{\pA} &=& \widetilde{\pT}^{-1}\widetilde{\pC} + \widetilde{\pE}_{2}, \qquad \widetilde{\pE}_{2} \sim \text{MN}\Lp \pzero,\widetilde{\pT}^{-1}\widetilde{\pQ}\widetilde{\pT}^{-1,\top},\pU\Rp.
	\end{eqnarray*}
	We can interpret the first equation as the likelihood and the second equation as the prior. 
	Using Bayes rule, the posterior distribution for the stacked state matrix is $\widetilde{\pA} \sim \text{MN}\Lp \widetilde{\pM},\widetilde{\pSigma},\pU\Rp$ with  $T \tilde{s} \times T\tilde{s}$ precision matrix $\widetilde{\pSigma}^{-1}  =  \widetilde{\pT}^{\top}\widetilde{\pQ}^{-1}\widetilde{\pT} + \widetilde{\pZ}^{\top}\widetilde{\pH}^{-1}\widetilde{\pZ}$.
	and $T \tilde{s} \times r$ posterior mean matrix $\widetilde{\pM} = 	\widetilde{\pSigma}\Lp \widetilde{\pT}^{\top}\widetilde{\pQ}^{-1}\widetilde{\pC} + \widetilde{\pZ}^{\top}\widetilde{\pH}^{-1}\Lb\widetilde{\pY}^{*} - \widetilde{\pD}^{*}\Rb\Rp$.
	
	To draw from this distribution efficiently, we take the following steps
	\begin{enumerate}
		\item Calculate the matrix $\widetilde{\pSigma}^{-1}$ and its Cholesky decomposition $\widetilde{\pSigma}^{-1} = \pL_{\sigma}\pL_{\sigma}^{\top}$.
		\item Calculate the Cholesky decomposition of $\pU = \pL_{\pU}\pL_{\pU}^{\top}$.
		\item Calculate the matrix $\widetilde{\pG}= \pL_{\sigma} \backslash \Lp \widetilde{\pT}^{\top}\widetilde{\pQ}^{-1}\widetilde{\pC} + \widetilde{\pZ}^{\top}\widetilde{\pH}^{-1}\Lb\widetilde{\pY}^{*} - \widetilde{\pD}^{*}\Rb\Rp / \pL_{\pU}^{\top}$ using forward substitution with $\pL_{\sigma}$ and backward substitution with $\pL_{\pU}^{\top}$. Here, $\pA \backslash \pB$ and $\pB / \pA$ denote the solutions $\pX$ of $\pA\pX = \pB$ and $\pX\pA = \pB$, respectively, computed by triangular solves rather than by forming $\pA^{-1}$ explicitly.
		
		\item A random $T\widetilde{s} \times r$ matrix $\pA^{d}$ drawn from the joint distribution is then obtained by drawing $\widetilde{\pE} \sim \text{MN}\Lp \pzero,\eye_{T\widetilde{s}},\eye_{r}\Rp$ and  calculating $\pA^{d} = \Lp \pL_{\sigma}^{\top}\backslash \Lb \widetilde{\pG} + \widetilde{\pE}\Rb \Rp \pL_{\pU}^{\top}.$
	\end{enumerate}
	In our experience, the precision sampler is typically faster than the simulation smoother for autogressive models and simple time series models. However, the simulation smoother is better for estimating unknown constant parameters in the matrices $\pC_{t}$ and $\pD_{t}$. 
	
	\subsection{Estimation of regression parameters}
	
	Let $\pX_t$ denote a matrix of observed covariates. Covariates can be incorporated by specifying the intercept matrices $\pD_t$ and/or $\pC_t$ in (\ref{obsrvation equation})--(\ref{transition equation}) as functions of $\pX_t$. For example, one may set $\pC_t = \pX_t \pbeta$, where $\pX_t$ is $s \times \ell$ and $\pbeta$ is an $\ell \times r$ matrix of unknown parameters. For Bayesian analysis, a conjugate prior can be specified as $\pbeta \sim \text{MN}(\underline{\pM}, \underline{\pP}, \pU)$ and $\pbeta$ can be drawn jointly with the latent states using the simulation smoother. For maximum likelihood estimation, a profile likelihood or concentrated estimator for $\pbeta$ can be derived analogously to \cite{deJong(91)}. 
	
	\subsection{Missing values} \label{Missing values}
	
	In empirical work, time series data may have missing values. The state space model (\ref{obsrvation equation})-(\ref{initial condition}) can accommodate missing values as long as the entire row or column vector of $\pY_{t}$ is treated as missing. While potentially restrictive, this is still empirically relevant for researchers working with mixed frequency data (quarterly, monthly, etc.) whose release dates are the same for all spatial units; e.g. government agencies often release economic data with this pattern.  Mixed frequency data is popular in economics; see, e.g. \cite{MarianoMurasawa(03)}, and \cite{SchorfheideSong(15)}.
	
	
	To handle missing values, we define an $m_{t} \times m$ matrix $\pS_{t}$ that selects out rows of the observed data at each date.\footnote{An entirely missing column can be handled symmetrically, by right-multiplying $\pY_t$ by an analogous $n_t\times n$ column-selection matrix $\pS_t^{c}$; this transforms $\pW_t\to\pS_t^{c}\pW_t$ and the column scale $\pU_{\pY,t}\to\pS_t^{c}\pU_{\pY,t}\pS_t^{c,\top}$, leaving $\pH_t$ unchanged.} We then left-multiply the original observation equation (\ref{obsrvation equation}) by $\pS_{t}$ to get  $\pY_{t}^{-} \ = \ \pD_{t}^{-} + \pZ_{t}^{-}\pA_{t}\pW_{t}^{\top} + \pE_{1t}^{-}$, with $\pE_{1t}^{-} \sim \text{MN}\Lp \pzero,\pH_{t}^{-},\pU_{\pY}\Rp$.
	The adjusted matrices are defined as $\pY_{t}^{-}  =  \pS_{t}\pY_{t}$, $\pD_{t}^{-} =  \pS_{t}\pD_{t}$, $\pZ_{t}^{-}  =  \pS_{t}\pZ_{t}$, and $\pH_{t}^{-} \ = \ \pS_{t}\pH_{t}\pS_{t}^{\top}$.
	The Kalman filter and associated algorithms can then be applied as usual but with system matrices whose dimension are changing over time.

	Entry-level (non-row-wise) missingness can also be handled by data augmentation, drawing the missing entries from their conditional Gaussian distribution within the Gibbs sampler and treating them as observed thereafter. This is the approach we use for the ragged-edge missing monthly observations in the empirical application (Section \ref{Gibbs sampler}). The frequentist analogue is the EM algorithm, which replaces each missing entry with its conditional expectation in the E-step. Both accommodate more general missingness patterns than row-selection. 
	
	\subsection{Efficient sampling of discrete indicators in Markov switching and mixture models} \label{Sampling discrete states}
	
	To add flexibility to the model (\ref{obsrvation equation})-(\ref{initial condition}), researchers often allow the parameters within the system matrices to be time-varying. A common approach is to make the parameters a function of a finite mixture or Markov-switching variable; see, e.g. \cite{Kim(94)}, \cite{KimNelsonBook(99)} and \cite{GiordaniKohnvanDijk(07)}. Let $s_{t}$ denote an indicator variable that can take on a finite number of values $s_{t}=j$ for $j = 1,\ldots,J$. Each discrete state $s_{t}=j$ corresponds to a different set of parameters. The system matrices are functions of the discrete state
	\begin{eqnarray}
		\pY_{t} & = & \pD\Lp s_{t}\Rp + \pZ\Lp s_{t}\Rp\pA_{t}\pW_{t}^{\top} + \pE_{1t}, \qquad \pE_{1t} \sim \text{MN}\Lp \pzero,\pH\Lp s_{t}\Rp,\pU_{\pY,t}\Rp, \label{obsrvation equation MS} \\
		\pA_{t+1} &=& \pC\Lp s_{t+1}\Rp + \pT\Lp s_{t+1}\Rp\pA_{t} + \pR\Lp s_{t+1}\Rp\pE_{2t}, \qquad \pE_{2t} \sim \text{MN}\Lp 0,\pQ\Lp s_{t+1}\Rp,\pU\Rp, \label{transition equation MS}
	\end{eqnarray}
	where $\pA_{1} \sim \text{MN}\Lp \pA_{1|0}(s_{1}),\pP_{1|0}\Lp s_{1}\Rp,\pU\Rp$.  One can generalize this further to make the system matrices a function of a vector of discrete states.\footnote{This extension is straightforward but notationally more complex. The algorithm for drawing the discrete states does not fundamentally change.} 
	
	\cite{GerlachCarterKohn(00)} and \cite{DoucetAndrieu(01)} develop algorithms for drawing from the conditional distribution $p \Lp s_{t}|\ps_{-t},\pY_{1:T},\ptheta\Rp \propto p\Lp \pY_{1:T}|\ps_{-t},s_{t},\ptheta\Rp p\Lp s_{t}|\ps_{-t},\ptheta\Rp$ of a single discrete state but with the continuous states integrated out. The notation $\ps_{-t}$ is a vector of all indicator variables but with $s_{t}$ omitted. For Bayesian estimation with MCMC, this improves mixing of the Markov chain because it is a collapsed Gibbs sampler. We extend their approach to the matrix state space model where the key result is the following proposition. 
	
	\begin{proposition} \label{proposition discrete state}
		Consider the state space model (\ref{obsrvation equation MS})-(\ref{transition equation MS}). The likelihood of $\pY_{1:T}$ conditional on the indicators $s_{1:T}$ is given up to proportionality by
		\begin{eqnarray*}	
			p\Lp \pY_{1:T}|s_{t},\ps_{-t},\ptheta\Rp 
			&\propto & \Lv \pF_{t}\Lp s_{1:t}\Rp\Rv^{-\frac{r}{2}}  \Lv \pP_{t|t}\Lp s_{1:t}\Rp\Rv^{-\frac{r}{2}} \Lv \pP_{t|T}\Lp s_{1:T}\Rp\Rv^{\frac{r}{2}}\Lv\pH\Lp s_{t}\Rp\Rv^{-\frac{(n-r)}{2}}\\
			& & \exp\Lp -\frac{1}{2}\text{tr}\Lb \pU^{-1}\pV_{t}\Lp s_{1:t}\Rp^{\top}\pF_{t}\Lp s_{1:t}\Rp^{-1}\pV_{t}\Lp s_{1:t}\Rp\Rb \Rp \\
			& & \exp\Lp -\frac{1}{2}\text{tr}\Lb
			\pH\Lp s_{t}\Rp^{-1} \Lb\pY_t^+ - \pD\Lp s_{t}\Rp^{+}\Rb\pPsi_{t}^{-1}\Lb\pY_t^{+}-\pD\Lp s_{t}\Rp^{+}\Rb^{\top}
			\Rb\Rp 
			\\
			& & \exp\Lp-\frac{1}{2}\text{tr}\Lb \pU^{-1}\pA_{t|t}\Lp s_{1:t}\Rp^{\top}\pP_{t|t}\Lp s_{1:t}\Rp^{-1}\pA_{t|t}\Lp s_{1:t}\Rp\Rb\Rp \\
			& & \exp\Lp \frac{1}{2}\text{tr}\Lb \pU^{-1}\pA_{t|T}\Lp s_{1:T}\Rp^{\top}\pP_{t|T}\Lp s_{1:T}\Rp^{-1}\pA_{t|T}\Lp s_{1:T}\Rp\Rb\Rp 
		\end{eqnarray*}
		where the constant of proportionality does not depend on $s_{t}$. 
	\end{proposition}
	
	The key insight of \cite{GerlachCarterKohn(00)} and \cite{DoucetAndrieu(01)} is an algorithm for calculating $p\Lp \pY_{1:T}|\ps_{-t},s_{t}=j,\ptheta\Rp$ for each state $s_{t} = j$ in a computationally efficient way. First, one runs a backwards pass of the Kalman information filter conditional on a previous MCMC draw of the discrete states $s_{1:T}$. Then, the algorithm iterates forward in time drawing the discrete states $s_{t}$ for $t=1,\ldots,T$ while evaluating $p\Lp \pY_{1:T}|s_{t}=j,\ps_{-t},\ptheta\Rp$ for each state $s_{t}=j$ for $j=1,\ldots,J$.
	
	The algorithm for drawing the discrete states $s_{t}$ from their conditional distribution $p \Lp s_{t}|\ps_{-t},\pY_{1:T},\ptheta\Rp$ sequentially forward for $t=1,\ldots,T$  is
	\begin{description}
		\item[1.] Given the current set of indicators $s_{1:T}$ from a previous MCMC iteration, run the backwards information filter. Initialize $\pB_{T|T} = \pzero$ and $\pPi_{T|T} = \pzero$. And, for $t=T-1
		\ldots,1$, calculate
		\begin{eqnarray*}
			\underline{\pPi}_{t+1|T} &=& \pPi_{t+1|T} + \pZ\Lp s_{t+1}\Rp ^{\top}\pH\Lp s_{t+1}\Rp^{-1}\pZ\Lp s_{t+1}\Rp  \\
			\underline{\pB}_{t+1|T}& = & \pB_{t+1|T} + \pZ\Lp s_{t+1}\Rp ^{\top}\pH\Lp s_{t+1}\Rp ^{-1}\Lp\pY_{t+1} - \pD\Lp s_{t+1}\Rp \Rp \pJ_{t+1}^{*,\top}  \\
			\pDelta_{t+1} &=& \eye_{s} + \underline{\pPi}_{t+1|T}\pR\Lp s_{t+1}\Rp \pQ\Lp s_{t+1}\Rp \pR\Lp s_{t+1}\Rp ^{\top} \\
			\pPi_{t|T}\Lp \ps_{t+1:T}\Rp &=&  \pT\Lp s_{t+1}\Rp ^{\top}\pDelta_{t+1}^{-1}\underline{\pPi}_{t+1|T}\pT\Lp s_{t+1}\Rp  \\
			\pB_{t|T}\Lp \ps_{t+1:T}\Rp &=& \pT\Lp s_{t+1}\Rp ^{\top}\pDelta_{t+1}^{-1}\Lp\underline{\pB}_{t+1|T}-\underline{\pPi}_{t+1|T}\pC\Lp s_{t+1}\Rp \Rp 
		\end{eqnarray*}
		During the backwards pass, store $\pB_{t|T}\Lp \ps_{t+1:T}\Rp $ and $\pPi_{t|T}\Lp \ps_{t+1:T}\Rp$.
		\vskip 0.1cm
		
		
		\item[2.] For $t=1,\ldots,T$ compute for each state $s_{t}=j$ for $j=1,\ldots,J$ at each date $t$
		\begin{description}
			\item[(2a.)] For $t=1$, the initial condition is $\pA_{1|0}\Lp s_{1}=j\Rp$ and $\pP_{1|0}\Lp s_{1}=j\Rp$. For $t=2,\ldots,T$, assume that we have already sampled $\ps_{1:t-1}$ and calculated $\pA_{t-1|t-1}\Lp\ps_{1:t-1}\Rp$ and $\pP_{t-1|t-1}\Lp \ps_{1:t-1}\Rp$ at a previous iteration. Then, we calculate the one-step ahead predictive distribution for each $s_{t} = j$
			\begin{eqnarray*}
				\pA_{t|t-1}\Lp \ps_{1:t}\Rp &=& \pT\Lp s_{t}\Rp\pA_{t-1|t-1}\Lp \ps_{1:t-1}\Rp + \pC\Lp s_{t}\Rp \\
				\pP_{t|t-1}\Lp \ps_{1:t}\Rp &=& \pT\Lp s_{t}\Rp\pP_{t-1|t-1}\Lp \ps_{1:t-1}\Rp\pT\Lp s_{t}\Rp^{\top} + \pR\Lp s_{t}\Rp\pQ\Lp s_{t}\Rp\pR\Lp s_{t}\Rp^{\top}
			\end{eqnarray*}

			\item[(2b.)] Calculate the prediction error and prediction error variance
			\begin{eqnarray*}
				\pV_{t}\Lp \ps_{1:t}\Rp &=& \pY_{t}^{*}-\pD_{t}\Lp \ps_{t}\Rp  \pJ_{t}^{*,\top}  - \pZ\Lp s_{t}\Rp\pA_{t|t-1}\Lp s_{1:t}\Rp\\
				\pF_{t}\Lp \ps_{1:t}\Rp &=&  \pZ\Lp s_{t}\Rp\pP_{t|t-1}\Lp s_{1:t}\Rp \pZ\Lp s_{t}\Rp^{\top} + \pH\Lp s_{t}\Rp
			\end{eqnarray*}
			and the filter
			\begin{eqnarray*}			
				\pK_{t}\Lp \ps_{1:t}\Rp &=& \pP_{t|t-1}\Lp \ps_{1:t}\Rp\pZ\Lp s_{t}\Rp^{\top}\pF_{t}\Lp \ps_{1:t}\Rp^{-1} \\
				\pA_{t|t}\Lp \ps_{1:t}\Rp & = &  \pA_{t|t-1}\Lp \ps_{1:t}\Rp + \pK_{t}\Lp \ps_{1:t}\Rp\pV_{t}\Lp \ps_{1:t}\Rp  \\
				\pP_{t|t}\Lp \ps_{1:t}\Rp & = & \pP_{t|t-1}\Lp \ps_{1:t}\Rp -  \pK_{t}\Lp \ps_{1:t}\Rp \pZ\Lp s_{t}\Rp\pP_{t|t-1}\Lp \ps_{1:t}\Rp
			\end{eqnarray*}
			
			\item[(2c.)] Combine the output of the backward information filter and the forward Kalman filter to calculate the smoothed estimate for each $s_{t} = j$
			\begin{eqnarray*}
				\pGamma_{t}\Lp \ps_{1:T}\Rp &=&   \eye_{s} + \pPi_{t|T}\Lp s_{t+1|T}\Rp\pP_{t|t}\Lp \ps_{1:t}\Rp \\
				\pA_{t|T}\Lp \ps_{1:T}\Rp &=& \pA_{t|t}\Lp \ps_{1:t}\Rp + \pP_{t|t}\Lp \ps_{1:t}\Rp \pGamma_{t}\Lp \ps_{1:T}\Rp^{-1}\Lb \pB_{t|T}\Lp \ps_{t+1:T}\Rp \right. \\
				& & \left. - \pPi_{t|T}\Lp \ps_{t+1:T}\Rp\pA_{t|t}\Lp \ps_{1:t}\Rp \Rb \\
				\pP_{t|T}\Lp \ps_{1:T}\Rp &=& \pP_{t|t}\Lp \ps_{1:t}\Rp - \pP_{t|t}\Lp \ps_{1:t}\Rp \pGamma_{t}\Lp \ps_{1:T}\Rp^{-1}\pPi_{t|T}\Lp s_{t+1|T}\Rp\pP_{t|t}\Lp \ps_{1:t}\Rp
			\end{eqnarray*}	
			
			\item[(2d.)] Draw $s_{t}$ from the discrete distribution
			\[ p\Lp s_{t}=j|\pY_{1:T},\ps_{-t},\ptheta\Rp \propto p\Lp \pY_{1:T}|s_{t}=j,\ps_{-t},\ptheta\Rp p\Lp s_{t}=j|\ps_{-t},\ptheta\Rp.\] 
			where $p\Lp \pY_{1:T}|s_{t}=j,\ps_{-t},\ptheta\Rp$ is given in Proposition \ref{proposition discrete state}. Save $\pA_{t|t}\Lp s_{t}=j\Rp$ and $\pP_{t|t}\Lp s_{t}=j\Rp$ for the next iteration and return to step \textbf{(2a.)} if $t<T$. 
		\end{description}
	\end{description} 
	While proving Proposition \ref{proposition discrete state}, we derived a second type of smoothing algorithm for the matrix state space model known as two-filter formula smoothing. This algorithm runs the matrix Kalman filter forward in time and the matrix backwards information filter recursively backwards, combining the two filters to calculate the smoothed distribution, see e.g \cite{Mayne(66)}.
	

	\section{Application: state level macroeconomic data}
	
	\subsection{Data}
	
	In our empirical application, we estimate the matrix bilinear dynamic factor model given by (\ref{Example 2a})--(\ref{Example 2c}) on a mixed-frequency data set of $m = 12$ economic variables for $n = 50$ U.S. states. Using the FRED SD database (see \cite{BokunJacksonKliesenOwyang(24)}), we extract monthly series on state-level unemployment and employment across sectors, along with three quarterly series: nominal personal income, real gross domestic product, and the FHFA home price index. The sample spans February 1990 through August 2025, yielding $T = 427$ observations. Additional details on data construction and transformations are provided in the online appendix.
	

	\subsection{Model} \label{Model application}

	We allow for common factors across both variables and U.S. states through the loading matrices $\pLambda \in \mathbb{R}^{m \times r_{\ell}}$ and $\pW \in \mathbb{R}^{n \times r_{r}}$. We adopt the orthogonal parameterization described in Example 2i, imposing $\pLambda^{\top}\pLambda = \eye$ and $\pW^{\top}\pW = \eye$ with $\pJ_{t}^{*} = \pW$ and $\pJ_{t}^{+} = \pW_{\perp}$. To address rotational indeterminacy, the covariance matrices $\pU$ and $\pOmega$ are restricted to be diagonal with eigenvalues ordered in descending order. 
	The observation right-scale matrix $\pU_{\pY,t}$ is defined in (\ref{U param}). 
	We allow $\pPsi$ to be dense and estimate it directly.
	
	We assume a diagonal baseline structure for the left scale matrix $\pSigma$
	to reduce dimensionality. To accommodate heavy-tailed measurement errors, we multiply these row variances by series-specific Student's $t$ scale mixtures, so that the observation left scale is
	$\pSigma_t=\mathrm{diag}(h_{1t}\sigma_1^2,\ldots,h_{mt}\sigma_m^2)$.
	Here, $h_{it} \sim \text{I.G.}\Lp \nu_i/2,\nu_i/2\Rp$ are latent scale inverse gamma variables that represent the Student's $t$ distribution as a normal mixture.
	We allow for additional time variation in the state equation by specifying the left scale matrix as $\pOmega_t=\kappa_t\pOmega$, where $\kappa_t$ is a scalar two-state Markov switching process. One regime sets $\kappa_t=1$ and the other sets $\kappa_t=5$, allowing the model to capture common extreme shocks such as those observed during COVID. To identify scale, we impose $\mathrm{tr}(\pOmega)/s = 1$ for the transition covariance. The autoregressive matrices $\pPhi_1$ and $\pPhi_2$ are restricted to lie in the stationary region.
	
	We denote by $\pY_{1:T}^{o}$ the observed monthly and quarterly data. Following \cite{MarianoMurasawa(03)}, the model is specified at a monthly frequency. Quarterly observations are treated as averages of the latent monthly variables, and missing quarterly values $\pY_{1:T}^{q}$ are imputed within the MCMC algorithm.
	
	\subsection{Bayesian estimation}
	
	\subsubsection{Priors} \label{Priors}
	
	We adopt weakly informative priors that respect the structural constraints of the model and facilitate efficient posterior simulation using a Gibbs sampler with parameter expanded data augmentation (PXDA), \cite{LiuWu(99)} and \cite{MengVanDyk(99)}. Full details of the prior specification and hyperparameter choices are provided in the online appendix.
	
	The factor loadings $\pLambda$ and $\pW$ are restricted to lie on the Stiefel manifold. We therefore place matrix von Mises--Fisher priors on each, following \cite{Hoff(09)}, which enforce orthonormality while allowing for flexible prior centering. We place a matrix normal prior on the autoregressive parameters
	$\pPhi = (\pPhi_{1}, \pPhi_{2}) \sim \mathrm{MN}(\underline{\pPhi}, \underline{\pV}, \pOmega),
	$ which centers the dynamics on a stationary VAR(2) process with diagonal shrinkage across lags and factors. The covariance matrices $\pOmega$, $\pU$, and $\pPsi$ are assigned inverse Wishart priors in their parameter-expanded forms, which lead to conditionally conjugate updates within the PXDA framework.
	The observation variances $\sigma_{i}^{2}$ are assigned inverse gamma priors, $
	\sigma_{i}^{2} \sim \mathrm{IG}(\underline{a}_i, \underline{b}_i),
	$ while the initial state follows
	$
	\pA_1 \sim \mathrm{MN}(\pzero, \underline{\pP}_1, \pU).
	$
	These priors are scaled using empirical moments of the data to ensure comparable magnitudes across series and states.

	\subsubsection{Gibbs sampler} \label{Gibbs sampler}
	
	The model in (\ref{Example 2a})--(\ref{Example 2c}) together with the priors leads to a partially collapsed Gibbs sampler that combines conjugate updates, PXDA, and Metropolis--Hastings steps on the Stiefel manifold. At a high level, each iteration proceeds as follows:
	\begin{description}
		\item[(1)] Draw the two-state Markov switching indicators $s_{t}$ for $t=1,\ldots,T$ that determine the transition-scale multiplier $\kappa_t$, with the latent states integrated out. 

		\item[(2)] Draw $(\pPsi,\pU,\pA_{1:T},\pY_{1:T}^{q})$ jointly using a partially collapsed step. The covariance matrices $\pPsi$ and $\pU$ are drawn using PXDA with the latent states integrated out via the Kalman filter. Conditional on these draws, the latent states are sampled using a simulation smoother that skips the systematically missing quarterly data. Finally, the missing quarterly observations are drawn from their conditional distributions.

		\item[(3)] Draw ragged-edge missing monthly observations from their conditional Gaussian distributions.

		\item[(4)] Draw the left factor loadings $\pLambda$ using a Metropolis--Hastings algorithm on the Stiefel manifold, targeting the posterior implied by its matrix von Mises--Fisher prior.

		\item[(5)] Draw the right factor loadings $\pW$ using a Metropolis--Hastings algorithm on the Stiefel manifold that accounts for both the factor space and its orthogonal complement. During this step, the parameters in $\pPsi$ are integrated out.

		\item[(6)] Draw the state innovation covariance matrix $\pOmega$ using a PXDA step, followed by a rotation and reparameterization of the state equation.

		\item[(7)] Draw the autoregressive parameters $\pPhi = (\pPhi_1,\pPhi_2)$ from a matrix normal distribution, imposing stationarity via rejection sampling.

		\item[(8)] Draw the observation variances $\sigma^{2}_{i}$ and the Student's $t$ scale mixtures $h_{it}$ used for the measurement errors.
	\end{description}

	Several steps of the algorithm are worth highlighting. First, several blocks integrate out the latent states using the Kalman filter, leading to a partially collapsed sampler that improves mixing. Second, the factor loadings are constrained to lie on the Stiefel manifold and are therefore sampled using Metropolis--Hastings steps rather than direct draws from their full conditional distributions. The combination of PXDA and collapsed steps substantially improves mixing in high-dimensional settings. Finally, the discrete states are drawn in step (1) using the algorithm of Section 3.7. Full details of all algorithmic steps are provided in the online appendix.
	
	\subsection{Empirical results}
	
	\begin{figure}[!t]
		\caption{Data and estimated conditional mean.} \label{fig:data_fit}
		\begin{center}
			\resizebox{0.80\textwidth}{!}{%
				\includegraphics{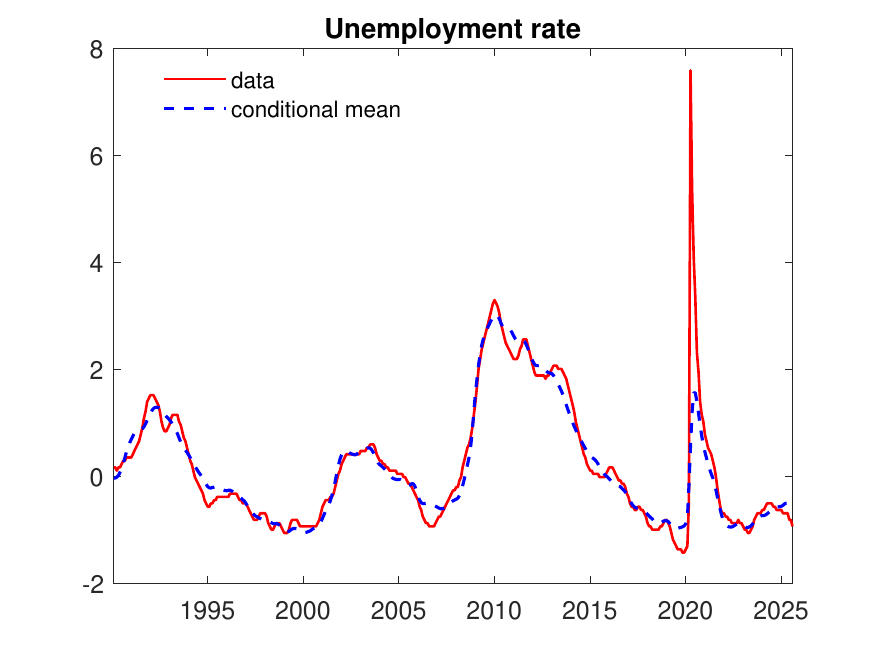}%
				\includegraphics{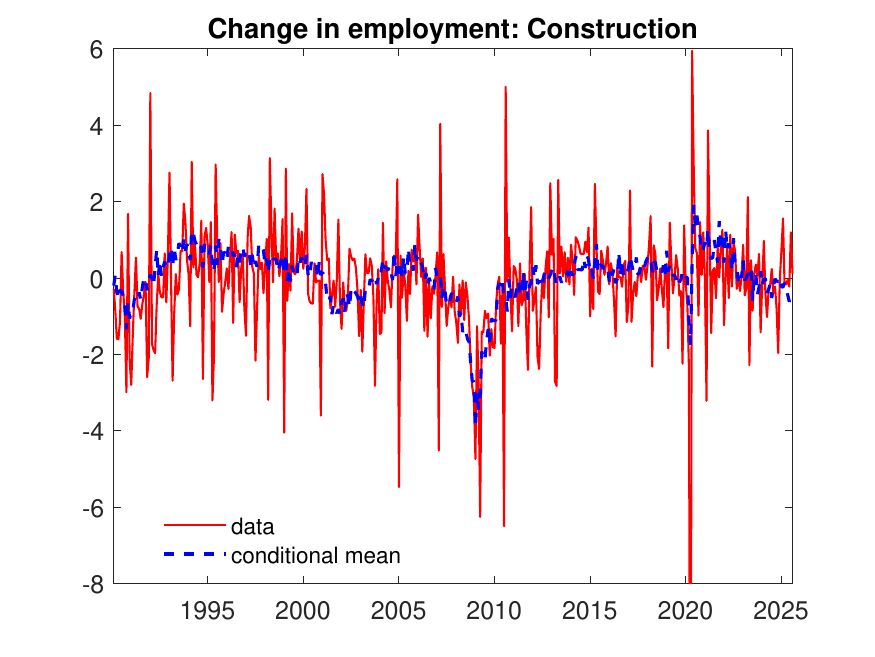}}
			\resizebox{0.80\textwidth}{!}{
				\includegraphics{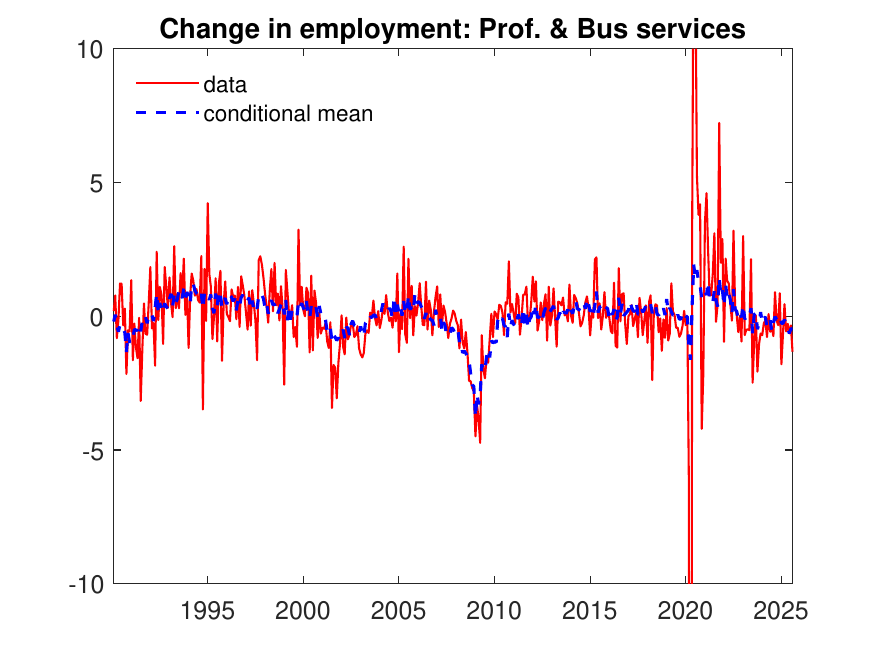}%
				\includegraphics{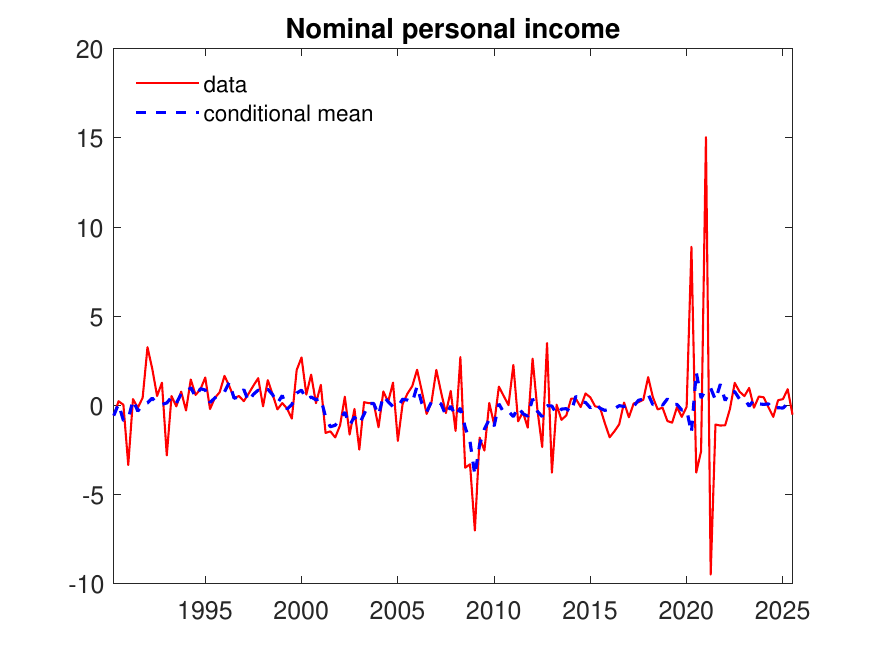}}
		\end{center}
		\textit{{\protect\footnotesize {Data and estimated conditional means for four series from Illinois. Top left: unemployment rate; Top right: change in employees, Construction; Bottom left: change in employees, Prof \& Business services; Bottom right: nominal personal income.}}}
	\end{figure}
	
	We compare WAIC across a grid of $(s,r)$ specifications, see Table~\ref{tab:waic}. WAIC keeps improving, with diminishing and increasingly noisy returns, as $r$ increases for each value of $s$. $p_{\text{waic}}$ fluctuates by several thousand across adjacent grid points, so WAIC alone does not identify a stable optimum. As the diagnostics below show, its improvement at large $r$ reflects the latent factors absorbing individual states' idiosyncratic dynamics rather than picking up common variation across states. We therefore do not select $(s,r)$ by minimizing WAIC directly, and instead combine it with two diagnostics calculated from the posterior draws. First, under $s=4$ the posterior mean of $\pOmega$'s fourth diagonal element is small relative to the other three and shrinks further with $r$, indicating a spurious fourth row factor. Therefore, we restrict attention to $s=2,3$. Second, for both $s=2$ and $s=3$ we track each column of $\pW$'s concentration ratio, which is the largest squared state loading divided by the column's total squared loading. For $r=1$ to $r=11$, every column's ratio stays below 0.2 for both $s$. From $r=14$ on, at least one column jumps above 0.4, dominated by a single state (Hawaii for $s=2$; persistently Texas for $s=3$ from $r=18$ on). This pattern replicates across independently reinitialized runs. We select $r=11$, the largest value before this appears. WAIC's real, non-noise improvement is concentrated in the move from $r=7$ to $r=9$--$11$ (roughly 5,000 points), while $r=9$ to $r=11$ is itself flat. Little genuine fit is sacrificed by stopping there. At the value of $r=11$, $s=2$ and $s=3$ are statistically tied on WAIC (613,273 versus 613,239). We focus on $s=3$ because it separates the panel into three economically distinct row factors -- unemployment, sector employment together with housing prices, and income/output. A model with $s=2$ cannot represent these factors without conflating two of them.

	\begin{table}[!t]
		\caption{WAIC across left-factor ($s$) and right-factor ($r$) specifications.} \label{tab:waic}
		\begin{center}
			\begin{tabular}{crrr}
				\toprule
				$r$ & $s=2$ & $s=3$ & $s=4$ \\
				\midrule
				7  & 620{,}066 & 618{,}033 & 618{,}527 \\
				9  & 614{,}629 & 613{,}187 & 614{,}236 \\
				11 & 613{,}273 & 613{,}239 & 613{,}354 \\
				14 & 604{,}247 & 603{,}106 & 609{,}915 \\
				16 & 597{,}263 & 597{,}510 & 606{,}927 \\
				18 & 597{,}207 & 598{,}783 & 602{,}451 \\
				20 & 594{,}151 & 594{,}732 & 599{,}441 \\
				22 & 589{,}687 & 591{,}284 & 595{,}375 \\
				\bottomrule
			\end{tabular}
		\end{center}
	\end{table}

	We estimate the model with a $3 \times 11$ latent factor matrix ($s=3$, $r=11$). Consequently, the
	600 observed state-level series are summarized by 33 dynamic factors. The purpose
	of the empirical exercise is to illustrate that the matrix state space structure
	can capture the main common movements in a large mixed-frequency panel while
	remaining computationally tractable.
	
	\begin{figure}[!t]
		\caption{Estimated factor loadings, most persistent state factor, and regime probabilities.} \label{fig:loadings_regime}
		\begin{center}
			\includegraphics[width=0.80\textwidth]{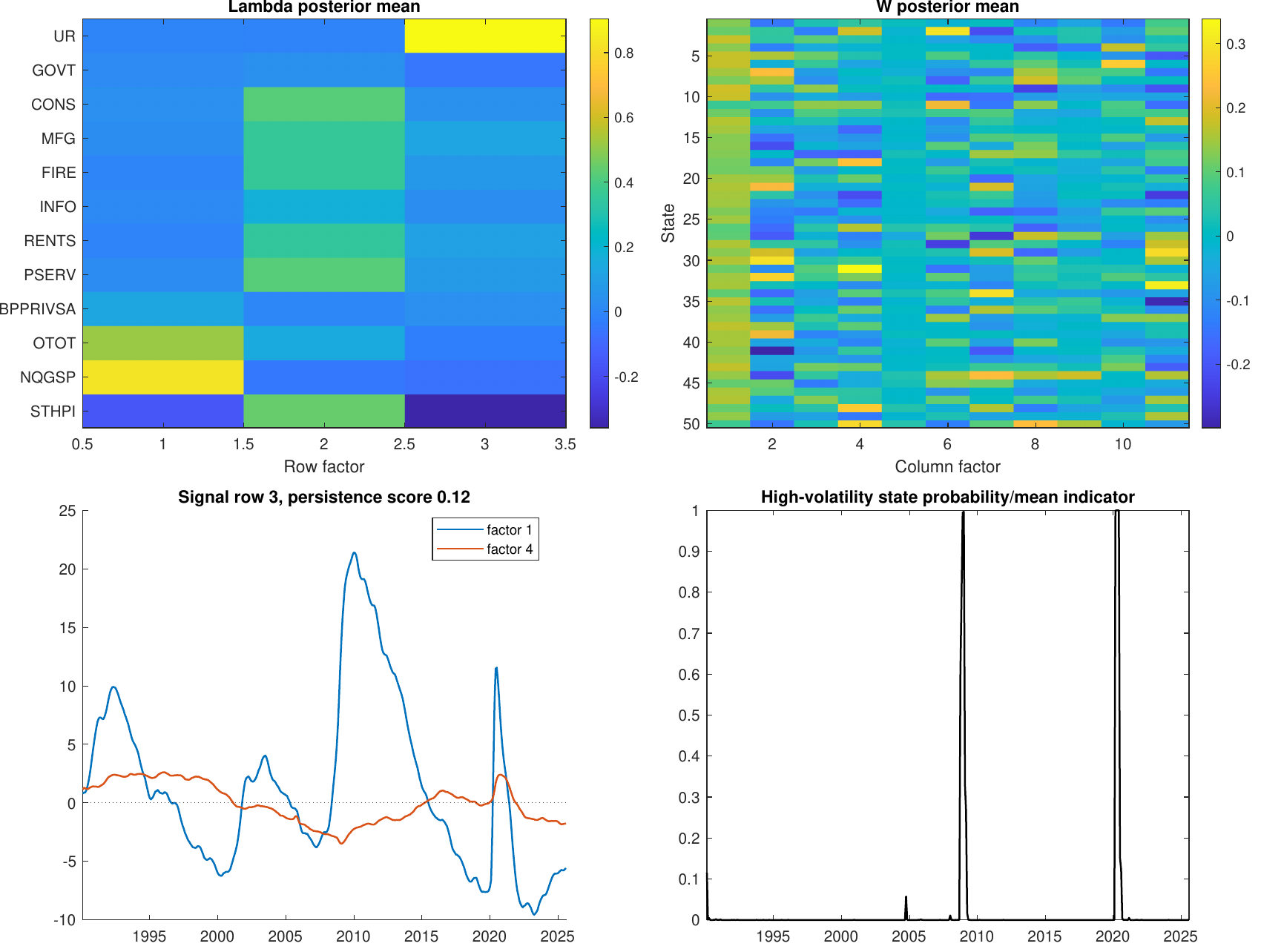}
		\end{center}
		\textit{{\protect\footnotesize {Estimated factor loadings, most persistent state factor, and regime probabilities. Top left: estimated left factor loadings $\pLambda$; Top right: estimated right factor loadings $\pW$; Bottom left: posterior mean of the most persistent row factor, selected by an AR(1)-times-posterior-scale score; Bottom right: smoothed probability of a high or low variance state.}}}
	\end{figure}
	
	Figure \ref{fig:data_fit} reports four representative series from Illinois together with their estimated conditional means from the posterior draws. The examples include the unemployment rate, changes in the total employees in the construction sector, changes in the total employees in the professional and business services sector, and quarterly nominal personal income. The fitted conditional means track the persistent movements in the data while smoothing through high-frequency
	idiosyncratic variation. This is especially clear in the employment series, where the observed monthly changes are volatile but the fitted component captures the lower-frequency movement. The model also captures the major business-cycle events in the sample, including the Great Recession and the sharp movements around the COVID period, while treating the largest transitory observations as measurement
	noise or heavy-tailed shocks rather than forcing them entirely into the common state.
	

	\begin{table}[!t]
		\caption{Integrated autocorrelation times (IACT) and effective sample sizes (ESS), $s=3$, $r=11$, 50{,}000 post-burn-in draws.} \label{tab:iact}
		\begin{center}
			\begin{tabular}{lrrr}
				\toprule
				Quantity & \# params & IACT & ESS \\
				\midrule
				$\text{diag}(\pOmega)$, factor 1 & 1 & 65.6 & 762 \\
				$\text{diag}(\pOmega)$, factor 2 & 1 & 65.9 & 759 \\
				$\text{diag}(\pOmega)$, factor 3 & 1 & 5.4 & 9{,}316 \\
				$\pLambda$ loading, factor 1 & 1 & 26.3 & 1{,}905 \\
				$\pLambda$ loading, factor 2 & 1 & 29.8 & 1{,}677 \\
				$\pLambda$ loading, factor 3 & 1 & 17.7 & 2{,}818 \\
				$\text{tr}(\pPsi)$ & 1 & 53.3 & 938 \\
				max eig$(\pPsi)$ & 1 & 39.8 & 1{,}258 \\
				$\text{diag}(\pU)$, all factors & 11 & 49.8--63.4 & 789--1{,}005 \\
				Idiosyncratic variances, all series & 12 & 9.6--41.3 & 1{,}209--5{,}222 \\
				$\pPhi_1,\pPhi_2$ coefficients & 18 & 1.2--21.5 & 2{,}330--41{,}201 \\
				$\pW$ anchor loadings, all columns & 11 & 1.1--75.7 & 660--44{,}089 \\
				\bottomrule
			\end{tabular}
		\end{center}
	\end{table}
	
	Figure \ref{fig:loadings_regime} summarizes the estimated factor structure and the probabilities for the two-state Markov switching variance. The posterior mean of the left loading matrix shows that the first row factor loads primarily on the two lower-frequency income and output series (nominal personal income and real gross state product), the second row factor loads broadly across the sector-level employment series together with the house price index, and the third row factor loads almost exclusively on the unemployment rate, with an offsetting loading on house prices. The posterior mean of the right loading matrix shows substantial heterogeneity across states, with the first column factor capturing a common national component (positive for all 50 states) and the remaining column factors capturing cross-state deviations from that common movement. The bottom-left panel plots the most persistent row factor by our AR(1)-times-scale ranking, the unemployment-loaded factor. It has a clear business-cycle interpretation, rising sharply during the Great Recession and again, briefly, during the COVID episode. The smoothed probability of the high-volatility transition state rises sharply and stays elevated for several months around exactly the 2008--09 financial crisis and the 2020 COVID episode. This is consistent with the role of $\kappa_t$ as a means of capturing common rare, large aggregate shocks.
	
	Table \ref{tab:iact} reports integrated autocorrelation times (IACT) and effective sample sizes (ESS), computed via \cite{Geyer(92)}'s initial monotone sequence estimator, for a representative set of parameters. Individually interpretable, low-dimensional quantities -- the diagonal of $\pOmega$, the $\pLambda$ anchor loadings, and the scale of $\pPsi$ -- are reported separately. The remaining blocks, which have no individual economic interpretation and are too numerous to report one by one. Instead, we summarize them by the range of IACT/ESS values across each block. The diagonal entries of $\pU$, the idiosyncratic variances, the autoregressive coefficients $\pPhi_1,\pPhi_2$, and the right factor loadings $\pW$. The slowest-mixing quantities are the two largest diagonal elements of $\pOmega$, with IACT $\approx 66$ and ESS $\approx 760$ out of 50{,}000 post-burn-in draws. Every other block mixes substantially faster. This confirms that the partially collapsed Gibbs sampler, combined with PXDA and the Metropolis--Hastings moves on the Stiefel manifold, delivers adequate mixing for this specification.
	
	Overall, the empirical results indicate that a small matrix of latent factors can summarize a large panel of state-level macroeconomic series. The row factors capture comovement across economic variables, the column factors capture cross-state dependence, and the Markov switching transition scale identifies periods in which common shocks are unusually large.  
	
	\section{Conclusion}
	
	We developed the Kalman filter, smoother, and posterior sampling algorithms for a large class of conditionally, linear matrix normal time series models.
	There are additional algorithms that could be derived for this class of time series models that we have omitted due to space constraints. These include disturbance smoothing and disturbance simulation smoothing, see \cite{Koopman(93)} and \cite{deJongShephard(95)}. One can also develop methods for exact treatment of initial conditions in non-stationary models, see \cite{deJong(91)} and \cite{Koopman(97)}. Finally, the matrix Kalman filter can be used within a particle filter, see \cite{ChenLiu(00)}. 
	
	\subsection{Declaration of generative AI and AI-assisted technologies in the manuscript preparation process}
	
	During the preparation of this work, the authors used Codex to proofread the paper and computer code. The authors reviewed and edited the output as needed and take full responsibility for the content of the published article. 
	
	\bibliographystyle{jf}
	\bibliography{creal}

\end{document}


\thispagestyle{empty}
\newlength{\oldparindent}
\oldparindent=\parindent
\parindent=0.0in

\vspace*{0.175in}

		\noindent{\Large\bf Online Appendix: Conditionally linear, matrix normal state space models}
		
		\vspace{0.075in}
		\noindent{%
			Drew D. Creal\footnote{
				Department of Economics, University of Illinois Urbana-Champaign;
				dcreal@illinois.edu.}, 
			Marcelo C. Medeiros,\footnote{
				Department of Economics, University of Illinois Urbana-Champaign;
				marcelom@illinois.edu.}
			and 
			Rodrigo Sarlo,\footnote{
				Department of Electrical Engineering, Pontifical Catholic University of Rio de Janeiro;
				rodrigosarlofilho@gmail.com.}
		}
		
		\vspace{0.075in}
		\noindent{This version: \today}
		
		{
			{\bf Abstract}
			
			\medskip
			This is the on-line appendix that derives the matrix Kalman filter, smoother, and simulation smoothing algorithms. It also contains more detailed information about the implementation of the MCMC algorithms for the empirical application in the paper. 
			\bigskip
			
			
			\medskip
			\noindent{\bf Keywords: }  matrix normal distribution, Kalman filter and smoother, simulation smoothing.
			
			\medskip
			
		}


\setcounter{footnote}{0}
\renewcommand{\thefootnote}{\arabic{footnote}}
\noindent \thispagestyle{empty}\addtocounter{page}{-1}\newpage{}%
\setcounter{equation}{0}


\appendix

\setcounter{section}{0} \renewcommand{\thesection}{Appendix \Alph{section}}
\setcounter{equation}{0}
\renewcommand{\theequation}{\Alph{section}.\arabic{equation}}

\clearpage

\section{Proofs of the propositions} \label{KFS derivation}

The proofs below adapt, to the matrix normal setting, proof techniques for the (vector) Kalman filter, smoother, and simulation smoother that are standard in the state space literature and are laid out in detail in \cite{DurbinKoopman(12)}. In particular, our use of the regression lemma to derive the filtering and smoothing distributions, the prediction-error decomposition of the log-likelihood, and the disturbance-smoother-based construction of the simulation smoother all follow the structure of their proofs.

\subsubsection{Derivation of the filtering distribution}

\emph{Proof of Proposition 1:} At date $t$, the likelihood and prior distribution for the collapsed  model are
\begin{eqnarray*}
	\pY_{t}^{*} & \sim & \text{MN}\Lp \pD_{t}\pJ_{t}^{*,\top} + \pZ_{t}\pA_{t},\pH_{t},\pU\Rp \\
	\pA_{t} & \sim & \text{MN}\Lp \pA_{t|t-1},\pP_{t|t-1},\pU\Rp  
\end{eqnarray*}
where $\pY_{t}^{*} = \pY_{t}\pJ_{t}^{*,\top}$. Next, we find the joint conditional distribution of $p\Lp \pY_{t}^{*}, \pA_{t}|\pY_{1:t-1},\ptheta\Rp$. The conditional mean of $\pY_{t}^{*}$ is
\begin{eqnarray*}
	\E\Lb \pY_{t}^{*}|\pY_{1:t-1}\Rb &=& \pD_{t}\pJ_{t}^{*,\top} + \pZ_{t}\pA_{t|t-1} 
\end{eqnarray*}
The conditional variance is
\begin{eqnarray*}
	\V\Lb \text{vec}\Lp\pY_{t}^{*}\Rp|\pY_{1:t-1}\Rb &=& \V\Lb \text{vec}\Lp \pZ_{t}\pA_{t}\Rp  + \text{vec}\Lp \pE_{1t} \Rp|\pY_{1:t-1}\Rb \\
	&=& \V\Lb \text{vec}\Lp \pZ_{t}\pA_{t}\Rp |\pY_{1:t-1}\Rb + \V\Lb\text{vec}\Lp \pE_{1t} \Rp |\pY_{1:t-1}\Rb \\
	&=& \V\Lb \Lp \eye \otimes \pZ_{t}\Rp  \text{vec}\Lp\pA_{t}\Rp |\pY_{1:t-1}\Rb + \Lp \pU \otimes \pH_{t}\Rp \\
	&=& \Lp \eye \otimes \pZ_{t}\Rp  \Lp \pU \otimes \pP_{t|t-1}\Rp \Lp \eye \otimes \pZ_{t}\Rp^{\top} + \Lp \pU \otimes \pH_{t}\Rp \\
	&=&  \pU \otimes   \pZ_{t}\pP_{t|t-1} \pZ_{t}^{\top} + \pH_{t}  
	\ = \ \pU \otimes \pF_{t}
\end{eqnarray*}
The covariance between $\pA_{t}$ and $\pY_{t}^{*}$ is
\begin{eqnarray*}
	\text{cov}\Lb \text{vec}\Lp\pY_{t}^{*}\Rp,\text{vec}\Lp \pA_{t}\Rp|\pY_{1:t-1}\Rb &=& \text{cov}\Lb \text{vec}\Lp \pZ_{t}\pA_{t} + \pE_{1t} \Rp,\text{vec}\Lp \pA_{t}\Rp|\pY_{1:t-1}\Rb \\
	&=& \text{cov}\Lb \text{vec}\Lp \pZ_{t}\pA_{t}\Rp + \text{vec}\Lp\pE_{1t} \Rp,\text{vec}\Lp \pA_{t}\Rp|\pY_{1:t-1}\Rb \\
	&=& \text{cov}\Lb \Lp\eye \otimes \pZ_{t}\Rp\text{vec}\Lp \pA_{t}\Rp,\text{vec}\Lp \pA_{t}\Rp|\pY_{1:t-1}\Rb \\
	& & + \text{cov}\Lb \text{vec}\Lp\pE_{1t}^{*} \Rp,\text{vec}\Lp \pA_{t}\Rp|\pY_{1:t-1}\Rb \\
	&=& \Lp\eye \otimes \pZ_{t}\Rp \V\Lb \text{vec}\Lp \pA_{t}\Rp|\pY_{1:t-1}\Rb \\
	&=& \Lp\eye \otimes \pZ_{t}\Rp \Lp \pU \otimes \pP_{t|t-1}\Rp \\
	&=& \pU\otimes \pZ_{t}\pP_{t|t-1} 
\end{eqnarray*}
Since $\pA_{t}$ and $\pE_{1t}^{*}$ are independent matrix normal random matrices with the same right scale matrix $\pU$, the stacked matrix $\Lp \pA_{t}^{\top},\pE_{1t}^{*,\top}\Rp^{\top}$ is itself matrix normal with right scale $\pU$ and block-diagonal left scale matrix, and $\Lp \pY_{t}^{*,\top},\pA_{t}^{\top}\Rp^{\top}$ is a linear function of this stacked matrix. Linear combinations of a matrix normal random matrix with right scale $\pU$ remain matrix normal with right scale $\pU$, so, conditional on past data $\pY_{1:t-1}$, the joint conditional distribution of the matrix $\Lp \pY_{t}^{*,\top},\pA_{t}^{\top}\Rp^{\top}$ is matrix normal with the moments computed above, i.e.\
\begin{eqnarray*}
	\Lp \begin{matrix}
		\pY_{t}^{*}	\\
		\pA_{t}
	\end{matrix}\Rp & \sim & \text{MN}\Lp  	\Lb \begin{matrix}
		\pD_{t}\pJ_{t}^{*,\top} + \pZ_{t}\pA_{t|t-1}	\\
		\pA_{t|t-1}
	\end{matrix}\Rb ,	\Lb \begin{matrix}
		\pF_{t} & \pZ_{t}\pP_{t|t-1}	\\
		\pP_{t|t-1}\pZ_{t}^{\top} & \pP_{t|t-1}
	\end{matrix}\Rb ,\pU\Rp 
\end{eqnarray*}
Next, suppose $\pY_{t}^{*}$ is observed while $\pA_{t}$ is latent. Applying the regression lemma, the filtering distribution of the latent matrix is 
\begin{eqnarray*}
	p\Lp\pA_{t}|\pY_{1:t}^{*},\ptheta\Rp & = & \text{MN}\Lp  \pA_{t|t},\pP_{t|t},\pU\Rp  
\end{eqnarray*}
where
\begin{eqnarray*}
	\pA_{t|t} & = &  \pA_{t|t-1} + \pP_{t|t-1}\pZ_{t}^{\top}\pF_{t}^{-1}\Lp \pY_{t}^{*} - \pD_{t}\pJ_{t}^{*,\top} - \pZ_{t}\pA_{t|t-1} \Rp  \\
	\pP_{t|t} & = & \pP_{t|t-1} -  \pP_{t|t-1}\pZ_{t}^{\top}\pF_{t}^{-1}\pZ_{t}\pP_{t|t-1} 
\end{eqnarray*}
We separate these equations in terms of the prediction errors and prediction error variances.


Next, we prove the one-step ahead predictive distribution.
Conditional on information $\pY_{1:t}$, we know that $\pA_{t} \sim \text{MN}\Lp \pA_{t|t},\pP_{t|t},\pU\Rp$. Linear combinations of matrix normals with a common second scale matrix $\pU$ are matrix normal. Therefore, the predictive distribution is matrix normal with mean
\begin{eqnarray*}
	\E\Lb\pA_{t+1}|\pY_{1:t} \Rb &=&  \pT_{t}	\E\Lb \pA_{t}|\pY_{1:t}\Rb + \pC_{t} + \E\Lb \pR_{t}\pE_{2,t}|\pY_{1:t}\Rb \\
	\pA_{t+1|t} &=&  \pT_{t}\pA_{t|t} + \pC_{t}
\end{eqnarray*}
The conditional variance is
\begin{eqnarray*}
	\V\Lb \text{vec}\Lp\pA_{t+1}\Rp|\pY_{1:t} \Rb &=& 	\V\Lb \text{vec}\Lp \pC_{t} + \pT_{t}\pA_{t} + \pR_{t}\pE_{2,t} \Rp|\pY_{1:t} \Rb \\
	&=& \V\Lb \text{vec}\Lp  \pT_{t}\pA_{t}\Rp|\pY_{1:t}\Rb + \V\Lb  \text{vec}\Lp  \pR_{t}\pE_{2,t} \Rp|\pY_{1:t} \Rb \\
	&=& \V\Lb \text{vec}\Lp  \pT_{t}\pA_{t}\Rp|\pY_{1:t}\Rb + \V\Lb  \Lp \eye \otimes \pR_{t}\Rp\text{vec}\Lp  \pE_{2,t} \Rp|\pY_{1:t} \Rb \\
	&=& \V\Lb \Lp \eye \otimes \pT_{t} \Rp\text{vec}\Lp  \pA_{t}\Rp|\pY_{1:t}\Rb + \Lp \eye \otimes \pR_{t}\Rp\Lp \pU\otimes \pQ_{t}\Rp \Lp \eye \otimes \pR_{t}\Rp^{\top}\\
	&=&  \Lp \eye\otimes \pT_{t} \Rp\V\Lb\text{vec}\Lp  \pA_{t}\Rp|\pY_{1:t}\Rb  \Lp \eye\otimes \pT_{t} \Rp^{\top} +  \Lp \pU\otimes \pR_{t}\pQ_{t}\pR_{t}^{\top}\Rp \\
	&=&  \Lp \eye\otimes \pT_{t} \Rp \Lp \pU\otimes \pP_{t|t}\Rp  \Lp \eye\otimes \pT_{t}^{\top} \Rp +  \Lp \pU\otimes \pR_{t}\pQ_{t}\pR_{t}^{\top}\Rp \\
	&=&  \Lp \pU\otimes \pT_{t}\pP_{t|t}\pT_{t}^{\top}\Rp +  \Lp \pU \otimes \pR_{t}\pQ_{t}\pR_{t}^{\top}\Rp \\
	&=&   \pU\otimes \Lp \pT_{t}\pP_{t|t}\pT_{t}^{\top} +  \pR_{t}\pQ_{t}\pR_{t}^{\top}\Rp 
\end{eqnarray*}
where $\pP_{t+1|t} \ = \ \pT_{t}\pP_{t|t}\pT_{t}^{\top} + \pR_{t}\pQ_{t}\pR_{t}^{\top}$.
This completes one iteration of the Kalman filter at an arbitrary date $t$, taking as given that $\pA_{t}|\pY_{1:t-1} \sim \text{MN}\Lp \pA_{t|t-1},\pP_{t|t-1},\pU\Rp$. The base case $t=1$ holds directly from the model's initial condition, $\pA_{1} \sim \text{MN}\Lp \pA_{1|0},\pP_{1|0},\pU\Rp$, and the proof is completed by induction over $t$.

\subsubsection{Derivation of the log-likelihood function} \label{loglik derivation}

By the standard prediction-error (chain-rule) decomposition of the likelihood of a time series,
\begin{eqnarray*}
	\log p\Lp \pY_{1:T}|\ptheta\Rp &=& \sum_{t=1}^{T}\log p\Lp \pY_{t}|\pY_{1:t-1},\ptheta\Rp.
\end{eqnarray*}

\textbf{The Jacobian of the collapsing transformation.} The map $\pY_{t}\mapsto\Lp \pY_{t}^{*},\pY_{t}^{+}\Rp$ defined by $\pY_{t}^{*}=\pY_{t}\pJ_{t}^{*,\top}$, $\pY_{t}^{+}=\pY_{t}\pJ_{t}^{+,\top}$ is linear and, since $\pJ_{t}=\Lp \pJ_{t}^{*,\top}\ \pJ_{t}^{+,\top}\Rp^{\top}$ is full rank ($n\times n$), invertible. In vec form,
\begin{eqnarray*}
	\Lp\begin{matrix} \text{vec}\Lp\pY_{t}^{*}\Rp \\ \text{vec}\Lp\pY_{t}^{+}\Rp\end{matrix}\Rp \ = \ \text{vec}\Lp \pY_{t}\pJ_{t}^{\top}\Rp \ = \ \Lp \pJ_{t}\otimes\eye_{m}\Rp\text{vec}\Lp\pY_{t}\Rp,
\end{eqnarray*}
using $\text{vec}\Lp\pA\pX\pB\Rp = \Lp \pB^{\top}\otimes\pA\Rp\text{vec}\Lp\pX\Rp$ with $\pA=\eye_m$, $\pX=\pY_t$, $\pB=\pJ_t^\top$. The Jacobian of this linear change of variables is
\begin{eqnarray*}
	\Lv \det\Lp\pJ_{t}\otimes\eye_{m}\Rp\Rv \ = \ \Lv \pJ_{t}\Rv^{m}\Lv \eye_{m}\Rv^{n} \ = \ \Lv \pJ_{t}\Rv^{m},
\end{eqnarray*}
using $\det\Lp\pA\otimes\pB\Rp = \Lv\pA\Rv^{q}\Lv\pB\Rv^{p}$ for $\pA$ ($p\times p$) and $\pB$ ($q\times q$). By the standard change-of-variables formula for densities,
\begin{eqnarray*}
	p\Lp\pY_{t}|\pY_{1:t-1},\ptheta\Rp \ = \ p\Lp \pY_{t}^{*},\pY_{t}^{+}|\pY_{1:t-1},\ptheta\Rp\,\Lv \pJ_{t}\Rv^{m}.
\end{eqnarray*}
Because $\pY_{t}^{+}$ does not depend on $\pA_{t}$ (condition (7)) and is generated from a disturbance $\pE_{1t}^{+}$ that is independent of $\pA_t$, of $\pE_{1t}^*$, and across dates, $\pY_{t}^{+}$ is independent of $\pY_{t}^{*}$ and of $\pY_{1:t-1}$; hence $p\Lp\pY_{t}^{*},\pY_{t}^{+}|\pY_{1:t-1},\ptheta\Rp = p\Lp \pY_{t}^{*}|\pY_{1:t-1},\ptheta\Rp\, p\Lp\pY_{t}^{+}|\ptheta\Rp$, and
\begin{eqnarray*}
	\log p\Lp\pY_{t}|\pY_{1:t-1},\ptheta\Rp \ = \ \log p\Lp \pY_{t}^{*}|\pY_{1:t-1},\ptheta\Rp + \log p\Lp\pY_{t}^{+}|\ptheta\Rp + m\log\Lv \pJ_{t}\Rv.
\end{eqnarray*}
Summing over $t=1,\ldots,T$ gives the decomposition
\begin{eqnarray*}
	\log p\Lp \pY_{1:T}|\ptheta\Rp \ = \ \log p\Lp \pY_{1:T}^{*}|\ptheta\Rp + \log p\Lp \pY_{1:T}^{+}|\ptheta\Rp + m\sum_{t=1}^{T}\log\Lv\pJ_{t}\Rv,
\end{eqnarray*}
where the Jacobian is applied \emph{independently to each of the $m$ rows} of $\pY_t$, and so carries multiplicity $m$.

\textbf{The likelihood contribution of $\pY_{1:T}^{*}$.} By Proposition 1, we have 
\begin{eqnarray*}
	\pY_{t}^{*}|\pY_{1:t-1}\sim\text{MN}\Lp \pD_{t}^{*}+\pZ_{t}\pA_{t|t-1},\pF_{t},\pU\Rp
\end{eqnarray*}
 so $\pV_{t}=\pY_{t}^{*}-\pD_{t}^{*}-\pZ_{t}\pA_{t|t-1}\sim\text{MN}\Lp\pzero,\pF_{t},\pU\Rp$. The density of a $\text{MN}\Lp\pzero,\pF_t,\pU\Rp$ random matrix is
\begin{eqnarray*}
	p\Lp \pV_t\Rp \ = \ \Lp 2\pi\Rp^{-mr/2}\Lv\pU\Rv^{-m/2}\Lv\pF_{t}\Rv^{-r/2}\exp\Lp -\frac{1}{2}\text{tr}\Lb\pU^{-1}\pV_{t}^{\top}\pF_{t}^{-1}\pV_{t}\Rb\Rp,
\end{eqnarray*}
so, using $\text{tr}\Lb\pU^{-1}\pV_{t}^{\top}\pF_{t}^{-1}\pV_{t}\Rb=\text{tr}\Lb\pF_{t}^{-1}\pV_{t}\pU^{-1}\pV_{t}^{\top}\Rb$,
\begin{eqnarray*}
	\log p\Lp \pY_{t}^{*}|\pY_{1:t-1},\ptheta\Rp \ = \ -\frac{mr}{2}\log\Lp2\pi\Rp - \frac{r}{2}\log\Lv\pF_{t}\Rv - \frac{m}{2}\log\Lv\pU\Rv - \frac{1}{2}\text{tr}\Lb\pF_{t}^{-1}\pV_{t}\pU^{-1}\pV_{t}^{\top}\Rb.
\end{eqnarray*}
Summing over $t$,
\begin{eqnarray*}
	\log p\Lp \pY_{1:T}^{*}|\ptheta\Rp &=& -\frac{Tmr}{2}\log\Lp2\pi\Rp - \frac{r}{2}\sum_{t=1}^{T}\log\Lv\pF_{t}\Rv - \frac{mT}{2}\log\Lv\pU\Rv \\
	& & - \frac{1}{2}\sum_{t=1}^{T}\text{tr}\Lb\pF_{t}^{-1}\pV_{t}\pU^{-1}\pV_{t}^{\top}\Rb,
\end{eqnarray*}
which is the stated formula.

\textbf{The likelihood contribution of $\pY_{1:T}^{+}$.} By the observation equation for $\pY_t^+$, $\pY_{t}^{+}-\pD_{t}^{+}\sim\text{MN}\Lp\pzero,\pH_{t},\pPsi_{t}\Rp$, independently across $t$. The same density formula, with $\pF_t\to\pH_t$, $\pU\to\pPsi_t$, and column dimension $r\to n-r$, gives
\begin{eqnarray*}
	\log p\Lp\pY_{t}^{+}|\ptheta\Rp & = & -\frac{m(n-r)}{2}\log(2\pi)-\frac{n-r}{2}\log\Lv\pH_{t}\Rv-\frac{m}{2}\log\Lv\pPsi_{t}\Rv \\
	& & -\frac{1}{2}\text{tr}\Lb\pH_{t}^{-1}\Lb\pY_{t}^{+}-\pD_{t}^{+}\Rb\pPsi_{t}^{-1}\Lb\pY_{t}^{+}-\pD_{t}^{+}\Rb^{\top}\Rb,
\end{eqnarray*}
and summing over $t$ gives the stated formula for $\log p\Lp\pY_{1:T}^{+}|\ptheta\Rp$.

\textbf{The special case $n=r$.} When $n=r$, $\pW_t$ ($n\times r$) is square and, under the orthonormal identification of Section 2.3, satisfies $\pW_t^\top\pW_t=\eye_r$, so $\pW_t$ is itself orthogonal and no genuine collapsing is needed. Taking $\pJ_t=\pW_t^\top$ (so $\pJ_t^*=\pW_t^\top$ and $\pJ_t^+$ is vacuous, $0\times n$) gives $\Lv\pJ_{t}\Rv=1$ and $\pY_t^+$ has zero columns, so both the Jacobian term and the $\pY_{1:T}^{+}$ term vanish, and $\log p\Lp\pY_{1:T}|\ptheta\Rp = \log p\Lp\pY_{1:T}^{*}|\ptheta\Rp$ reduces to the simple expression stated at the start of the main paper's log-likelihood section.

\textbf{The Jacobian in the orthogonal and GLS parameterizations.} When $\pJ_t$ is orthonormal (Example 2i), $\Lv\pJ_t\Rv=1$ trivially. In the GLS parameterization (Example 1i), $\pJ_t \pJ_t^\top$ has determinant
\begin{eqnarray*}
	\Lv\pJ_{t}\Rv^{2} \ = \ \Lv\pJ_{t}\pJ_{t}^{\top}\Rv \ = \ \Lv\pU_{\pY,t}\Rv^{-1}\Lv\pJ_{t}\pU_{\pY,t}\pJ_{t}^{\top}\Rv,
\end{eqnarray*}
using $\pJ_{t}\pJ_{t}^{\top} = \pJ_{t}\pU_{\pY,t}^{-1}\pU_{\pY,t}\pJ_{t}^{\top}$ and $\Lv\pA\pB\Rv=\Lv\pA\Rv\Lv\pB\Rv$. The matrix $\pJ_{t}\pU_{\pY,t}\pJ_{t}^{\top}$ is block diagonal,
\begin{eqnarray*}
	\pJ_{t}\pU_{\pY,t}\pJ_{t}^{\top} \ = \ \Lb\begin{matrix} \pJ_{t}^{*}\pU_{\pY,t}\pJ_{t}^{*,\top} & \pJ_{t}^{*}\pU_{\pY,t}\pJ_{t}^{+,\top} \\ \pJ_{t}^{+}\pU_{\pY,t}\pJ_{t}^{*,\top} & \pJ_{t}^{+}\pU_{\pY,t}\pJ_{t}^{+,\top}\end{matrix}\Rb \ = \ \Lb\begin{matrix} \pU & \pzero \\ \pzero & \pJ_{t}^{+}\pU_{\pY,t}\pJ_{t}^{+,\top}\end{matrix}\Rb,
\end{eqnarray*}
using condition (8) for the top-left block and condition (9) (and its transpose) for the off-diagonal blocks, so $\Lv\pJ_{t}\pU_{\pY,t}\pJ_{t}^{\top}\Rv = \Lv\pU\Rv\cdot\Lv\pJ_{t}^{+}\pU_{\pY,t}\pJ_{t}^{+,\top}\Rv$. Under the normalization $\Lv\pJ_t^{+}\pU_{\pY,t}\pJ_t^{+,\top}\Rv=1$, this gives $\Lv\pJ_{t}\Rv^{2}=\Lv\pU_{\pY,t}\Rv^{-1}\Lv\pU\Rv$, i.e.\ $\Lv\pJ_{t}\Rv=\Lv\pU_{\pY,t}\Rv^{-1/2}\Lv\pU\Rv^{1/2}$.

\textbf{Basis invariance.} The complement $\pJ_t^+$ is not unique: any $(n-r)\times n$ matrix satisfying $\pJ_t^+\pW_t = \pzero$ and $\pJ_t^*\pU_{\pY,t}\pJ_t^{+,\top} = \pzero$ works, with $\pJ_t = \Lp \pJ_t^{*,\top} \ \pJ_t^{+,\top}\Rp^{\top}$ full rank. We show the log-likelihood does not depend on which valid $\pJ_t^+$ is chosen, even though $\pPsi_t$ itself does. Let $\pB$ be any nonsingular $(n-r)\times(n-r)$ matrix and set $\widetilde{\pJ}_t^+ = \pB\pJ_t^+$. Since $\pJ_t^*$ depends only on $\pW_t$ and $\pU_{\pY,t}$, not on the choice of $\pJ_t^+$, $\pY_t^* = \pY_t\pJ_t^{*,\top}$ and hence $\pU$ (via condition (8)) are unaffected by this replacement. That $\widetilde{\pJ}_t^+$ remains a valid complement follows immediately: $\widetilde{\pJ}_t^+\pW_t = \pB\pJ_t^+\pW_t = \pB\pzero = \pzero$, and $\pJ_t^*\pU_{\pY,t}\widetilde{\pJ}_t^{+,\top} = \pJ_t^*\pU_{\pY,t}\pJ_t^{+,\top}\pB^\top = \pzero\pB^\top = \pzero$.

For the transformed complement, $\widetilde{\pY}_t^+ = \pY_t\pJ_t^{+,\top}\pB^\top = \pY_t^+\pB^\top$, so $\widetilde{\pD}_t^+ = \pD_t^+\pB^\top$ and $\widetilde{\pY}_t^+ - \widetilde{\pD}_t^+ = \Lp\pY_t^+-\pD_t^+\Rp\pB^\top$. A linear transformation of a matrix normal random matrix by right-multiplication is matrix normal with the right scale transformed congruently, so $\widetilde{\pY}_t^+ - \widetilde{\pD}_t^+ \sim \text{MN}\Lp\pzero,\pH_t,\pB\pPsi_t\pB^\top\Rp$, confirming $\widetilde{\pPsi}_t = \pB\pPsi_t\pB^\top$.

We verify that the two terms of the factorization depending on the complement -- $\log p\Lp\widetilde{\pY}_t^+|\ptheta\Rp$ and $m\log\Lv\widetilde{\pJ}_t\Rv$ -- change by exactly offsetting amounts. First, $\Lv\widetilde{\pPsi}_t\Rv = \Lv\pB\pPsi_t\pB^\top\Rv = \Lv\pB\Rv^2\Lv\pPsi_t\Rv$, so
\begin{eqnarray*}
\log p\Lp\widetilde{\pY}_t^+|\ptheta\Rp & = & -\frac{m(n-r)}{2}\log(2\pi) - \frac{n-r}{2}\log\Lv\pH_t\Rv - \frac{m}{2}\log\Lv\pPsi_t\Rv \\
& & - m\log\Lv\pB\Rv - \frac{1}{2}\text{tr}\Lb \pH_t^{-1}\Lp\widetilde{\pY}_t^+-\widetilde{\pD}_t^+\Rp\widetilde{\pPsi}_t^{-1}\Lp\widetilde{\pY}_t^+-\widetilde{\pD}_t^+\Rp^\top\Rb. 
\end{eqnarray*}
The trace term is unchanged. Using $\widetilde{\pPsi}_t^{-1} = \Lp\pB^\top\Rp^{-1}\pPsi_t^{-1}\pB^{-1}$ and $\widetilde{\pY}_t^+-\widetilde{\pD}_t^+ = \Lp\pY_t^+-\pD_t^+\Rp\pB^\top$,
\begin{eqnarray*}
	\pH_t^{-1}\Lp\widetilde{\pY}_t^+-\widetilde{\pD}_t^+\Rp\widetilde{\pPsi}_t^{-1}\Lp\widetilde{\pY}_t^+-\widetilde{\pD}_t^+\Rp^\top & = & \pH_t^{-1}\Lp\pY_t^+-\pD_t^+\Rp\pB^\top\Lp\pB^\top\Rp^{-1}\pPsi_t^{-1}\pB^{-1}\pB\Lp\pY_t^+-\pD_t^+\Rp^\top \\
	& = & \pH_t^{-1}\Lp\pY_t^+-\pD_t^+\Rp\pPsi_t^{-1}\Lp\pY_t^+-\pD_t^+\Rp^\top,
\end{eqnarray*}
so $\log p\Lp\widetilde{\pY}_t^+|\ptheta\Rp = \log p\Lp\pY_t^+|\ptheta\Rp - m\log\Lv\pB\Rv$.

Second, $\widetilde{\pJ}_t = \Lb\begin{matrix}\eye_r & \pzero \\ \pzero & \pB\end{matrix}\Rb\pJ_t$, so $\Lv\widetilde{\pJ}_t\Rv = \Lv\pB\Rv\cdot\Lv\pJ_t\Rv$ and $m\log\Lv\widetilde{\pJ}_t\Rv = m\log\Lv\pB\Rv + m\log\Lv\pJ_t\Rv$. Adding the two
\begin{eqnarray*}
	\log p\Lp\widetilde{\pY}_t^+|\ptheta\Rp + m\log\Lv\widetilde{\pJ}_t\Rv & = & \Lb\log p\Lp\pY_t^+|\ptheta\Rp - m\log\Lv\pB\Rv\Rb + \Lb m\log\Lv\pB\Rv + m\log\Lv\pJ_t\Rv\Rb \\
	&  = & \log p\Lp\pY_t^+|\ptheta\Rp + m\log\Lv\pJ_t\Rv.
\end{eqnarray*} 
The $\pB$-dependence cancels exactly, date by date, and hence in the sum over $t=1,\ldots,T$: the total log-likelihood is invariant to the choice of basis for the complement.

\subsubsection{Collapsing the observation equation from the left} \label{JK collapse}

Section \ref{loglik derivation} collapses the column dimension of the observation equation from $n$ to $r$, since $\pA_t$ only enters through $\pY_t\pJ_t^{*,\top}$. A parallel collapse is available on the row side, following \cite{JungbackerKoopman(15)}, whenever $m$ is large relative to $s$, the row dimension of $\pA_t$ ($\pZ_t$ is $m\times s$): the $m\times r$ matrix $\pY_t^{*}$ can be replaced by an $s\times r$ statistic that is sufficient for $\pA_t$, at the cost of a state-free remainder term in the log-likelihood. Since $s$ is typically a small multiple of the number of factors while $m$ counts the series, this turns an $m\times m$ matrix inversion at every date of the Kalman filter into an $s\times s$ one.

\textbf{Setup.}  By Proposition 1, $\pY_{t}^{*} = \pD_{t}^{*} + \pZ_{t}\pA_{t} + \pE_{1t}^{*}$ with $\pE_{1t}^{*}\sim\text{MN}\Lp\pzero,\pH_{t},\pU\Rp$, where $\pD_t^*=\pD_t\pJ_t^{*,\top}$. Assume $\pZ_{t}$ ($m\times s$) has full column rank $s\le m$ (Remark 3 relaxes this). Define the $s\times m$ GLS collapsing matrix
\begin{eqnarray*}
	\pZ_{t}^{\#} \ = \ \Lp \pZ_{t}^{\top}\pH_{t}^{-1}\pZ_{t}\Rp^{-1}\pZ_{t}^{\top}\pH_{t}^{-1}, \qquad \text{so that } \pZ_{t}^{\#}\pZ_{t}=\eye_{s},
\end{eqnarray*}
and let $\pC_{t}^{\perp}$ be any $(m-s)\times m$ matrix satisfying $\pC_{t}^{\perp}\pZ_{t}=\pzero$ such that $\pO_{t} := \Lp\pZ_{t}^{\#,\top}\ \ \pC_{t}^{\perp,\top}\Rp^{\top}$ is invertible (e.g.\ $\pC_t^\perp$ any basis of the orthogonal complement of the column space of $\pZ_t$). Define the collapsed observation and its complement
\begin{eqnarray*}
	\pY_{t}^{*\#} \ = \ \pZ_{t}^{\#}\pY_{t}^{*} \quad (s\times r), \qquad \pY_{t}^{*\perp} \ = \ \pC_{t}^{\perp}\pY_{t}^{*} \quad \Lp(m-s)\times r\Rp.
\end{eqnarray*}
Substituting the observation equation and using $\pZ_{t}^{\#}\pZ_{t}=\eye_{s}$, $\pC_{t}^{\perp}\pZ_{t}=\pzero$,
\begin{eqnarray*}
	\pY_{t}^{*\#} & = & \pZ_{t}^{\#}\pD_{t}^{*} + \pA_{t} + \pZ_{t}^{\#}\pE_{1t}^{*}, \\
	\pY_{t}^{*\perp} & = & \pC_{t}^{\perp}\pD_{t}^{*} + \pC_{t}^{\perp}\pE_{1t}^{*};
\end{eqnarray*}
the state $\pA_{t}$ has dropped out of $\pY_{t}^{*\perp}$ entirely. Since $\pE_{1t}^{*}\sim\text{MN}\Lp\pzero,\pH_{t},\pU\Rp$ and $\pZ_{t}^{\#},\pC_{t}^{\perp}$ act on the left, the standard matrix-normal transformation rule gives $\pZ_{t}^{\#}\pE_{1t}^{*}\sim\text{MN}\Lp\pzero,\pF_{t}^{\#},\pU\Rp$ with
\begin{eqnarray*}
	\pF_{t}^{\#} \ := \ \pZ_{t}^{\#}\pH_{t}\pZ_{t}^{\#,\top} \ = \ \Lp\pZ_{t}^{\top}\pH_{t}^{-1}\pZ_{t}\Rp^{-1},
\end{eqnarray*}
and $\pC_{t}^{\perp}\pE_{1t}^{*}\sim\text{MN}\Lp\pzero,\pC_{t}^{\perp}\pH_{t}\pC_{t}^{\perp,\top},\pU\Rp$. The two are uncorrelated: $\pZ_{t}^{\#}\pH_{t}\pC_{t}^{\perp,\top} = \Lp\pZ_{t}^{\top}\pH_{t}^{-1}\pZ_{t}\Rp^{-1}\pZ_{t}^{\top}\pC_{t}^{\perp,\top} = \pzero$ using $\pC_{t}^{\perp}\pZ_{t}=\pzero$. Since $\Lp\pA_{t}^{\top},\pE_{1t}^{*,\top}\Rp^{\top}$ is jointly matrix normal with right scale $\pU$ (as in Section \ref{KFS derivation}) and $\Lp\pY_{t}^{*\#,\top},\pY_{t}^{*\perp,\top}\Rp^{\top}$ is a linear function of it, $\pY_{t}^{*\#}$ and $\pY_{t}^{*\perp}$ are independent given $\pY_{1:t-1}$, and $\pY_{t}^{*\perp}$ is independent of $\pA_t$.

\textbf{Left collapse.} \emph{(a) Filtering equivalence.} $p\Lp\pA_{t}\mid \pY_{t}^{*},\pY_{1:t-1}\Rp = p\Lp\pA_{t}\mid \pY_{t}^{*\#},\pY_{1:t-1}\Rp$, and this filtering distribution is obtained by applying the update of Proposition 1 with $\pZ_t\to\eye_s$, $\pD_t^*\to\pZ_t^{\#}\pD_t^*$, $\pH_t\to\pF_t^{\#}$:
\begin{eqnarray*}
	\pA_{t|t} & = & \pA_{t|t-1} + \pP_{t|t-1}\Lp\pP_{t|t-1}+\pF_{t}^{\#}\Rp^{-1}\Lp\pY_{t}^{*\#}-\pZ_{t}^{\#}\pD_{t}^{*}-\pA_{t|t-1}\Rp, \\
	\pP_{t|t} & = & \pP_{t|t-1} - \pP_{t|t-1}\Lp\pP_{t|t-1}+\pF_{t}^{\#}\Rp^{-1}\pP_{t|t-1},
\end{eqnarray*}
replacing every $m\times m$ inversion in the original filter with an $s\times s$ one, at every date $t=1,\ldots,T$.

\emph{(b) Likelihood decomposition.}
\begin{eqnarray*}
	\log p\Lp \pY_{t}^{*}\mid\pY_{1:t-1},\ptheta\Rp \ = \ \log p\Lp \pY_{t}^{*\#}\mid\pY_{1:t-1},\ptheta\Rp + \log p\Lp \pY_{t}^{*\perp}\mid\ptheta\Rp + r\log\Lv\pO_{t}\Rv,
\end{eqnarray*}
where the first term on the right uses the collapsed $s\times r$ filter of part (a) and $\pY_t^{*\perp}\sim\text{MN}\Lp\pC_{t}^{\perp}\pD_{t}^{*},\pC_{t}^{\perp}\pH_{t}\pC_{t}^{\perp,\top},\pU\Rp$ is evaluated in closed form, without any filtering.

\emph{Proof:} (a) $\Lp\pY_{t}^{*\#},\pY_{t}^{*\perp}\Rp$ is an invertible function of $\pY_t^*$, and $\pY_{t}^{*\perp}$ is independent of both $\pA_t$ and $\pY_t^{*\#}$, so $p\Lp\pA_{t}\mid\pY_{t}^{*},\pY_{1:t-1}\Rp = p\Lp\pA_{t}\mid\pY_{t}^{*\#},\pY_{t}^{*\perp},\pY_{1:t-1}\Rp = p\Lp\pA_{t}\mid\pY_{t}^{*\#},\pY_{1:t-1}\Rp$. And $\pY_{t}^{*\#}=\pZ_{t}^{\#}\pD_{t}^{*}+\pA_{t}+\pZ_{t}^{\#}\pE_{1t}^{*}$, with $\pZ_{t}^{\#}\pE_{1t}^{*}\sim\text{MN}\Lp\pzero,\pF_{t}^{\#},\pU\Rp$, is an observation equation of the same form assumed in Proposition 1, so its filtering update applies directly with the stated substitutions.

(b) The map $\pY_{t}^{*}\mapsto\Lp\pY_{t}^{*\#},\pY_{t}^{*\perp}\Rp = \pO_{t}\pY_{t}^{*}$ is linear and invertible, with vec-Jacobian $\Lv\det\Lp\pO_{t}\otimes\eye_{r}\Rp\Rv=\Lv\pO_{t}\Rv^{r}$ by the same argument as Section \ref{loglik derivation}, so $p\Lp\pY_{t}^{*}\mid\pY_{1:t-1},\ptheta\Rp = p\Lp\pY_{t}^{*\#},\pY_{t}^{*\perp}\mid\pY_{1:t-1},\ptheta\Rp\,\Lv\pO_{t}\Rv^{r}$. Independence of $\pY_{t}^{*\#}$ and $\pY_{t}^{*\perp}$ factors the joint density, and taking logs gives the decomposition. \hfill$\blacksquare$

Combining part (b) with the right-hand decomposition of Section \ref{loglik derivation} and summing over $t$ gives a fully collapsed log-likelihood,
\begin{eqnarray*}
	\log p\Lp\pY_{1:T}\mid\ptheta\Rp & = & \log p\Lp \pY_{1:T}^{*\#}\mid\ptheta\Rp + \sum_{t=1}^{T}\log p\Lp\pY_{t}^{*\perp}\mid\ptheta\Rp + \log p\Lp\pY_{1:T}^{+}\mid\ptheta\Rp \\
	& & + \sum_{t=1}^{T}\Lb r\log\Lv\pO_{t}\Rv + m\log\Lv\pJ_{t}\Rv\Rb,
\end{eqnarray*}
in which $\log p\Lp \pY_{1:T}^{*\#}\mid\ptheta\Rp$ is the only term requiring a Kalman filter, and that filter is now $s\times r$ rather than $m\times r$.

\remark{Invariance to the choice of $\pC_t^\perp$} Any two valid choices $\pC_{t}^{\perp}$ and $\widetilde{\pC}_{t}^{\perp}=\pM\pC_{t}^{\perp}$, for invertible $\pM$ ($(m-s)\times(m-s)$), give the same total $\log p\Lp\pY_t^*\mid\pY_{1:t-1},\ptheta\Rp$ in part (b). Writing $\widetilde{\pY}_t^{*\perp}=\pM\pY_t^{*\perp}$, the density transforms as $\log p\Lp\widetilde{\pY}_t^{*\perp}\mid\ptheta\Rp = \log p\Lp\pY_t^{*\perp}\mid\ptheta\Rp - r\log\Lv\pM\Rv$ (the same Kronecker-Jacobian rule, now for an $(m-s)\times(m-s)$ block), while $\widetilde{\pO}_t=\Lp\pZ_t^{\#,\top}\ \ \pM^{\top}\pC_t^{\perp,\top}\Rp^{\top}$ has $\Lv\widetilde{\pO}_t\Rv=\Lv\pM\Rv\Lv\pO_t\Rv$ (row-block scaling of a block matrix), so $r\log\Lv\widetilde{\pO}_t\Rv = r\log\Lv\pM\Rv+r\log\Lv\pO_t\Rv$. The $\mp r\log\Lv\pM\Rv$ terms cancel, so the total is unchanged; only the sufficient statistic $\pY_t^{*\#}$, not the specific complement basis, matters.

%

\subsubsection{Derivation of the smoothed distribution} \label{KS derivation}

\emph{Proof of Proposition 2:}  The proof uses the matrix regression lemma applied to the conditional joint distribution of $\pV_{t:T}$ and $\pA_{t}$ given the data $\pY_{1:t-1}$. First, we derive the moments of the conditional joint distribution. The mean and variance of $\pA_{t}$ are
\begin{eqnarray*}
	\E\Lb \pA_{t}|\pY_{1:t-1}\Rb &=& \pA_{t|t-1} \\
	\V\Lb \pA_{t}|\pY_{1:t-1}\Rb &=&  \pU \otimes \pP_{t|t-1}
\end{eqnarray*}
The mean of any prediction error $\pV_{t}$ for any date $t$ is
\begin{eqnarray*}
	\E\Lb \pV_{t}|\pY_{1:t-1}\Rb &=& \E\Lb \pY_{t}^{*} - \pD_{t}\pJ_{t}^{*,\top} - \pZ_{t}\pA_{t|t-1} |\pY_{1:t-1}\Rb \\
	& = &  \E\Lb  \pZ_{t}\pA_{t} + \pE_{1t}^{*} - \pZ_{t}\pA_{t|t-1} |\pY_{1:t-1}\Rb \\
	&=& \E\Lb  \pE_{1t}^{*}|\pY_{1:t-1}\Rb \ = \ \pzero
\end{eqnarray*}
The variance of any prediction error $\pV_{t}$ is
\begin{eqnarray*}
	\V\Lb \text{vec}\Lp\pV_{t}\Rp|\pY_{1:t-1}\Rb &=& \V\Lb \text{vec}\Lp\pY_{t}^{*} - \pD_{t}\pJ_{t}^{*,\top} - \pZ_{t}\pA_{t|t-1}\Rp |\pY_{1:t-1}\Rb \\
	&=& \V\Lb \text{vec}\Lp\pZ_{t}\pA_{t}\Rp |\pY_{1:t-1}\Rb + \V\Lb  \text{vec}\Lp\pE_{1t}^{*}\Rp |\pY_{1:t-1}\Rb \\
	&=& \Lp \eye \otimes \pZ_{t}\Rp\V\Lb \text{vec}\Lp\pA_{t}\Rp |\pY_{1:t-1}\Rb \Lp \eye \otimes \pZ_{t}^{\top}\Rp +  \pU \otimes \pH_{t} \\
	&=& \Lp \eye \otimes \pZ_{t}\Rp \Lp \pU \otimes \pP_{t|t-1}\Rp\Lp \eye \otimes \pZ_{t}^{\top}\Rp +  \pU \otimes \pH_{t} \\
	&=& \pU \otimes \Lp \pZ_{t}\pP_{t|t-1}\pZ_{t}^{\top} + \pH_{t}\Rp \\
	&=& \pU \otimes \pF_{t} 
\end{eqnarray*}
Throughout the remainder of the proof we use that $\pE_{2t}$ and $\pE_{1t}^{*}$ are, by the model's assumptions, independent of $\pA_{t}$ and of each other, and independent across dates. This is what allows every cross-moment below that involves an $\pE_{2j}$ or $\pE_{1j}^{*}$ with $j\ge t$ to be evaluated using only $\pP_{t|t-1}$ and the recursion matrices.

Next, we define the state estimation error as
\begin{eqnarray*}
	\pX_{t+1} &=& \pA_{t+1} - \pA_{t+1|t} \\
	&=& \pT_{t}\pA_{t} + \pC_{t} + \pR_{t}\pE_{2t} - \pT_{t}\pA_{t|t-1} - \pC_{t} - \pK_{t}\pV_{t} \\
	&=& \pT_{t}\pX_{t} + \pR_{t}\pE_{2t} - \pK_{t}\pZ_{t}\pX_{t} - \pK_{t}\pE_{1t}^{*} \\
	&=& \pL_{t}\pX_{t} + \pR_{t}\pE_{2t} - \pK_{t}\pE_{1t}^{*}
\end{eqnarray*}
This implies that we can write the model in error form
\begin{eqnarray*}
	\pV_{t} &=& \pZ_{t}\pX_{t} + \pE_{1t}^{*} \\
	\pX_{t+1} &=& \pL_{t}\pX_{t} + \pR_{t}\pE_{2t} - \pK_{t}\pE_{1t}^{*}
\end{eqnarray*}
The covariance between $\pA_{t}$ and $\pV_{j}$ for any date $j=t,\ldots,T$ is
\begin{eqnarray*}
	\text{cov}\Lb \text{vec}\Lp\pA_{t}\Rp,\text{vec}\Lp\pV_{j}\Rp|\pY_{1:t-1}\Rb &=& \E\Lb \text{vec}\Lp\pA_{t}\Rp \text{vec}\Lp\pV_{j}\Rp^{\top}  |\pY_{1:t-1}\Rb \\
	&=& \E\Lb \text{vec}\Lp\pA_{t}\Rp \text{vec}\Lp\pZ_{j}\pX_{j} + \pE_{1j}^{*}\Rp^{\top}  |\pY_{1:t-1}\Rb \\
	&=& \E\Lb \text{vec}\Lp\pA_{t}\Rp \text{vec}\Lp\pX_{j}\Rp^{\top} |\pY_{1:t-1}\Rb\Lp \eye \otimes \pZ_{j}^{\top}\Rp
\end{eqnarray*}
The expected second moment is then computed as
\begin{eqnarray*}
	\E\Lb \text{vec}\Lp\pA_{t}\Rp \text{vec}\Lp\pX_{t}\Rp^{\top} |\pY_{1:t-1}\Rb &=& \pP_{t|t-1} \\
	\E\Lb \text{vec}\Lp\pA_{t}\Rp \text{vec}\Lp\pX_{t+1}\Rp^{\top} |\pY_{1:t-1}\Rb &=& \E\Lb \text{vec}\Lp\pA_{t}\Rp \text{vec}\Lp \pL_{t}\pX_{t} + \pR_{t}\pE_{2t} - \pK_{t}\pE_{1t} \Rp^{\top} |\pY_{1:t-1}\Rb \\
	&=& \E\Lb \text{vec}\Lp\pA_{t}\Rp \text{vec}\Lp \pX_{t} \Rp^{\top} |\pY_{1:t-1}\Rb \Lp \eye \otimes \pL_{t}^{\top}\Rp \\
	& =& \Lp \pU \otimes \pP_{t|t-1}\Rp\Lp \eye \otimes \pL_{t}^{\top}\Rp  \\
	& =& \Lp \pU \otimes \pP_{t|t-1}\pL_{t}^{\top}\Rp
\end{eqnarray*}
By iterated substitution, one can show that
\begin{eqnarray*}
	\E\Lb \text{vec}\Lp\pA_{t}\Rp \text{vec}\Lp\pX_{t+2}\Rp^{\top} |\pY_{1:t-1}\Rb & =&\Lp \pU \otimes \pP_{t|t-1}\pL_{t}^{\top}\pL_{t+1}^{\top}\Rp
\end{eqnarray*}
and more generally
\begin{eqnarray*}
	\E\Lb \text{vec}\Lp\pA_{t}\Rp \text{vec}\Lp\pX_{T}\Rp^{\top} |\pY_{1:t-1}\Rb & =&\Lp \pU \otimes \pP_{t|t-1}\pL_{t}^{\top}\pL_{t+1}^{\top}\ldots\pL_{T-1}^{\top}\Rp
\end{eqnarray*}
From these conditional expectations, the covariance between $\pA_{t}$ and $\pV_{j}$, $j=t,\ldots,T$, is
\begin{eqnarray*}
	\text{cov}\Lb\text{vec}\Lp\pA_{t}\Rp,\text{vec}\Lp\pV_{j}\Rp|\pY_{1:t-1}\Rb &=& \Lp \pU \otimes \pP_{t|t-1}\pL_{t}^{\top}\pL_{t+1}^{\top}\ldots\pL_{j-1}^{\top}\Rp\Lp \eye\otimes \pZ_{j}^{\top}\Rp \\
	&=& \pU \otimes \Lp \pP_{t|t-1}\pL_{t}^{\top}\pL_{t+1}^{\top}\ldots\pL_{j-1}^{\top}\pZ_{j}^{\top}\Rp,
\end{eqnarray*}
with the convention that the product $\pL_{t}^{\top}\ldots\pL_{j-1}^{\top}$ is the identity when $j=t$.

To apply the matrix regression lemma to the joint distribution of $\pA_{t}$ and the stacked prediction errors $\Lp \pV_{T}^{\top},\ldots,\pV_{t}^{\top}\Rp^{\top}$, we also need the cross-covariances $\text{cov}\Lp \pV_{i},\pV_{j}\Rp$ for $i\ne j$. These vanish: for $t\le i<j\le T$, $\pV_{j}$ is a fixed invertible linear function of $\pY_{j}$ given $\pY_{1:j-1}$, and $\pY_{1:j-1}$ carries exactly the same information as $\Lp \pY_{1:t-1},\pV_{t},\ldots,\pV_{j-1}\Rp$, since each $\pV_{k}$ is recovered from $\pY_{k}$ and $\pY_{1:k-1}$ (and vice versa) by construction of the filter. Because $\E\Lb \pV_{j}|\pY_{1:j-1}\Rb=\pzero$ (shown above), the law of iterated expectations gives, for $i \in \{t,\ldots,j-1\}$,
\begin{eqnarray*}
	\text{cov}\Lb\text{vec}\Lp\pV_{i}\Rp,\text{vec}\Lp\pV_{j}\Rp|\pY_{1:t-1}\Rb &=& \E\Lb\text{vec}\Lp\pV_{i}\Rp \E\Lb\text{vec}\Lp\pV_{j}\Rp^{\top}|\pY_{1:t-1},\pV_{t},\ldots,\pV_{j-1}\Rb|\pY_{1:t-1}\Rb \\
	&=& \pzero,
\end{eqnarray*}
using $\E\Lb\pV_{j}|\pY_{1:t-1},\pV_{t},\ldots,\pV_{j-1}\Rb=\E\Lb\pV_{j}|\pY_{1:j-1}\Rb=\pzero$. Hence, conditional on $\pY_{1:t-1}$, the prediction errors $\pV_{t},\ldots,\pV_{T}$ are mutually uncorrelated, and the joint distribution of $\pA_{t}$ and $\Lp \pV_{T}^{\top},\ldots,\pV_{t}^{\top}\Rp^{\top}$ is matrix normal with right scale $\pU$, mean $\Lp \pA_{t|t-1}^{\top},\pzero,\ldots,\pzero\Rp^{\top}$, and \emph{block-diagonal} left scale matrix
\begin{eqnarray*}
	\Lb \begin{matrix}
		\pF_{T} & & & \\
		 & \ddots & & \\
		 & & \pF_{t} & \\
		 & & & \pP_{t|t-1}
	\end{matrix}\Rb.
\end{eqnarray*}

Applying the matrix regression lemma with $\pA_{t}$ latent and $\Lp \pV_{T},\ldots,\pV_{t}\Rp$ jointly observed, the block-diagonal structure of the left scale matrix means the regression coefficient on the stacked prediction errors is itself block diagonal, so the smoothed mean reduces to a sum of the individual per-date terms computed above:
\begin{eqnarray*}
	\pA_{t|T} & = & \E\Lb \pA_{t}|\pY_{1:T}\Rb \\
	& = & \pA_{t|t-1} + \sum_{j=t}^{T} \pP_{t|t-1}\pL_{t}^{\top}\ldots\pL_{j-1}^{\top}\pZ_{j}^{\top}\pF_{j}^{-1}\pV_{j} \\
	& = & \pA_{t|t-1} + \pP_{t|t-1}\pG_{t-1}.
\end{eqnarray*}
The smoothed mean is therefore a combination of the one-step-ahead predictor of the state and all future prediction errors, discounted back to date $t$ through the matrices $\pL_{t},\ldots,\pL_{T-1}$. Peeling off the $j=t$ term from the sum defining $\pG_{t-1}$ shows that it satisfies the backward recursion
\begin{eqnarray*}
	\pG_{t-1} &=& \pZ_{t}^{\top}\pF_{t}^{-1}\pV_{t} + \pL_{t}^{\top}\underbrace{\sum_{j=t+1}^{T}\pL_{t+1}^{\top}\ldots\pL_{j-1}^{\top}\pZ_{j}^{\top}\pF_{j}^{-1}\pV_{j}}_{=\ \pG_{t}} \\
	& = & \pZ_{t}^{\top}\pF_{t}^{-1}\pV_{t} + \pL_{t}^{\top}\pG_{t},
\end{eqnarray*}
starting from $\pG_{T} = \pzero_{s \times r}$, which is the empty sum obtained when $t=T+1$.

By the same block-diagonal argument, the smoothed scale matrix is
\begin{eqnarray*}
	\pP_{t|T} &=& \pP_{t|t-1} - \sum_{j=t}^{T}\pP_{t|t-1}\pL_{t}^{\top}\ldots\pL_{j-1}^{\top}\pZ_{j}^{\top}\pF_{j}^{-1}\pZ_{j}\pL_{j-1}\ldots\pL_{t}\pP_{t|t-1} \\
	&  = &  \pP_{t|t-1} - \pP_{t|t-1}\pN_{t-1}\pP_{t|t-1},
\end{eqnarray*}
and peeling off the $j=t$ term as before shows that $\pN_{t-1}$ satisfies the backward recursion
\begin{eqnarray*}
	\pN_{t-1} &=& \pZ_{t}^{\top}\pF_{t}^{-1}\pZ_{t} + \pL_{t}^{\top}\pN_{t}\pL_{t},
\end{eqnarray*}
with initial condition $\pN_{T} \ = \ \pzero_{s\times s}$.

\subsubsection{Derivation of the simulation smoother} \label{Sim smoo derivation}

\emph{Proof of Proposition 3:} Let $\pA^{d} = \Lp \pA_{1}^{d,\top},\ldots,\pA_{T}^{d,\top}\Rp^{\top}$, $\widehat{\pA} = \Lp \widehat{\pA}_{1|T}^{\top}, \ldots,\widehat{\pA}_{T|T}^{\top}\Rp^{\top}$ and $\pA^{\dagger} = \Lp \pA_{1}^{\dagger,\top},\ldots,\pA_{T}^{\dagger,\top}\Rp^{\top}$ denote $Ts \times r$ matrices that collect all time periods of the state posterior draw, state smoothed estimates, and state random draw. We also collect all data in $Tm \times r$ matrices $\pY^{*} = \Lp \pY_{1}^{*\top},\ldots,\pY_{T}^{*\top}\Rp^{\top}$ and $\pY^{*\dagger} = \Lp \pY_{1}^{*\dagger,\top},\ldots,\pY_{T}^{*\dagger,\top}\Rp^{\top}$. Next, we note that the joint unconditional distribution of the state matrices and data are 
\begin{eqnarray*}
	\Lp\begin{matrix}
		\pA  \\
		\pY^{*} \\
		\pA^{\dagger} \\
		\pY^{*\dagger}
	\end{matrix}\Rp & \sim & \text{MN}\Lp \Lb\begin{matrix}
		\pM_{a}  \\
		\pM_{y}^{*} \\
		\pzero \\
		\pzero
	\end{matrix}\Rb,\Lb\begin{matrix}
		\pSigma_{aa} & \pSigma_{ay} & \pzero & \pzero  \\
		\pSigma_{ya} & \pSigma_{yy} & \pzero & \pzero \\
		\pzero & \pzero & \pSigma_{aa} & \pSigma_{ay}\\
		\pzero &\pzero  &	\pSigma_{ya} & \pSigma_{yy}
	\end{matrix}\Rb,\pU\Rp 
\end{eqnarray*}
The matrices of means, variances, and covariances are functions of the parameters in the system matrices $\pD_{t}, \pZ_{t}, \pH_{t}$, etc. The dagger copy $\pA^{\dagger}$ and its implied pseudo-data $\pY^{*\dagger}$ are simulated using the \emph{same} transition, loading, and variance matrices $\pT_{t},\pZ_{t},\pR_{t},\pQ_{t},\pH_{t}$ as the actual model, so that $(\pA^{\dagger},\pY^{*\dagger})$ shares the scale matrices $\pSigma_{aa},\pSigma_{ay},\pSigma_{yy}$ with $(\pA,\pY^{*})$; only the constants $\pC_{t}$, $\pD_{t}^{*}$ and the mean of the initial state $\pA_{1|0}$ are set to zero when generating the dagger copy, which is why $\pA^{\dagger}$ and $\pY^{*\dagger}$ have mean zero while the variance of the initial state, $\pP_{1|0}$, is left unchanged. Finally, the random matrices $\pA^{\dagger}$ and $\pY^{*\dagger}$ are independent of $\pA$ and $\pY^{*}$, since they are built from an independent draw of the disturbances (and initial state), resulting in a block diagonal left scale matrix.

Define $\widehat{\pA} = \E\Lb \pA|\pY^{*}\Rb$ as the smoothed mean of the actual data (Proposition 2, stacked over $t$), and $\widehat{\pA}^{\dagger} = \E\Lb \pA^{\dagger}|\pY^{*\dagger}\Rb$ as the smoothed mean obtained by applying the \emph{same} smoothing formula to the simulated pseudo-data $\pY^{*\dagger}$; since $(\pA^{\dagger},\pY^{*\dagger})$ shares the scale matrices of $(\pA,\pY^{*})$ but has mean zero, $\widehat{\pA}^{\dagger} = \pSigma_{ay}\pSigma_{yy}^{-1}\pY^{*\dagger}$. Following \cite{DurbinKoopman(02)}, the simulated draw is
\begin{eqnarray*}
	\pA^{d} \ := \ \widehat{\pA} - \widehat{\pA}^{\dagger} + \pA^{\dagger}.
\end{eqnarray*}
Because $\widehat{\pA}$ and $\widehat{\pA}^{\dagger}$ are both evaluations of the same linear map $\pY \mapsto \pM_{a}+\pSigma_{ay}\pSigma_{yy}^{-1}\Lp \pY - \pM_{y}^{*}\Rp$ (at $\pY=\pY^{*}$, and at $\pY=\pY^{*\dagger}$ using $\pM_{y}^{*}=\pzero$ for the dagger copy), the two smoothing passes can be combined into a single application of the regression formula to the pseudo-data $\widehat{\pY} := \pY^{*}-\pY^{*\dagger}$, so that $\pA^{d}$ can be written in stacked matrix form as
\begin{eqnarray*}
	\pA^{d} \ = \  \pM_{a} + \pSigma_{ay}\pSigma_{yy}^{-1}\Lp \widehat{\pY} - \pM_{y}^{*}\Rp + \pA^{\dagger} & = & \pM_{a} + \pSigma_{ay}\pSigma_{yy}^{-1}\Lp \pY^{*} - \pY^{*\dagger} - \pM_{y}^{*}\Rp + \pA^{\dagger} \\
	&=& \pM_{a} + \pSigma_{ay}\pSigma_{yy}^{-1}\Lp \pY^{*} - \pM_{y}^{*}\Rp + \pA^{\dagger} \\
	& &  - \pSigma_{ay}\pSigma_{yy}^{-1}\pY^{*\dagger}
\end{eqnarray*}
In practice this means the smoothing recursion of Proposition 2 need only be run once, on the pseudo-data $\widehat{\pY}$, rather than separately on $\pY^{*}$ and $\pY^{*\dagger}$.
Next, we take conditional expectations of the draw $\pA^{d}$  given by
\begin{eqnarray}
	\E\Lp \pA^{d}|\pY^{*}\Rp &=& \pM_{a} + \pSigma_{ay}\pSigma_{yy}^{-1}\Lp \pY^{*} - \pM_{y}^{*}\Rp + \E\Lp \pA^{\dagger}|\pY^{*}\Rp  - \pSigma_{ay}\pSigma_{yy}^{-1}\E\Lp \pY^{*\dagger} |\pY^{*}\Rp \nonumber \\
	&=& \pM_{a} + \pSigma_{ay}\pSigma_{yy}^{-1}\Lp \pY^{*} - \pM_{y}^{*}\Rp + \E\Lp \pA^{\dagger}\Rp  - \pSigma_{ay}\pSigma_{yy}^{-1}\E\Lp \pY^{*\dagger} \Rp \nonumber \\
	&=& \pM_{a} + \pSigma_{ay}\pSigma_{yy}^{-1}\Lp \pY^{*} - \pM_{y}^{*}\Rp \label{cond mean SS}
\end{eqnarray}
The conditional variance is
\begin{eqnarray}
	\V\Lb\text{vec}\Lp\pA^{d}\Rp|\pY^{*}\Rb &=&  \V\Lb \text{vec}\Lp\pA^{\dagger}\Rp |\pY^{*}\Rb  + \V\Lb\text{vec}\Lp\pSigma_{ay}\pSigma_{yy}^{-1}\pY^{*\dagger}\Rp |\pY^{*}\Rb \nonumber \\
	& & - 2\text{C}\Lb \text{vec}\Lp\pA^{\dagger}\Rp,\text{vec}\Lp\pSigma_{ay}\pSigma_{yy}^{-1}\pY^{*\dagger}\Rp |\pY^{*}\Rb \nonumber \\
	&=&  \V\Lb \text{vec}\Lp\pA^{\dagger}\Rp \Rb  + \Lp\eye \otimes \pSigma_{ay}\pSigma_{yy}^{-1}\Rp\V\Lb  \text{vec}\Lp\pY^{*\dagger}\Rp\Rb \Lp\eye \otimes \pSigma_{yy}^{-1}\pSigma_{ay}^{\top}\Rp \nonumber \\
	& & - 2\text{C}\Lb \text{vec}\Lp\pA^{\dagger}\Rp,\text{vec}\Lp\pY^{*\dagger}\Rp\Rb \Lp\eye \otimes \pSigma_{yy}^{-1}\pSigma_{ay}^{\top}\Rp  \nonumber \\
	&=&  \Lp \pU \otimes \pSigma_{aa}\Rp  + \Lp\eye \otimes \pSigma_{ay}\pSigma_{yy}^{-1}\Rp \Lp \pU \otimes \pSigma_{yy}\Rp \Lp\eye \otimes \pSigma_{yy}^{-1}\pSigma_{ay}^{\top}\Rp \nonumber \\
	& & - 2 \Lp \pU \otimes \pSigma_{ay}\Rp \Lp\eye \otimes \pSigma_{yy}^{-1}\pSigma_{ay}^{\top}\Rp \nonumber \\
	&=&  \pU \otimes \Lp\pSigma_{aa}- \pSigma_{ay}\pSigma_{yy}^{-1}\pSigma_{ay}^{\top}\Rp \label{cond var SS}
\end{eqnarray}
Since $\pA^{d}$ is a linear function of the jointly matrix normal random matrices $\pA^{\dagger}$ and $\pY^{*\dagger}$ (conditional on $\pY^{*}$, which enters only as a fixed shift), $\pA^{d}|\pY^{*}$ is itself matrix normal with right scale $\pU$; a matrix normal distribution is fully characterized by its mean and left and right scale matrices. The conditional mean (\ref{cond mean SS}) and variance (\ref{cond var SS}) of the draw $\pA^{d}$ are equivalent to the mean and variance of the joint smoothing distribution for the transformed model derived in Proposition 2. Therefore, $\pA^{d}$ and $\pA|\pY^{*}$ have the same matrix normal distribution, so $\pA^{d}$ is a valid random draw from the smoothing distribution.

\subsubsection{Derivation of the algorithm for sampling discrete states} \label{discrete state simulation derivation}

In this sub-section, all distributions are functions of $\ptheta$, which we omit from the notation. 
 
\emph{Proof of the proposition:} The joint likelihood is the integral over all latent states but conditional on the discrete indicators
\begin{eqnarray*}
	p\Lp \pY_{1:T}|\ps_{1:T}\Rp  & = & \int \ldots \int \prod_{j=1}^{T} p\Lp \pY_{j}|\pA_{j},s_{j}\Rp \prod_{j=2}^{T}p\Lp \pA_{j}|\pA_{j-1},s_{j}\Rp p\Lp \pA_{1}|s_{1}\Rp d\pA_{1},\ldots, d\pA_{T} 
\end{eqnarray*}
The likelihood can always be written in terms of the forward-backward decomposition, see Definition 3.1.6 in \cite{CappeMoulinesRyden(05)}.
\begin{eqnarray*}
	p\Lp \pY_{1:T}|\ps_{1:T}\Rp  & = & \int \alpha_{t}\Lp \pA_{t}|\pY_{1:t},s_{1:t}\Rp\beta_{t}\Lp \pA_{t}|\pY_{t+1:T},s_{t+1:T}\Rp d\pA_{t}
\end{eqnarray*}
where the forward measure $\alpha_{t}\Lp \pA_{t}|\pY_{1:t},s_{1:t}\Rp$ and backwards function $\beta_{t}\Lp \pA_{t}|\pY_{t+1:T},s_{t+1:T}\Rp$ split the original multiple integral at a given date $t$ as
\begin{eqnarray*}
	\alpha_{t}\Lp \pA_{t}|\pY_{1:t},s_{1:t}\Rp  & = & \int \ldots \int \prod_{j=1}^{t} p\Lp \pY_{j}|\pA_{j},s_{j}\Rp \prod_{j=2}^{t}p\Lp \pA_{j}|\pA_{j-1},s_{j}\Rp \\
	& &  p\Lp \pA_{1}|\ps_{1}\Rp d\pA_{1},\ldots,d\pA_{t-1} \\
	\beta_{t}\Lp \pA_{t}|\pY_{t+1:T},s_{t+1:T}\Rp & = & \int \ldots \int \prod_{j=t+1}^{T} p\Lp \pY_{j}|\pA_{j},s_{j}\Rp \\
	& & \prod_{j=t+1}^{T}p\Lp \pA_{j}|\pA_{j-1},s_{j}\Rp d\pA_{t+1},\ldots,d\pA_{T}
\end{eqnarray*}
From Proposition 3.2.1 of \cite{CappeMoulinesRyden(05)}, the forward measures and backward functions have a recursive representation
\begin{eqnarray*}
	\alpha_{t}\Lp \pA_{t}|\pY_{1:t},s_{1:t}\Rp  & = &  \int \alpha_{t-1}\Lp \pA_{t-1}|\pY_{1:t-1},s_{1:t-1}\Rp p\Lp \pY_{t}|\pA_{t},s_{t}\Rp \\
	 & & p\Lp \pA_{t}|\pA_{t-1},s_{t}\Rp  d\pA_{t-1} \\ 
	\beta_{t}\Lp \pA_{t}|\pY_{t+1:T},s_{t+1:T}\Rp & = & \int \beta_{t+1}\Lp \pA_{t+1}|\pY_{t+2:T},s_{t+2:T}\Rp p\Lp \pY_{t+1}|\pA_{t+1},s_{t+1}\Rp  \\
	& & p\Lp \pA_{t+1}|\pA_{t},s_{t+1}\Rp d\pA_{t+1}
\end{eqnarray*} 
Once renormalized, the forward measure is equivalent to the filtered distribution
\begin{eqnarray*}
	p\Lp \pA_{t}|\pY_{1:t},s_{1:t}\Rp  & = & \frac{\alpha_{t}\Lp \pA_{t}|\pY_{1:t},s_{1:t}\Rp}{p\Lp \pY_{1:t}|s_{1:t}\Rp}
\end{eqnarray*} 
and the forward recursion for recursively calculating $\alpha_{t}\Lp \pA_{t}|\pY_{1:t},s_{1:t}\Rp$ is the Kalman filter.

The recursive equations for the backward functions have an analytical solution, which is the backwards information filter. The backwards information filter combined with the forward filtering distribution can be used to calculate the smoothed distribution
\begin{eqnarray*}
	p\Lp \pA_{t}|\pY_{1:T},s_{1:T}\Rp  & = & \frac{p\Lp \pA_{t}|\pY_{1:t},s_{1:t} \Rp \beta_{t}\Lp \pA_{t}|\pY_{t+1:T},s_{t+1:T}\Rp}{\int p\Lp \pA_{t}|\pY_{1:t},s_{1:t}\Rp \beta_{t}\Lp \pA_{t}|\pY_{t+1:T},s_{t+1:T}\Rp d\pA_{t}}
\end{eqnarray*} 
This is Bayes rule where the backward function $\beta_{t}\Lp \pA_{t}|\pY_{t+1:T},s_{t+1:T}\Rp$ is the conditional likelihood and the filtering distribution is the prior.

The likelihood can be written as
\begin{eqnarray*}
	p\Lp \pY_{1:T}|\ps_{1:T}\Rp  & = & \int \alpha_{t}\Lp \pA_{t}|\pY_{1:t},s_{1:t}\Rp\beta_{t}\Lp \pA_{t}|\pY_{t+1:T}, s_{t+1:T}\Rp d\pA_{t} \\
	&=& \int \int  \alpha_{t-1}\Lp \pA_{t-1}|\pY_{1:t-1},s_{1:t-1}\Rp p\Lp \pA_{t}|\pA_{t-1},s_{t}\Rp p\Lp \pY_{t}|\pA_{t},s_{t}\Rp \\
	& &  \beta_{t}\Lp \pA_{t}|\pY_{t+1:T},s_{t+1:T}\Rp d\pA_{t-1} d\pA_{t} \\
	& = & \int \int p\Lp \pY_{1:t-1}|s_{1:t-1}\Rp p\Lp \pA_{t-1}|\pY_{1:t-1},s_{1:t-1}\Rp p\Lp \pA_{t}|\pA_{t-1},s_{t}\Rp \\ 
	& & p\Lp \pY_{t}|\pA_{t},s_{t}\Rp\beta_{t|T}\Lp \pA_{t}|\pY_{t+1:T},s_{t+1:T}\Rp d\pA_{t-1} d\pA_{t} \\
	& \propto & \int \int  p\Lp \pA_{t-1}|\pY_{1:t-1},s_{1:t-1}\Rp p\Lp \pA_{t}|\pA_{t-1},s_{t}\Rp p\Lp \pY_{t}|\pA_{t},s_{t}\Rp \\
	& &  \beta_{t}\Lp \pA_{t}|\pY_{t+1:T},s_{t+1:T}\Rp d\pA_{t-1} d\pA_{t} 
\end{eqnarray*}
This is up to proportionality due to dropping the previous likelihood $p\Lp \pY_{1:t-1}|s_{1:t-1}\Rp$ since it is not a function of $s_{t}$. Integrating out $\pA_{t-1}$, we get the predictive distribution
\begin{eqnarray*}
	p\Lp \pY_{1:T}|\ps_{1:T}\Rp  & \propto & \int  p\Lp \pA_{t}|\pY_{1:t-1},s_{1:t}\Rp p\Lp \pY_{t}|\pA_{t},s_{t}\Rp \beta_{t}\Lp \pA_{t}|\pY_{t+1:T},s_{t+1:T}\Rp  d\pA_{t} 
\end{eqnarray*}
Next, we note that $p\Lp \pY_{t}|\pY_{1:t-1},s_{1:t} \Rp p\Lp \pA_{t}|\pY_{1:t},s_{1:t}\Rp = p\Lp \pA_{t}|\pY_{1:t-1},s_{1:t}\Rp p\Lp \pY_{t}|\pA_{t},s_{t}\Rp$ and therefore 
\begin{eqnarray*}	
	p\Lp \pY_{1:T}|\ps_{1:T}\Rp & \propto & \int  p\Lp \pY_{t}|\pY_{1:t-1},s_{1:t} \Rp p\Lp \pA_{t}|\pY_{1:t},s_{1:t}\Rp \beta_{t}\Lp \pA_{t}|\pY_{t+1:T},s_{t+1:T}\Rp  d\pA_{t} \\
	& \propto & p\Lp \pY_{t}|\pY_{1:t-1},s_{1:t} \Rp  \int p\Lp \pA_{t}|\pY_{1:t},s_{1:t}\Rp \beta_{t}\Lp \pA_{t}|\pY_{t+1:T},s_{t+1:T}\Rp d\pA_{t} 
\end{eqnarray*}
From the definition of the backwards function and Bayes rule, we know that
\begin{eqnarray*}		
	\frac{p\Lp \pA_{t}|\pY_{1:t},s_{1:t} \Rp \beta_{t}\Lp \pA_{t}|\pY_{t+1:T},s_{t+1:T}\Rp}{p\Lp \pA_{t}|\pY_{1:T},s_{1:T}\Rp} &= &   \int p\Lp \pA_{t}|\pY_{1:t},s_{1:t}\Rp \beta_{t}\Lp \pA_{t}|\pY_{t+1:T},s_{t+1:T}\Rp d\pA_{t} 
\end{eqnarray*}
and this equality holds when the left-hand side is evaluated at any value $\pA_{t}$. We choose $\pA_{t} = \pzero$. Then, we get
\begin{eqnarray*}		
p\Lp \pY_{1:T}|\ps_{1:T}\Rp	& \propto &  p\Lp \pY_{t}|\pY_{1:t-1},s_{1:t} \Rp  \frac{p\Lp \pA_{t}= \pzero|\pY_{1:t},s_{1:t} \Rp \beta_{t}\Lp \pA_{t}= \pzero|\pY_{t+1:T},s_{t+1:T}\Rp}{p\Lp \pA_{t}= \pzero|\pY_{1:T},s_{1:T}\Rp}  \\
	& \propto & p\Lp \pY_{t}|\pY_{1:t-1},s_{1:t} \Rp \frac{p\Lp \pA_{t}= \pzero|\pY_{1:t},s_{1:t} \Rp }{p\Lp \pA_{t}= \pzero|\pY_{1:T},s_{1:T}\Rp}  
\end{eqnarray*}
where the backwards function in the numerator can be dropped since it is not a function of $s_{t}$. It remains to evaluate $p\Lp \pY_{t}|\pY_{1:t-1},s_{1:t}\Rp$ and the two densities of $\pA_{t}$ at $\pA_{t}=\pzero$. By the collapsing transformation, $\pY_{t} = \Lp \pY_{t}^{*},\pY_{t}^{+}\Rp$ where $\pY_{t}^{*}$ and $\pY_{t}^{+}$ are independent, $\pY_{t}^{*}$ is the only component that depends on $\pA_{t}$, and $\pY_{t}^{+}$ is independent of $\pY_{1:t-1}$ given $s_{t}$; hence $p\Lp \pY_{t}|\pY_{1:t-1},s_{1:t}\Rp = p\Lp \pY_{t}^{*}|\pY_{1:t-1},s_{1:t}\Rp\, p\Lp \pY_{t}^{+}|s_{t}\Rp$, the first factor matrix normal with mean $\pD\Lp s_{t}\Rp\pJ_{t}^{*,\top}+\pZ\Lp s_{t}\Rp\pA_{t|t-1}$ and scale matrices $\pF_{t}\Lp s_{1:t}\Rp,\pU$ (Proposition 1), and the second matrix normal with mean $\pD\Lp s_{t}\Rp^{+}$ and scale matrices $\pH\Lp s_{t}\Rp,\pPsi_{t}$. Plugging in these expressions together with the filtered and smoothed densities of $\pA_{t}$ evaluated at $\pA_{t}=\pzero$, we get
\begin{eqnarray*}
	p\Lp \pY_{1:T}|s_{1:T}\Rp &\propto & \Lv \pF_{t}\Lp s_{1:t}\Rp\Rv^{-\frac{r}{2}}  \Lv \pP_{t|t}\Lp s_{1:t}\Rp\Rv^{-\frac{r}{2}} \Lv \pP_{t|T}\Lp s_{1:T}\Rp\Rv^{\frac{r}{2}}\Lv\pH\Lp s_{t}\Rp\Rv^{-\frac{(n-r)}{2}}\\
	& & \exp\Lp -\frac{1}{2}\text{tr}\Lb \pU^{-1}\pV_{t}\Lp s_{1:t}\Rp^{\top}\pF_{t}\Lp s_{1:t}\Rp^{-1}\pV_{t}\Lp s_{1:t}\Rp\Rb \Rp \\
	& & \exp\Lp -\frac{1}{2}\text{tr}\Lb
	\pH\Lp s_{t}\Rp^{-1} \Lb\pY_t^+ - \pD\Lp s_{t}\Rp^{+}\Rb\pPsi_{t}^{-1}\Lb\pY_t^{+}-\pD\Lp s_{t}\Rp^{+}\Rb^{\top}
	\Rb\Rp\\
	& & \exp\Lp-\frac{1}{2}\text{tr}\Lb \pU^{-1}\pA_{t|t}\Lp s_{1:t}\Rp^{\top}\pP_{t|t}\Lp s_{1:t}\Rp^{-1}\pA_{t|t}\Lp s_{1:t}\Rp\Rb\Rp \\
	& & \exp\Lp \frac{1}{2}\text{tr}\Lb \pU^{-1}\pA_{t|T}\Lp s_{1:T}\Rp^{\top}\pP_{t|T}\Lp s_{1:T}\Rp^{-1}\pA_{t|T}\Lp s_{1:T}\Rp\Rb\Rp 
\end{eqnarray*}
This completes the proof.

\subsubsection{Derivation of the backward information filter} \label{backward information filter}

In this section, we derive the backwards information filter for the matrix state space model. All distributions are functions of $\ptheta$, which we omit from the notation. We have also shifted the timing of the system matrices in the transition equation, for example defining $\pT\Lp s_{t+1}\Rp = \pT_{t+1}$, $\pC\Lp s_{t+1}\Rp = \pC_{t+1}$, etc.

The backwards function $\beta_{t}\Lp \pA_{t}|\pY_{t+1:T},s_{t+1:T}\Rp$ can be defined recursively as 
\begin{eqnarray*}
	\beta_{t}\Lp \pA_{t}|\pY_{t+1:T},s_{t+1:T}\Rp & = & \int \beta_{t+1}\Lp \pA_{t+1}|\pY_{t+2:T},s_{t+2:T}\Rp p\Lp \pY_{t+1}|\pA_{t+1},s_{t+1}\Rp  \\
	& & p\Lp \pA_{t+1}|\pA_{t},s_{t+1}\Rp d\pA_{t+1}
\end{eqnarray*} 
This is a function of $\pA_{t}$ that integrates future states from the model conditional on the future data $\pY_{t+1:T}$. For the matrix normal state space model, the backwards function can always be written up to a constant of proportionality as the kernel of a matrix normal distribution. The proof uses the information parameterization of the distribution 
\begin{eqnarray*}
	\beta_{t}\Lp \pA_{t}|\pY_{t+1:T},s_{t+1:T}\Rp & = & C \exp\Lp -\frac{1}{2}\text{tr}\Lb \pU^{-1}\Lp\pA_{t}^{\top}\pPi_{t|T}\pA_{t} - \pA_{t}^{\top}\pB_{t|T}  - \pB_{t|T}^{\top}\pA_{t} \Rp\Rb\Rp 
\end{eqnarray*} 
where $\pPi_{t|T}$ is the information matrix and $\pB_{t|T}$ is the scaled mean. The function $\beta_{t}\Lp \cdot \Rp$ is not a probability distribution since it does not integrate to one over $\pA_{t}$. The constant of proportionality $C$ will depend on the future data $\pY_{t+1:T}$, model parameters $\ptheta$, and future discrete states $s_{t+1:T}$ but not the future continuous states.

At time $t=T$, we set $\pB_{T|T} = 0$ and $\pPi_{T|T} = 0$ meaning that $\beta_{T} = 1$ since there are no future observations.

Define $\pM_{t+1} = \pT_{t+1}\pA_{t} + \pC_{t+1}$. We note that $\pA_{t+1} = \pM_{t+1} + \pR_{t+1}\pE_{2t}$. The proof of the information filter needs to recursively integrate out the shock $\pE_{2t}$ instead of $\pA_{t+1}$ since $\pR_{t+1}$ may be singular. The backwards function can be written as
\begin{footnotesize}
	\begin{eqnarray*}
		\beta_{t+1}\Lp \pA_{t},\pE_{2t}|\pY_{t+2:T},s_{t+2:T}\Rp & = & C \exp\Lp -\frac{1}{2}\text{tr}\Lb \pU^{-1}\Lp\pA_{t+1}^{\top}\pPi_{t+1|T}\pA_{t+1} - \pA_{t+1}^{\top}\pB_{t+1|T}  - \pB_{t+1|T}^{\top}\pA_{t+1} \Rp\Rb\Rp \\
		& = & C \exp\Lp -\frac{1}{2}\text{tr}\Lb \pU^{-1}\Lp \Lb\pM_{t+1} + \pR_{t+1}\E_{2t}\Rb^{\top}\pPi_{t+1|T}\Lb\pM_{t+1} + \pR_{t+1}\E_{2t}\Rb \Rp\Rb\Rp \\
		& &  \exp\Lp -\frac{1}{2}\text{tr}\Lb \pU^{-1}\Lp - \Lb\pM_{t+1} + \pR_{t+1}\E_{2t}\Rb^{\top}\pB_{t+1|T}  - \pB_{t+1|T}^{\top}\Lb\pM_{t+1} + \pR_{t+1}\E_{2t}\Rb\Rp\Rb\Rp \\
		& = & C \exp\Lp -\frac{1}{2}\text{tr}\Lb \pU^{-1}\Lp \pM_{t+1}^{\top}\pPi_{t+1|T}\pM_{t+1} \Rp\Rb\Rp \\
		& & \exp\Lp -\frac{1}{2}\text{tr}\Lb \pU^{-1}\Lp  \E_{2t}^{\top}\pR_{t+1}^{\top}\pPi_{t+1|T}\pR_{t+1}\E_{2t} \Rp\Rb\Rp \\
		& &  \exp\Lp -\frac{1}{2}\text{tr}\Lb \pU^{-1}\Lp \pM_{t+1}^{\top}\pPi_{t+1|T}\pR_{t+1}\E_{2t}  +  \E_{2t}^{\top}\pR_{t+1}^{\top}\pPi_{t+1|T}\pM_{t+1} \Rp\Rb\Rp \\
		& &  \exp\Lp -\frac{1}{2}\text{tr}\Lb \pU^{-1}\Lp - \pM_{t+1}^{\top}\pB_{t+1|T}  - \pB_{t+1|T}^{\top}\pR_{t+1}\E_{2t}\Rp\Rb\Rp \\
		& &  \exp\Lp -\frac{1}{2}\text{tr}\Lb \pU^{-1}\Lp -  \E_{2t}^{\top}\pR_{t+1}^{\top}\pB_{t+1|T}  - \pB_{t+1|T}^{\top}\pM_{t+1} \Rp\Rb\Rp \\
	\end{eqnarray*} 
\end{footnotesize}

At date $t$, we define
\begin{footnotesize}
	\begin{eqnarray*}
		\beta_{t}\Lp \pA_{t}|\pY_{t+1:T},s_{t+1:T}\Rp & = & \int 	\beta_{t+1}\Lp \pA_{t},\pE_{2t}|\pY_{t+2:T},s_{t+2:T}\Rp p\Lp \pY_{t+1}|\pA_{t},\pE_{2t},s_{t+1}\Rp p\Lp \pE_{2t}|s_{t+1},\ptheta\Rp d\pE_{2t} \\
		& = & C\int  \Lp\frac{1}{2\pi}\Rp^{mr/2} \Lv \pH_{t+1}\Rv^{-\frac{r}{2}}\Lv \pU\Rv^{-\frac{m}{2}} \\
		&  & \exp\Lp -\frac{1}{2}\text{tr}\Lb \pU^{-1}\Lp \pM_{t+1}^{\top}\pPi_{t+1|T}\pM_{t+1} \Rp\Rb\Rp \\
		& & \exp\Lp -\frac{1}{2}\text{tr}\Lb \pU^{-1}\Lp  \E_{2t}^{\top}\pR_{t+1}^{\top}\pPi_{t+1|T}\pR_{t+1}\E_{2t} \Rp\Rb\Rp \\
		& &  \exp\Lp -\frac{1}{2}\text{tr}\Lb \pU^{-1}\Lp \pM_{t+1}^{\top}\pPi_{t+1|T}\pR_{t+1}\E_{2t}  +  \E_{2t}^{\top}\pR_{t+1}^{\top}\pPi_{t+1|T}\pM_{t+1} \Rp\Rb\Rp \\
		& &  \exp\Lp -\frac{1}{2}\text{tr}\Lb \pU^{-1}\Lp - \pM_{t+1}^{\top}\pB_{t+1|T}  - \pB_{t+1|T}^{\top}\pM_{t+1}  \Rp\Rb\Rp \\
		& &  \exp\Lp -\frac{1}{2}\text{tr}\Lb \pU^{-1}\Lp -  \E_{2t}^{\top}\pR_{t+1}^{\top}\pB_{t+1|T} - \pB_{t+1|T}^{\top}\pR_{t+1}\E_{2t} \Rp\Rb\Rp \\
		&& \exp\Lp-\frac{1}{2}\text{tr}\Lb \pU^{-1}\Lp \pY_{t+1}^{*}-\pZ_{t+1}\pM_{t+1}-\pZ_{t+1}\pR_{t+1}\pE_{2t}\Rp^{\top}\pH_{t+1}^{-1}\Lp \pY_{t+1}^{*}-\pZ_{t+1}\pM_{t+1}-\pZ_{t+1}\pR_{t+1}\pE_{2t}\Rp \Rb\Rp \\
		& &  \Lp\frac{1}{2\pi}\Rp^{qr/2} \Lv \pQ_{t+1}\Rv^{-\frac{r}{2}}\Lv \pU\Rv^{-\frac{q}{2}}\\
		& & \exp\Lp-\frac{1}{2}\text{tr}\Lb  \pU^{-1} \pE_{2,t}^{\top} \pQ_{t+1}^{-1}\pE_{2t}\Rb\Rp  d\pE_{2t} \\
		& = &C\int  \Lp\frac{1}{2\pi}\Rp^{mr/2} \Lv \pH_{t+1}\Rv^{-\frac{r}{2}}\Lv \pU\Rv^{-\frac{m}{2}} \Lp\frac{1}{2\pi}\Rp^{qr/2} \Lv \pQ_{t+1}\Rv^{-\frac{r}{2}}\Lv \pU\Rv^{-\frac{q}{2}}\\
		&  & \exp\Lp -\frac{1}{2}\text{tr}\Lb \pU^{-1}\Lp \pM_{t+1}^{\top}\Lb \pPi_{t+1|T} +\pZ_{t+1}^{\top}\pH_{t+1}^{-1}\pZ_{t+1}\Rb\pM_{t+1} \Rp\Rb\Rp \\
		& &  \exp\Lp -\frac{1}{2}\text{tr}\Lb \pU^{-1}\Lp - \pM_{t+1}^{\top}\Lb \pB_{t+1|T} +\pZ_{t+1}^{\top}\pH_{t+1}^{-1}\pY_{t+1}^{*}\Rb - \Lb \pB_{t+1|T} +\pZ_{t+1}^{\top}\pH_{t+1}^{-1}\pY_{t+1}^{*} \Rb^{\top}\pM_{t+1}  \Rp\Rb\Rp \\
		&& \exp\Lp-\frac{1}{2}\text{tr}\Lb \pU^{-1} \Lp \pY_{t+1}^{*,\top}\pH_{t+1}^{-1}\pY_{t+1}^{*} \Rp\Rb\Rp \\
		&& \exp\Lp-\frac{1}{2}\text{tr}\Lb \pU^{-1} \Lp -\Lb  \pB_{t+1|T} + \pY_{t+1}^{*,\top}\pH_{t+1}^{-1}\pZ_{t+1}\Rb\pR_{t+1}\pE_{2t} +\pM_{t+1}^{\top}\Lb \pPi_{t+1|T} + \pZ_{t+1}^{\top}\pH_{t+1}^{-1}\pZ_{t+1}\Rb\pR_{t+1}\pE_{2t}\Rp \Rb\Rp \\
		&& \exp\Lp-\frac{1}{2}\text{tr}\Lb \pU^{-1} \Lp -\pE_{2t}^{\top}\pR_{t+1}^{\top}\Lb \pB_{t+1|T} +  \pZ_{t+1}^{\top}\pH_{t+1}^{-1}\pY_{t+1}^{*}\Rb +\pE_{2t}^{\top}\pR_{t+1}^{\top}\Lb \pPi_{t+1|T} + \pZ_{t+1}^{\top}\pH_{t+1}^{-1}\pZ_{t+1}\Rb\pM_{t+1}\Rp \Rb\Rp \\
		&& \exp\Lp-\frac{1}{2}\text{tr}\Lb \pU^{-1} \Lp \pE_{2t}\pR_{t+1}^{\top}\Lb \pPi_{t+1|T} + \pZ_{t+1}^{\top}\pH_{t+1}^{-1}\pZ_{t+1}\Rb \pR_{t+1}\pE_{2t}\Rp \Rb\Rp \\
		& & \exp\Lp-\frac{1}{2}\text{tr}\Lb  \pU^{-1} \pE_{2t}^{\top} \pQ_{t+1}^{-1}\pE_{2t}\Rb\Rp  d\pE_{2t} \\
	\end{eqnarray*}
\end{footnotesize}
where we define 
\begin{eqnarray*}
	\widetilde{\pPi}_{t+1|T} &=& \pPi_{t+1|T} +   \pZ_{t+1}^{\top}\pH_{t+1}^{-1}\pZ_{t+1} \\
	\widetilde{\pB}_{t+1|T} & =&  \pB_{t+1|T} +  \pZ_{t+1}^{\top}\pH_{t+1}^{-1}\pY_{t+1}^{*}
\end{eqnarray*}
Using these, we can write
\begin{footnotesize}
	\begin{eqnarray*}
		\beta_{t}\Lp \pA_{t}|\pY_{t+1:T},s_{t+1:T}\Rp & = &C\int  \Lp\frac{1}{2\pi}\Rp^{mr/2} \Lv \pH_{t+1}\Rv^{-\frac{r}{2}}\Lv \pU\Rv^{-\frac{m}{2}} \Lp\frac{1}{2\pi}\Rp^{qr/2} \Lv \pQ_{t+1}\Rv^{-\frac{r}{2}}\Lv \pU\Rv^{-\frac{q}{2}}\\
		&  & \exp\Lp -\frac{1}{2}\text{tr}\Lb \pU^{-1}\Lp \pM_{t+1}^{\top}\widetilde{\pPi}_{t+1|T}\pM_{t+1} \Rp\Rb\Rp \\
		& &  \exp\Lp -\frac{1}{2}\text{tr}\Lb \pU^{-1}\Lp - \pM_{t+1}^{\top}\widetilde{\pB}_{t+1|T} - \widetilde{\pB}_{t+1|T}^{\top}\pM_{t+1}  \Rp\Rb\Rp \\
		&& \exp\Lp-\frac{1}{2}\text{tr}\Lb \pU^{-1} \Lp \pY_{t+1}^{*,\top}\pH_{t+1}^{-1}\pY_{t+1}^{*} \Rp\Rb\Rp \\
		&& \exp\Lp-\frac{1}{2}\text{tr}\Lb \pU^{-1} \Lp -\widetilde{\pB}_{t+1|T}^{\top}\pR_{t+1}\pE_{2t} +\pM_{t+1}^{\top}\widetilde{\pPi}_{t+1|T}\pR_{t+1}\pE_{2t}\Rp \Rb\Rp \\
		&& \exp\Lp-\frac{1}{2}\text{tr}\Lb \pU^{-1} \Lp -\pE_{2t}^{\top}\pR_{t+1}^{\top}\widetilde{\pB}_{t+1|T} +\pE_{2t}^{\top}\pR_{t+1}^{\top}\widetilde{\pPi}_{t+1|T}\pM_{t+1}\Rp \Rb\Rp \\
		&& \exp\Lp-\frac{1}{2}\text{tr}\Lb \pU^{-1} \Lp \pE_{2t}\pR_{t+1}^{\top}\widetilde{\pPi}_{t+1|T} \pR_{t+1}\pE_{2t}\Rp \Rb\Rp \\
		& & \exp\Lp-\frac{1}{2}\text{tr}\Lb  \pU^{-1} \pE_{2t}^{\top} \pQ_{t+1}^{-1}\pE_{2t}\Rb\Rp  d\pE_{2t} \\
		& = &C\int  \Lp\frac{1}{2\pi}\Rp^{mr/2} \Lv \pH_{t+1}\Rv^{-\frac{r}{2}}\Lv \pU\Rv^{-\frac{m}{2}} \Lp\frac{1}{2\pi}\Rp^{qr/2} \Lv \pQ_{t+1}\Rv^{-\frac{r}{2}}\Lv \pU\Rv^{-\frac{q}{2}}\\	
		&  & \exp\Lp -\frac{1}{2}\text{tr}\Lb \pU^{-1}\Lp  \pM_{t+1}^{\top}\widetilde{\pPi}_{t+1|T}\pM_{t+1} \Rp\Rb\Rp \\
		&& \exp\Lp-\frac{1}{2}\text{tr}\Lb \pU^{-1} \Lp \pY_{t+1}^{*,\top}\pH_{t+1}^{-1}\pY_{T+1}^{*} - \widetilde{\pB}_{t+1|T}^{\top}\pM_{t+1} -\pM_{t+1}^{\top}\widetilde{\pB}_{t+1|T}\Rp \Rb\Rp \\
		&& \exp\Lp-\frac{1}{2}\text{tr}\Lb \pU^{-1} \Lp -\Lb\widetilde{\pB}_{t+1|T} -\widetilde{\pPi}_{t+1|T}\pM_{t+1}\Rb^{\top} \pR_{t+1}\pE_{2t}\Rp \Rb\Rp \\
		&& \exp\Lp-\frac{1}{2}\text{tr}\Lb \pU^{-1} \Lp -\pE_{2t}^{\top}\pR_{t+1}^{\top}\Lb\widetilde{\pB}_{t+1|T} - \widetilde{\pPi}_{t+1|T}\pM_{t+1}\Rb\Rp \Rb\Rp \\
		&& \exp\Lp-\frac{1}{2}\text{tr}\Lb \pU^{-1}\Lp \pE_{2t}^{\top}\Lb \pQ_{t+1}^{-1} + \pR_{t+1}^{\top}\widetilde{\pPi}_{t+1|T}\pR_{t+1} \Rb\pE_{2t}\Rp \Rb\Rp d\pE_{2t}
	\end{eqnarray*}
\end{footnotesize}
Next, we define the following
\begin{eqnarray*}
	\widetilde{\pJ}_{t+1}^{-1} &=& \pQ_{t+1}^{-1} + \pR_{t+1}^{\top}\widetilde{\pPi}_{t+1|T}\pR_{t+1} \\
	\widetilde{\pK}_{t+1} &=& 	\widetilde{\pJ}_{t+1}\pR_{t+1}^{\top}\Lb\widetilde{\pB}_{t+1|T} - \widetilde{\pPi}_{t+1|T}\pM_{t+1}\Rb
\end{eqnarray*}
Using these definitions, we write 
\begin{footnotesize}
	\begin{eqnarray*}
		\beta\Lp \pA_{t}|\pY_{t+1:T},s_{t+1:T}\Rp &= & C\exp\Lp-\frac{1}{2}\text{tr}\Lb \pU^{-1} \Lp \pY_{t+1}^{*,\top}\pH_{t+1}^{-1}\pY_{t+1}^{*} - \widetilde{\pB}_{t+1|T}^{\top}\pM_{t+1} -\pM_{t+1}^{\top}\widetilde{\pB}_{t+1|T}+\pM_{t+1}^{\top}\widetilde{\pPi}_{t+1|T}\pM_{t+1}\Rp\Rb\Rp \\
		&  &  \Lp\frac{1}{2\pi}\Rp^{mr/2} \Lv \pH_{t+1}\Rv^{-\frac{r}{2}}\Lv \pU\Rv^{-\frac{m}{2}} \Lp\frac{1}{2\pi}\Rp^{qr/2} \Lv \pQ_{t+1}\Rv^{-\frac{r}{2}}\Lv \pU\Rv^{-\frac{q}{2}}\\	
		&& \int \exp\Lp-\frac{1}{2}\text{tr}\Lb \pU^{-1} \Lp \pE_{2t}^{\top}	\widetilde{\pJ}_{t+1}^{-1}\pE_{2t}-\pE_{2t}^{\top}\widetilde{\pJ}_{t}^{-1}\widetilde{\pK}_{t+1}-\widetilde{\pK}_{t+1}^{\top}\widetilde{\pJ}_{t+1}^{-1}\pE_{2t}\Rp \Rb\Rp  d\pE_{2t}
	\end{eqnarray*}
\end{footnotesize}
Complete the square by adding and subtracting $\widetilde{\pK}_{t+1}^{\top}\widetilde{\pJ}_{t+1}^{-1}\widetilde{\pK}_{t+1}$.
\begin{footnotesize}
	\begin{eqnarray*}
		\beta\Lp \pA_{t}|\pY_{t+1:T},s_{t+1:T}\Rp &= & C\exp\Lp-\frac{1}{2}\text{tr}\Lb \pU^{-1} \Lp \pY_{t+1}^{*,\top}\pH_{t+1}^{-1}\pY_{t+1}^{*} - \widetilde{\pB}_{t+1|T}^{\top}\pM_{t+1} -\pM_{t+1}^{\top}\widetilde{\pB}_{t+1|T}+\pM_{t+1}^{\top}\widetilde{\pPi}_{t+1|T}\pM_{t+1}\Rp\Rb\Rp \\
		&  &  \Lp\frac{1}{2\pi}\Rp^{mr/2} \Lv \pH_{t+1}\Rv^{-\frac{r}{2}}\Lv \pU\Rv^{-\frac{m}{2}} \Lp\frac{1}{2\pi}\Rp^{qr/2} \Lv \pQ_{t+1}\Rv^{-\frac{r}{2}}\Lv \pU\Rv^{-\frac{q}{2}}\\	
		&& \exp\Lp-\frac{1}{2}\text{tr}\Lb \pU^{-1} \Lp -\widetilde{\pK}_{t+1}^{\top}\widetilde{\pJ}_{t+1}^{-1}\widetilde{\pK}_{t+1}\Rp \Rb\Rp \\
		&& \int \exp\Lp-\frac{1}{2}\text{tr}\Lb \pU^{-1} \Lp \pE_{2t}-\widetilde{\pK}_{t+1}\Rp^{\top} 	\widetilde{\pJ}_{t+1}^{-1}\Lp \pE_{2t}-\widetilde{\pK}_{t+1}\Rp \Rb\Rp  d\pE_{2t}
	\end{eqnarray*}
\end{footnotesize}
We get the function
\begin{footnotesize}
	\begin{eqnarray*}
		\beta\Lp \pA_{t}|\pY_{t+1:T},s_{t+1:T}\Rp & =&  C\Lp\frac{1}{2\pi}\Rp^{mr/2} \Lv \pH_{t+1}\Rv^{-\frac{r}{2}}\Lv \pU\Rv^{-\frac{m}{2}} \Lv \pQ_{t+1}\Rv^{-\frac{r}{2}}\Lv \widetilde{\pJ}_{t+1}\Rv^{\frac{r}{2}} \\	 
		&& \exp\Lp-\frac{1}{2}\text{tr}\Lb \pU^{-1} \Lp \pY_{t+1}^{*,\top}\pH_{t+1}^{-1}\pY_{t+1}^{*} - \widetilde{\pB}_{t+1|T}^{\top}\pM_{t+1} -\pM_{t+1}^{\top}\widetilde{\pB}_{t+1|T}+\pM_{t+1}^{\top}\widetilde{\pPi}_{t+1|T}\pM_{t+1}\Rp\Rb\Rp \\
		&& \exp\Lp-\frac{1}{2}\text{tr}\Lb \pU^{-1} \Lp -\widetilde{\pK}_{t+1}^{\top}\widetilde{\pJ}_{t+1}^{-1}\widetilde{\pK}_{t+1}\Rp \Rb\Rp 
	\end{eqnarray*}
\end{footnotesize}
Completing the proof amounts to writing this function in terms of  $\pA_{t}$.
\begin{footnotesize}
	\begin{eqnarray*}
		\beta\Lp \pA_{t}|\pY_{t+1:T},s_{t+1:T}\Rp & =&  C\Lp\frac{1}{2\pi}\Rp^{mr/2} \Lv \pH_{t+1}\Rv^{-\frac{r}{2}}\Lv \pU\Rv^{-\frac{m}{2}} \Lv \pQ_{t+1}\Rv^{-\frac{r}{2}}\Lv \widetilde{\pJ}_{t+1}\Rv^{\frac{r}{2}} \\	 
		&& \exp\Lp-\frac{1}{2}\text{tr}\Lb \pU^{-1} \Lp \pY_{t+1}^{*,\top}\pH_{t+1}^{-1}\pY_{t+1}^{*}\Rp\Rb\Rp \\
		&& \exp\Lp-\frac{1}{2}\text{tr}\Lb \pU^{-1} \Lp  - \widetilde{\pB}_{t+1|T}^{\top}\Lb\pT_{t+1}\pA_{t} + \pC_{t+1}\Rb -\Lb\pT_{t+1}\pA_{t} + \pC_{t+1}\Rb^{\top}\widetilde{\pB}_{t+1|T}\Rp\Rb\Rp \\
		&& \exp\Lp-\frac{1}{2}\text{tr}\Lb \pU^{-1} \Lp \Lb\pT_{t+1}\pA_{t} + \pC_{t+1}\Rb^{\top}\widetilde{\pPi}_{t+1|T}\Lb\pT_{t+1}\pA_{t} + \pC_{t+1}\Rb\Rp\Rb\Rp \\
		&& \exp\Lp-\frac{1}{2}\text{tr}\Lb \pU^{-1} \Lp -\Lb\widetilde{\pB}_{t+1|T} - \widetilde{\pPi}_{t+1|T}\pM_{t+1}\Rb^{\top}\pR_{t+1}\widetilde{\pJ}_{t+1}\pR_{t+1}^{\top}\Lb\widetilde{\pB}_{t+1|T} - \widetilde{\pPi}_{t+1|T}\pM_{t+1}\Rb  \Rp \Rb\Rp 
	\end{eqnarray*}
\end{footnotesize}
Expanding terms
\begin{footnotesize}
	\begin{eqnarray*}
		\beta\Lp \pA_{t}|\pY_{t+1:T},s_{t+1:T}\Rp & =&  C\Lp\frac{1}{2\pi}\Rp^{mr/2} \Lv \pH_{t+1}\Rv^{-\frac{r}{2}}\Lv \pU\Rv^{-\frac{m}{2}} \Lv \pQ_{t+1}\Rv^{-\frac{r}{2}}\Lv \widetilde{\pJ}_{t+1}\Rv^{\frac{r}{2}} \\	 
		&& \exp\Lp-\frac{1}{2}\text{tr}\Lb \pU^{-1} \Lp \pY_{t+1}^{*,\top}\pH_{t+1}^{-1}\pY_{t+1}^{*}\Rp\Rb\Rp \\
		&& \exp\Lp-\frac{1}{2}\text{tr}\Lb \pU^{-1} \Lp  - \widetilde{\pB}_{t+1|T}^{\top}\pC_{t+1} - \pC_{t+1}^{\top}\widetilde{\pB}_{t+1|T}\Rp\Rb\Rp \\
		&& \exp\Lp-\frac{1}{2}\text{tr}\Lb \pU^{-1} \Lp  - \widetilde{\pB}_{t+1|T}^{\top}\pT_{t+1}\pA_{t+1}  -\pA_{t}^{\top}\pT_{t+1}^{\top}\widetilde{\pB}_{t+1|T}\Rp\Rb\Rp \\
		&& \exp\Lp-\frac{1}{2}\text{tr}\Lb \pU^{-1} \Lp  \pC_{t+1}^{\top}\widetilde{\pPi}_{t+1|T} \pC_{t+1}\Rp\Rb\Rp \\
		&& \exp\Lp-\frac{1}{2}\text{tr}\Lb \pU^{-1} \Lp \pA_{t}^{\top}\pT_{t+1}^{\top}\widetilde{\pPi}_{t+1|T}\pT_{t+1}\pA_{t}\Rp\Rb\Rp \\
		&& \exp\Lp-\frac{1}{2}\text{tr}\Lb \pU^{-1} \Lp  \pC_{t+1}^{\top}\widetilde{\pPi}_{t+1|T}\pT_{t+1}\pA_{t} \Rp\Rb\Rp \\
		&& \exp\Lp-\frac{1}{2}\text{tr}\Lb \pU^{-1} \Lp \pA_{t}^{\top}\pT_{t+1}^{\top}\widetilde{\pPi}_{t+1|T} \pC_{t+1}\Rp\Rb\Rp \\		
		&& \exp\Lp-\frac{1}{2}\text{tr}\Lb \pU^{-1} \Lp \pM_{t+1}^{\top}\widetilde{\pPi}_{t+1|T}\pR_{t+1}\widetilde{\pJ}_{t+1}\pR_{t+1}^{\top}\widetilde{\pB}_{t+1|T} +\widetilde{\pB}_{t+1|T}^{\top}\pR_{t+1}\widetilde{\pJ}_{t+1}\pR_{t+1}^{\top} \widetilde{\pPi}_{t+1|T}\pM_{t+1} \Rp \Rb\Rp \\
		&& \exp\Lp-\frac{1}{2}\text{tr}\Lb \pU^{-1} \Lp - \pM_{t+1}^{\top}\widetilde{\pPi}_{t+1|T}\pR_{t+1}\widetilde{\pJ}_{t+1}\pR_{t+1}^{\top}\widetilde{\pPi}_{t+1|T}\pM_{t+1} \Rp \Rb\Rp \\
		&& \exp\Lp-\frac{1}{2}\text{tr}\Lb \pU^{-1} \Lp -\widetilde{\pB}_{t+1|T}^{\top}\pR_{t+1}\widetilde{\pJ}_{t+1}\pR_{t+1}^{\top}\widetilde{\pB}_{t+1|T}\Rp \Rb\Rp 
	\end{eqnarray*}
\end{footnotesize}
Again, substituting the definition of $\pM_{t+1}$ and combining like terms, we get
\begin{footnotesize}
	\begin{eqnarray*}
		\beta\Lp \pA_{t}|\pY_{t+1:T},s_{t+1:T}\Rp & =&  C\Lp\frac{1}{2\pi}\Rp^{mr/2} \Lv \pH_{t+1}\Rv^{-\frac{r}{2}}\Lv \pU\Rv^{-\frac{m}{2}} \Lv \pQ_{t+1}\Rv^{-\frac{r}{2}}\Lv \widetilde{\pJ}_{t+1}\Rv^{\frac{r}{2}} \\	 
		&& \exp\Lp-\frac{1}{2}\text{tr}\Lb \pU^{-1} \Lp \pY_{t+1}^{*,\top}\pH_{t+1}^{-1}\pY_{t+1}^{*} + \pC_{t+1}^{\top}\widetilde{\pPi}_{t+1|T} \pC_{t+1}\Rp\Rb\Rp \\		
		&& \exp\Lp-\frac{1}{2}\text{tr}\Lb \pU^{-1} \Lp -\widetilde{\pB}_{t+1|T}^{\top}\pR_{t+1}\widetilde{\pJ}_{t+1}\pR_{t+1}^{\top}\widetilde{\pB}_{t+1|T}  - \widetilde{\pB}_{t+1|T}^{\top}\pC_{t+1} - \pC_{t+1}^{\top}\widetilde{\pB}_{t+1|T}\Rp\Rb\Rp \\
		&& \exp\Lp-\frac{1}{2}\text{tr}\Lb \pU^{-1} \Lp  - \widetilde{\pB}_{t+1|T}^{\top}\pT_{t+1}\pA_{t}  + \pC_{t+1}^{\top}\widetilde{\pPi}_{t+1|T}\pT_{t+1}\pA_{t} \Rp\Rb\Rp \\
		&& \exp\Lp-\frac{1}{2}\text{tr}\Lb \pU^{-1} \Lp \pA_{t}^{\top}\pT_{t+1}^{\top}\widetilde{\pPi}_{t+1|T}\pT_{t+1}\pA_{t}\Rp\Rb\Rp \\
		&& \exp\Lp-\frac{1}{2}\text{tr}\Lb \pU^{-1} \Lp \pA_{t}^{\top}\pT_{t+1}^{\top}\widetilde{\pPi}_{t+1|T} \pC_{t+1}-\pA_{t}^{\top}\pT_{t+1}^{\top}\widetilde{\pB}_{t+1|T}\Rp\Rb\Rp \\		
		&& \exp\Lp-\frac{1}{2}\text{tr}\Lb \pU^{-1} \Lp \Lb\pT_{t+1}\pA_{t}+ \pC_{t+1}\Rb^{\top}\widetilde{\pPi}_{t+1|T}\pR_{t+1}\widetilde{\pJ}_{t+1}\pR_{t+1}^{\top}\widetilde{\pB}_{t+1|T} +\widetilde{\pB}_{t+1|T}^{\top}\pR_{t+1}\widetilde{\pJ}_{t+1}\pR_{t+1}^{\top} \widetilde{\pPi}_{t+1|T}\Lb\pT_{t+1}\pA_{t}+ \pC_{t+1}\Rb \Rp \Rb\Rp \\
		&& \exp\Lp-\frac{1}{2}\text{tr}\Lb \pU^{-1} \Lp - \Lb\pT_{t+1}\pA_{t}+ \pC_{t+1}\Rb^{\top}\widetilde{\pPi}_{t+1|T}\pR_{t+1}\widetilde{\pJ}_{t+1}\pR_{t+1}^{\top}\widetilde{\pPi}_{t+1|T}\Lb\pT_{t+1}\pA_{t}+ \pC_{t+1}\Rb \Rp \Rb\Rp \\
	\end{eqnarray*}
\end{footnotesize}
Expanding and simplifying, we get
\begin{footnotesize}
	\begin{eqnarray*}
		\beta\Lp \pA_{t}|\pY_{t+1:T},s_{t+1:T}\Rp & =&  C\Lp\frac{1}{2\pi}\Rp^{mr/2} \Lv \pH_{t+1}\Rv^{-\frac{r}{2}}\Lv \pU\Rv^{-\frac{m}{2}} \Lv \pQ_{t+1}\Rv^{-\frac{r}{2}}\Lv \widetilde{\pJ}_{t+1}\Rv^{\frac{r}{2}} \\	 
		&& \exp\Lp-\frac{1}{2}\text{tr}\Lb \pU^{-1} \Lp \pY_{t+1}^{*,\top}\pH_{t+1}^{-1}\pY_{t+1}^{*} + \pC_{t+1}^{\top}\widetilde{\pPi}_{t+1|T} \pC_{t+1}\Rp\Rb\Rp \\		
		&& \exp\Lp-\frac{1}{2}\text{tr}\Lb \pU^{-1} \Lp -\widetilde{\pB}_{t+1|T}^{\top}\pR_{t+1}\widetilde{\pJ}_{t+1}\pR_{t+1}^{\top}\widetilde{\pB}_{t+1|T}  - \widetilde{\pB}_{t+1|T}^{\top}\pC_{t+1} - \pC_{t+1}^{\top}\widetilde{\pB}_{t+1|T}\Rp\Rb\Rp \\
		&& \exp\Lp-\frac{1}{2}\text{tr}\Lb \pU^{-1} \Lp  \pC_{t+1}^{\top}\widetilde{\pPi}_{t+1|T}\pR_{t+1}\widetilde{\pJ}_{t+1}\pR_{t+1}^{\top}\widetilde{\pB}_{t+1|T} +\widetilde{\pB}_{t+1|T}^{\top}\pR_{t+1}\widetilde{\pJ}_{t+1}\pR_{t+1}^{\top} \widetilde{\pPi}_{t+1|T}\pC_{t+1} \Rp \Rb\Rp \\		
		&& \exp\Lp-\frac{1}{2}\text{tr}\Lb \pU^{-1} \Lp - \pC_{t+1}^{\top}\widetilde{\pPi}_{t+1|T}\pR_{t+1}\widetilde{\pJ}_{t+1}\pR_{t+1}^{\top}\widetilde{\pPi}_{t+1|T}\pC_{t+1} \Rp \Rb\Rp  \\
		&& \exp\Lp-\frac{1}{2}\text{tr}\Lb \pU^{-1} \Lp \pA_{t}^{\top}\pT_{t+1}^{\top}\widetilde{\pPi}_{t+1|T}\pT_{t+1}\pA_{t}\Rp\Rb\Rp \\
		&& \exp\Lp-\frac{1}{2}\text{tr}\Lb \pU^{-1} \Lp - \pA_{t}^{\top}\pT_{t+1}^{\top}\widetilde{\pPi}_{t+1|T}\pR_{t+1}\widetilde{\pJ}_{t+1}\pR_{t+1}^{\top}\widetilde{\pPi}_{t+1|T}\pT_{t+1}\pA_{t} \Rp \Rb\Rp \\
		&& \exp\Lp-\frac{1}{2}\text{tr}\Lb \pU^{-1} \Lp \pA_{t}^{\top}\pT_{t+1}^{\top}\widetilde{\pPi}_{t+1|T} \pC_{t+1}-\pA_{t}^{\top}\pT_{t+1}^{\top}\widetilde{\pB}_{t+1|T}\Rp\Rb\Rp \\		
		&& \exp\Lp-\frac{1}{2}\text{tr}\Lb \pU^{-1} \Lp \pA_{t}^{\top}\pT_{t+1}^{\top}\widetilde{\pPi}_{t+1|T}\pR_{t+1}\widetilde{\pJ}_{t+1}\pR_{t+1}^{\top}\widetilde{\pB}_{t+1|T} +\widetilde{\pB}_{t+1|T}^{\top}\pR_{t+1}\widetilde{\pJ}_{t+1}\pR_{t+1}^{\top} \widetilde{\pPi}_{t+1|T}\pT_{t+1}\pA_{t} \Rp \Rb\Rp \\
		&& \exp\Lp-\frac{1}{2}\text{tr}\Lb \pU^{-1} \Lp -  \pC_{t+1}^{\top}\widetilde{\pPi}_{t+1|T}\pR_{t+1}\widetilde{\pJ}_{t+1}\pR_{t+1}^{\top}\widetilde{\pPi}_{t+1|T}\pT_{t+1}\pA_{t} \Rp \Rb\Rp \\
		&& \exp\Lp-\frac{1}{2}\text{tr}\Lb \pU^{-1} \Lp  - \widetilde{\pB}_{t+1|T}^{\top}\pT_{t+1}\pA_{t}  + \pC_{t+1}^{\top}\widetilde{\pPi}_{t+1|T}\pT_{t+1}\pA_{t} \Rp\Rb\Rp \\	
		&& \exp\Lp-\frac{1}{2}\text{tr}\Lb \pU^{-1} \Lp - \pA_{t}^{\top}\pT_{t+1}^{\top}\widetilde{\pPi}_{t+1|T}\pR_{t+1}\widetilde{\pJ}_{t+1}\pR_{t+1}^{\top}\widetilde{\pPi}_{t+1|T}\pC_{t+1} \Rp \Rb\Rp \\
	\end{eqnarray*}
\end{footnotesize}
We define the backward filter's information matrix as
\begin{footnotesize}
	\begin{eqnarray*}
		\pPi_{t|T} & =& \pT_{t+1}^{\top}\widetilde{\pPi}_{t+1|T}\pT_{t+1} - \pT_{t+1}^{\top}\widetilde{\pPi}_{t+1|T}\pR_{t+1}\widetilde{\pJ}_{t+1}\pR_{t+1}^{\top}\widetilde{\pPi}_{t+1|T}\pT_{t+1} \\
		& =& \pT_{t+1}^{\top}\Lb \widetilde{\pPi}_{t+1|T} - \widetilde{\pPi}_{t+1|T}\pR_{t+1}\widetilde{\pJ}_{t+1}\pR_{t+1}^{\top}\widetilde{\pPi}_{t+1|T}\Rb\pT_{t+1} \\
		& =& \pT_{t+1}^{\top}\Lb \widetilde{\pPi}_{t+1|T} - \widetilde{\pPi}_{t+1|T}\pR_{t+1}\Lp \pQ_{t+1}^{-1} + \pR_{t+1}^{\top}\widetilde{\pPi}_{t+1|T}\pR_{t+1}\Rp^{-1}\pR_{t+1}^{\top}\widetilde{\pPi}_{t+1|T}\Rb\pT_{t+1} \\
		& =& \pT_{t+1}^{\top}\Lb \widetilde{\pPi}_{t+1|T}^{-1} + \pR_{t+1}\pQ_{t+1}\pR_{t+1}^{\top}\Rb^{-1}\pT_{t+1} \\
		& =& \pT_{t+1}^{\top}\Lb \eye + \widetilde{\pPi}_{t+1|T}\pR_{t+1}\pQ_{t+1}\pR_{t+1}^{\top}\Rb^{-1}\widetilde{\pPi}_{t+1|T} \pT_{t+1} \\
		& =& \pT_{t+1}^{\top}\pDelta_{t+1}^{-1}\widetilde{\pPi}_{t+1|T} \pT_{t+1}
	\end{eqnarray*}
\end{footnotesize}
where $\pDelta_{t+1} = \eye + \widetilde{\pPi}_{t+1|T}\pR_{t+1}\pQ_{t+1}\pR_{t+1}^{\top}$.
The third to fourth line uses the Woodbury matrix inversion lemma. We plug this definition in and combine like terms.
\begin{footnotesize}
	\begin{eqnarray*}
		\beta\Lp \pA_{t}|\pY_{t+1:T},s_{t+1:T}\Rp & =&  C\Lp\frac{1}{2\pi}\Rp^{mr/2} \Lv \pH_{t+1}\Rv^{-\frac{r}{2}}\Lv \pU\Rv^{-\frac{m}{2}} \Lv \pQ_{t+1}\Rv^{-\frac{r}{2}}\Lv \widetilde{\pJ}_{t+1}\Rv^{\frac{r}{2}} \\	 
		&& \exp\Lp-\frac{1}{2}\text{tr}\Lb \pU^{-1} \Lp \pY_{t+1}^{*,\top}\pH_{t+1}^{-1}\pY_{t+1}^{*} + \pC_{t+1}^{\top}\widetilde{\pPi}_{t+1|T} \pC_{t+1}\Rp\Rb\Rp \\		
		&& \exp\Lp-\frac{1}{2}\text{tr}\Lb \pU^{-1} \Lp -\widetilde{\pB}_{t+1|T}^{\top}\pR_{t+1}\widetilde{\pJ}_{t+1}\pR_{t+1}^{\top}\widetilde{\pB}_{t+1|T}  - \widetilde{\pB}_{t+1|T}^{\top}\pC_{t+1} - \pC_{t+1}^{\top}\widetilde{\pB}_{t+1|T}\Rp\Rb\Rp \\
		&& \exp\Lp-\frac{1}{2}\text{tr}\Lb \pU^{-1} \Lp  \pC_{t+1}^{\top}\widetilde{\pPi}_{t+1|T}\pR_{t+1}\widetilde{\pJ}_{t+1}\pR_{t+1}^{\top}\widetilde{\pB}_{t+1|T} +\widetilde{\pB}_{t+1|T}^{\top}\pR_{t+1}\widetilde{\pJ}_{t+1}\pR_{t+1}^{\top} \widetilde{\pPi}_{t+1|T}\pC_{t+1} \Rp \Rb\Rp \\		
		&& \exp\Lp-\frac{1}{2}\text{tr}\Lb \pU^{-1} \Lp - \pC_{t+1}^{\top}\widetilde{\pPi}_{t+1|T}\pR_{t+1}\widetilde{\pJ}_{t+1}\pR_{t+1}^{\top}\widetilde{\pPi}_{t+1|T}\pC_{t+1} \Rp \Rb\Rp  \\
		&& \exp\Lp-\frac{1}{2}\text{tr}\Lb \pU^{-1} \Lp \pA_{t}^{\top}\pPi_{t|T}\pA_{t}\Rp\Rb\Rp \\
		&& \exp\Lp-\frac{1}{2}\text{tr}\Lb \pU^{-1} \Lp -\pA_{t}^{\top}\Lb \pT_{t+1}^{\top} - \pT_{t+1}^{\top}\widetilde{\pPi}_{t+1|T}\pR_{t+1}\widetilde{\pJ}_{t+1}\pR_{t+1}^{\top}\Rb\Lb \widetilde{\pB}_{t+1|T}-\widetilde{\pPi}_{t+1|T}\pC_{t+1}\Rb\Rp\Rb\Rp \\
		&& \exp\Lp-\frac{1}{2}\text{tr}\Lb \pU^{-1} \Lp -\Lb\widetilde{\pB}_{t+1|T}-\widetilde{\pPi}_{t+1|T}\pC_{t+1}  \Rb^{\top}\Lb \pT_{t+1} -  \pR_{t+1}\widetilde{\pJ}_{t+1}\pR_{t+1}^{\top} \widetilde{\pPi}_{t+1|T}\pT_{t+1} \Rb \pA_{t}   \Rp\Rb\Rp 
	\end{eqnarray*}
\end{footnotesize}
From our earlier definition, we know that
\begin{footnotesize}
	\begin{eqnarray*}
		\Lb \pT_{t+1}^{\top} - \pT_{t+1}^{\top}\widetilde{\pPi}_{t+1|T}\pR_{t+1}\widetilde{\pJ}_{t+1}\pR_{t+1}^{\top}\Rb &=& \pT_{t+1}^{\top}\Lb \eye  - \widetilde{\pPi}_{t+1|T}\pR_{t+1}\widetilde{\pJ}_{t+1}\pR_{t+1}^{\top}\Rb \\
		&=& \pT_{t+1}^{\top}\Lb \eye + \widetilde{\pPi}_{t+1|T}\pR_{t+1}\pQ_{t+1}\pR_{t+1}^{\top}\Rb^{-1}
	\end{eqnarray*}
\end{footnotesize}
Using this, we define
\begin{footnotesize}
	\begin{eqnarray*}
		\pB_{t|T} &=& \pT_{t+1}^{\top}\pDelta_{t+1}^{-1}\Lp\widetilde{\pB}_{t+1|T}-\widetilde{\pPi}_{t+1|T}\pC_{t+1}\Rp 
	\end{eqnarray*}
\end{footnotesize}
and using this definition we get
\begin{footnotesize}
	\begin{eqnarray*}
		\beta\Lp \pA_{t}|\pY_{t+1:T},s_{t+1:T}\Rp & =&  C\Lp\frac{1}{2\pi}\Rp^{mr/2} \Lv \pH_{t+1}\Rv^{-\frac{r}{2}}\Lv \pU\Rv^{-\frac{m}{2}} \Lv \pQ_{t+1}\Rv^{-\frac{r}{2}}\Lv \widetilde{\pJ}_{t+1}\Rv^{\frac{r}{2}} \\	 
		&& \exp\Lp-\frac{1}{2}\text{tr}\Lb \pU^{-1} \Lp \pY_{t+1}^{*,\top}\pH_{t+1}^{-1}\pY_{t+1}^{*}  \Rp\Rb\Rp \\		
		&& \exp\Lp-\frac{1}{2}\text{tr}\Lb \pU^{-1} \Lp \pC_{t+1}^{\top}\Lb \eye + \widetilde{\pPi}_{t+1|T}\pR_{t+1}\widetilde{\pJ}_{t+1}\pR_{t+1}^{\top}\Rb \widetilde{\pPi}_{t+1|T}\pC_{t+1} \Rp \Rb\Rp  \\
		&& \exp\Lp-\frac{1}{2}\text{tr}\Lb \pU^{-1} \Lp -\widetilde{\pB}_{t+1|T}^{\top}\pR_{t+1}\widetilde{\pJ}_{t+1}\pR_{t+1}^{\top}\widetilde{\pB}_{t+1|T}  - \widetilde{\pB}_{t+1|T}^{\top}\pC_{t+1} \right. \right. \right. \\
		& &  \left. \left. \left. +\widetilde{\pB}_{t+1|T}^{\top}\pR_{t+1}\widetilde{\pJ}_{t+1}\pR_{t+1}^{\top} \widetilde{\pPi}_{t+1|T}\pC_{t+1}\Rp\Rb\Rp \\
		&& \exp\Lp-\frac{1}{2}\text{tr}\Lb \pU^{-1} \Lp  - \pC_{t+1}^{\top}\Lb \eye +  \widetilde{\pPi}_{t+1|T}\pR_{t+1}\widetilde{\pJ}_{t+1}\pR_{t+1}^{\top}\Rb\widetilde{\pB}_{t+1|T}  \Rp \Rb\Rp \\
		&& \exp\Lp-\frac{1}{2}\text{tr}\Lb \pU^{-1} \Lp \pA_{t}^{\top}\pPi_{t|T}\pA_{t}-\pA_{t}^{\top}\pB_{t|T}^{\top}-\pB_{t|T}^{\top}\pA_{t} \Rp\Rb\Rp 
	\end{eqnarray*}
\end{footnotesize}
This is the backwards function at an arbitrary date $t$. The last line is the kernel of a matrix normal distribution and all other terms are part of the constant. The proof of the backwards information filter is completed by induction over $t$.

\section{Empirical application} \label{Empirical application}

\subsection{Model}

We provide further details on the model as well as the Gibbs sampling algorithm used to estimate it in the paper. The model is complicated by the fact that we have mixed-frequency data with heavy-tailed measurement errors and time-varying transition-equation volatility. There are several types of missing data patterns. First, the quarterly series are only observed every three months. In this case, all $n = 50$ states are not observed. In addition, the data state level real GDP only beginnings in 2006, and is missing prior to this period. Finally, there are a small number of monthly data points for some U.S. states, where we do not observe employment numbers. We impute these during the MCMC algorithm. For notation, we use $\pY_{t}^{o}$ to denote all the observed data at a given date (including monthly and quarterly values). We use $\pY_{t}^{q}$ to denote missing quarterly values that are imputed during the MCMC algorithm. Note that in some months, all quarterly values are observed and there are no missing values. We use $\pY_{t}^{m}$ to denote all of the observed values $\pY_{t}^{o}$ plus the small number of monthly values that are imputed.

We specify the latent dynamics at a monthly frequency. We then employ the standard approach of \cite{MarianoMurasawa(03)} for modeling the quarterly series.
The factor dynamics follow a matrix VAR(2) process
\begin{eqnarray}
	\pFcal_{t+1} &=& \pPhi_{1}\pFcal_{t} + \pPhi_{2}\pFcal_{t-1} + \pE_{2t},  \qquad \pE_{2t} \sim \text{MN}\Lp \pzero,\pOmega_{t},\pU\Rp. \label{Example 2c}
\end{eqnarray}
We allow for episodes of common, extreme shocks to the factors -- such as those observed during COVID -- by specifying $\pOmega_{t} = \kappa_{t}\pOmega$, where $\kappa_{t}$ is a latent two-state Markov switching process taking values $1$ (with stationary probability $p$) or $5$ (with probability $1-p$). We set $p = 0.995$ and do not estimate it; $\kappa_t$ itself, together with the underlying Markov chain, is drawn during the MCMC algorithm using the algorithm of Sections \ref{discrete state simulation derivation}--\ref{backward information filter}.

Let $\py_{i,t}^{\top}$ denote a $1 \times n$ vector for the $i$-th row of $\pY_{t}$. For monthly series that we observe at every date $t$, the observation equation is
\begin{eqnarray}
	\py_{i,t}^{\top} &=& \plambda_{i}^{\top}\pFcal_{t}\pW^{\top} + \pe_{it}^{\top},
\end{eqnarray}
with no observation intercept, $\pD_t=\pzero$. To accommodate heavy-tailed measurement errors, each row's variance is multiplied by a series-specific Student's $t$ scale mixture: $\pe_{it}^{\top}\sim\text{N}\Lp\pzero, h_{it}\sigma_i^2\pU_{\pY,t}\Rp$ (row-wise; the full matrix-normal form is given below), where $h_{it}\sim\text{I.G.}\Lp \nu_i/2,\nu_i/2\Rp$ are latent scale variables representing the Student's $t$ distribution as a normal mixture, and $\nu_i$ is fixed (not estimated). Both $\sigma_i^2$ and $\Ls h_{it}\Rs_{t=1}^{T}$ are drawn during the MCMC algorithm.

For quarterly series, we model them as the average of the state variables over the past three months
\begin{eqnarray}
	\py_{i,t}^{\top} &=& \sum_{j=1}^{3}\frac{1}{3}\plambda_{i}^{\top}\pFcal_{t-j+1}\pW^{\top}  + \pe_{it}^{\top},
\end{eqnarray}
The model can be written in the matrix state space representation but with system matrices whose dimensions change depending on whether they are monthly or quarterly. The system matrices for the transition equation are
\begin{eqnarray*}
	\pA_{t} &=& \Lp\begin{matrix}
		\pFcal_{t} \\
		\pFcal_{t-1} \\
		\pFcal_{t-2}
	\end{matrix}\Rp \\
	\pT_{t} &=& \Lp\begin{matrix}
		\pPhi_{1} &  \pPhi_{2} & \pzero\\
		\eye & \pzero & \pzero \\
		\pzero & \eye & \pzero
	\end{matrix}\Rp \\ 
	\pC_{t} &=& \Lp\begin{matrix}
		\pzero \\
		\pzero \\
		\pzero
	\end{matrix}\Rp  \\ 
	\pR_{t} &=& \Lp\begin{matrix}
		\eye  \\
		\pzero \\
		\pzero 
	\end{matrix}\Rp \\
	\pQ_{t} &=& \pOmega_{t}
\end{eqnarray*}	
We place the quarterly series last in the observation matrix $\pY_{t}$.
To define the observation equation, we define factor loadings matrices
\begin{eqnarray*}
	\widetilde{\pLambda}_{\ell,1} &=& \Lp\begin{matrix}
		\plambda_{1}^{\top} \\
		\plambda_{2}^{\top} \\
		\vdots \\
		\frac{1}{3}\plambda_{m-1}^{\top} \\
		\frac{1}{3}\plambda_{m}^{\top} \\ 
	\end{matrix}\Rp \\
	\widetilde{\pLambda}_{2} &=& \Lp\begin{matrix}
		\pzero^{\top} \\
	\pzero^{\top} \\
		\vdots \\
		\frac{1}{3}\plambda_{m-1}^{\top} \\
		\frac{1}{3}\plambda_{m}^{\top} \\ 
	\end{matrix}\Rp \\
\end{eqnarray*}
Then, the system matrices for the observation equation are 
\begin{eqnarray*}
	\pZ_{t} &=& \Lp\begin{matrix}
		\widetilde{\pLambda}_{1} &  \widetilde{\pLambda}_{2} & \widetilde{\pLambda}_{2} \\
	\end{matrix}\Rp \\
	\pD_{t} &=& \pzero \\
	\pH_{t} &=& \pSigma_{t} \\
	\pW_{t} &=& \pW  \\
\end{eqnarray*}
The initial conditions are a matrix of zeros $\pA_{1|0} = \pzero$ and a diagonal matrix for $\pP_{1|0}$ with fixed entries.

To model conditional heteroskedasticty, we specify the left scale matrix of the observation equation as $\pSigma_{t} = \text{diag}\Lp h_{1t}\sigma^{2}_{1},\ldots,h_{mt}\sigma^{2}_{m}\Rp$, where $\sigma_i^2$ is a series-specific baseline variance and $h_{it}\sim\text{I.G.}\Lp \nu_i/2,\nu_i/2\Rp$ is a latent Student's $t$ scale mixture drawn fresh at every date $t$ (not a persistent Markov chain). During the MCMC algorithm we draw $\sigma_i^2$ for $i=1,\ldots,m$ and the full sequence $\Ls h_{it}\Rs_{t=1}^{T}$ for each series.

The matrix $\pU_{\pY,t}$ is constant over time and is specified as
\begin{eqnarray}
	\pU_{\pY} &=& \pW \pU \pW^\top + \pW_{\perp} \pPsi \pW_{\perp}^\top, \label{U param}
\end{eqnarray}
The identifiable object here is the $n \times n$ matrix $\pW_{\perp} \pPsi \pW_{\perp}^\top$.

The identifying restrictions in the model are the following
\begin{itemize}
	\item The right factor loadings $\pW$ are orthonormal $\pW^{\top}\pW = \eye$. This fixes the scale. We must also impose a sign restriction. For each factor $j$, you must pick one ``anchor'' variable $i_{j}$ and impose that $\pW_{ij} > 0$. The sign restriction prevents the latent factors from flipping sign during the MCMC algorithm.
	\item $\pU$ is a diagonal matrix with eigenvalues ordered in descending order. This prevents the factors from re-ordering themselves and rotating.
	\item The left factor loadings $\pLambda$ are orthonormal $\pLambda^{\top}\pLambda = \eye$. This fixes the scale. We must also impose a sign restriction. For each factor $j$, you must pick one ``anchor'' variable $i_{j}$ and impose that $\pLambda_{ij} > 0$. The sign restriction prevents the latent factors from flipping sign during the MCMC algorithm.
	\item $\pOmega$ is a diagonal matrix with eigenvalues ordered in descending order. This prevents the factors from re-ordering themselves and rotating.
	\item To identify the covariance matrix $\pU \otimes \pOmega$ of the transition equation and, through the shared $\pU$, the covariance matrix $\pU_{\pY,t}\otimes\pSigma_t$ of the measurement equation, we need only \emph{one} restriction on the overall scale (Section 2.3 of the main paper). We impose $\text{tr}\Lp \pU\Rp/r = 1$.
\end{itemize}

\subsection{Priors}

We use weakly informative but structured priors that are scaled to the data.

\begin{itemize}
	
	\item \textbf{Factor loadings.}  
	We place matrix von Mises--Fisher priors on the loading matrices:
	\[
	\pLambda \sim \mathrm{vMF}(\underline{\pM}_{\lambda}), 
	\quad
	\pW \sim \mathrm{vMF}(\underline{\pM}_{W}),
	\]
	which center the loadings around economically meaningful directions while enforcing orthonormality.
	
	\item \textbf{Observation variances.}
	For each series $i=1,\ldots,m$ (including $i=1$: no series is fixed, since the single scale restriction $\text{tr}\Lp\pU\Rp/r=1$ already identifies the model), the baseline variance follows an inverse gamma prior:
	\[
	\sigma_{i}^{2} \sim \mathrm{IG}(\underline{a}_i, \underline{b}_i).
	\]
	The scale parameters are calibrated using the sample variance of each series, ensuring that the prior is appropriately scaled while remaining moderately tight so that common factors capture cross-sectional comovement. The Student's $t$ scale mixtures $h_{it}$ have prior $h_{it}\sim\text{I.G.}\Lp \nu_i/2,\nu_i/2\Rp$, with $\nu_i$ fixed rather than estimated.
	
	\item \textbf{State innovation covariance.}  
	We use parameter-expanded inverse Wishart priors:
	\[
	\widetilde{\pOmega} \sim \mathrm{IW}(\underline{\kappa}, \underline{\pGamma}),
	\quad
	\widetilde{\pU} \sim \mathrm{IW}(\underline{\nu}, \underline{\pS}),
	\]
	and
	\[
	\widetilde{\pPsi} \sim \mathrm{IW}(\underline{\nu}_{\psi}, \underline{\pS}_{\psi}).
	\]
	The hyperparameters are chosen so that the prior means are diagonal and scaled to match the variability of the data. For $\pPsi$, we further set $\underline{\pS}_\psi \propto \eye_{n-r}$ (isotropic): beyond weak informativeness, this specific choice is what makes the Metropolis--Hastings step for $\pW$ (Step 5 below) basis-invariant when $\pPsi$ is integrated out analytically, since the resulting log-kernel depends on $\underline{\pS}_\psi$ only through $\log\Lv \pQ(\pW)^\top \pG_\star \pQ(\pW) + \underline{\pS}_\psi \Rv$, which is unaffected by the choice of orthonormal basis $\pQ(\pW)$ only when $\underline{\pS}_\psi$ is proportional to the identity.

	\item \textbf{Autoregressive parameters.}  
	We place a matrix normal prior on
	\[
	\pPhi = (\pPhi_1, \pPhi_2) \sim \mathrm{MN}(\underline{\pPhi}, \underline{\pV}, \pOmega),
	\]
	where $\underline{\pPhi}$ is centered on a stationary AR(2) process and $\underline{\pV}$ is diagonal, implying independent shrinkage across lags and factors.
	
	\item \textbf{Initial state.}  
	The initial state is
	\[
	\pA_1 \sim \mathrm{MN}(\pzero, \underline{\pP}_1, \pU),
	\]
	where $\underline{\pP}_1$ is scaled using the empirical variance of the data.
	
	\item \textbf{Transition-scale mixture.}
	The Markov-switching indicator underlying $\kappa_t\in\Ls 1,5\Rs$ has a fixed stationary probability $p=0.995$ of $\kappa_t=1$; the transition probabilities and the two values $\Ls 1,5\Rs$ are fixed, not estimated.

\end{itemize}

\subsection{Gibbs sampling algorithm}

The model and priors leads to a partially collapsed Gibbs sampler that includes several parameter expanded data augmented (PXDA) steps. Conditional on parameters, the model is linear and Gaussian, allowing for exact simulation smoothing and collapsed likelihood evaluation via the Kalman filter. The implementation needs to be careful after each PXDA step to rotate all parameters and latent states back to their original representation and to also impose any sign restrictions on the model.

\begin{description}
	\item[1.] Draw the discrete indicators $\{s_t\}_{t=1}^{T}$ governing the transition-scale multiplier $\kappa_t$.

	Recall that the state innovations satisfy
	\begin{eqnarray}
		\pE_{2t} \sim \text{MN}(\pzero, \kappa_t \pOmega, \pU),
	\end{eqnarray}
	where $\kappa_t = \kappa_S(s_t)$ is governed by a two-state indicator $s_t\in\Ls 1,2\Rs$ with $\kappa_S(1)=1$, $\kappa_S(2)=5$, evolving according to a fixed (not estimated) transition matrix whose stationary distribution assigns probability $p=0.995$ to state $1$.

	We draw $\Ls s_t\Rs_{t=1}^{T}$ sequentially, integrating out the latent states $\pA_{1:T}$ exactly at every candidate value of $s_t$, using the algorithm of Section \ref{discrete state simulation derivation} specialized to $J=2$ states: a single backward pass of the information filter of Section \ref{backward information filter} produces $\pPi_{t|T}\Lp s_{t+1:T}\Rp$ and $\pB_{t|T}\Lp s_{t+1:T}\Rp$ for $t=T-1,\ldots,1$, which depend only on future states and so need not be recomputed as $s_t$ is redrawn; the forward pass then evaluates, for each candidate $s_t=j$, the filtered moments $\pA_{t|t}(j),\pP_{t|t}(j)$ from the one-step Kalman update using $\kappa_S(j)\pOmega$, combines them with $\pPi_{t|T},\pB_{t|T}$ to obtain the smoothed moments $\pA_{t|T}(j),\pP_{t|T}(j)$, and evaluates the density ratio of Proposition 4 to obtain $p\Lp \pY_{1:T}\mid s_t=j,\ps_{-t}\Rp$ up to a constant not depending on $j$. Normalizing across $j=1,2$ and drawing $s_t$ from the resulting two-point distribution completes the sweep for date $t$.

	This step is drawn first in the sweep, ahead of everything else. The reason is a partially-collapsed-Gibbs validity issue: Step 2 below draws $(\pPsi,\pU,\pA_{1:T})$ using sufficient statistics that themselves depend on $\{s_t\}$ through $\kappa_t\pOmega$ in the Kalman filter recursion, and $\pA_{1:T}$ (drawn via simulation smoother inside that step) has a genuine, non-removable dependence on the regime path.

	\item[2.] Draw from the joint full conditional $p\Lp \pA_{1:T},\pU,\pPsi,\pY_{1:T}^{q}|\pY_{1:T}^{o},\ldots,\Rp$ which is a partially collapsed Gibbs step. The order of the draws is important for maintaining the partially collapsed structure.
	\begin{itemize}
		\item Draw the matrix $\pPsi$. We integrate out the latent states using the Kalman filter. $\pPsi$ governs the state-free complement $\pY_t^+$ (Section \ref{loglik derivation}), not the collapsed observation $\pY_t^*$, so its sufficient statistic is built from $\pY_t^+$ directly rather than from the state-dependent prediction errors $\pV_t$. Let $\pE_t^+ = \pY_t^+ - \pD_t^+$; the likelihood contribution can be summarized by the sufficient statistic
		\begin{eqnarray}
			\pS_{\psi} = \sum_{t=1}^{T} \pE_t^{+,\top} \pH_t^{-1} \pE_t^{+},
		\end{eqnarray}
		an $(n-r)\times(n-r)$ matrix.
		
		This draw is conditional on the current (pre-sweep) $\pW$, and is used to keep $\pU_{\pY}$ current for any steps between here and the $\pW$ update below. It is provisional: since the $\pW$ update integrates $\pPsi$ out rather than conditioning on it, $\pPsi$ is redrawn again, conditional on the new $\pW$, once $\pW$ has been updated (Step 5 below) -- the value obtained here is not reused after that point.

		We employ a parameter expanded data augmentation (PXDA) step. Let $\widetilde{\pPsi}$ denote the expanded covariance matrix. Combining the likelihood with the inverse Wishart prior
		\[
		\widetilde{\pPsi} \sim \text{IW}(\underline{\nu}_{\psi}, \underline{\pS}_{\psi}),
		\]
		the posterior distribution is
		\begin{eqnarray}
			\widetilde{\pPsi} \mid \pY_{1:T}, \ldots \sim 
			\text{IW}(\overline{\nu}_{\psi}, \overline{\pS}_{\psi}),
		\end{eqnarray}
		where
		\begin{eqnarray}
			\overline{\nu}_{\psi} &=& \underline{\nu}_{\psi} + \sum_{t=1}^{T} m_t, \\
			\overline{\pS}_{\psi} &=& \underline{\pS}_{\psi} + \pS_{\psi}.
		\end{eqnarray}
		Let the eigenvalue decomposition of $\widetilde{\pPsi}$ be
		\[
		\widetilde{\pPsi} = \pR_{\perp} \pD_{\perp} \pR_{\perp}^{\top},
		\]
		where $\pD_{\perp} = \text{diag}(d_1,\ldots,d_{n-r})$ with $d_1 \geq \cdots \geq d_{n-r}$.
		
		We set
		\begin{eqnarray}
			\pPsi = \pD_{\perp},
		\end{eqnarray}
		and interpret $\pR_{\perp}$ as a rotation matrix in the orthogonal complement.
		
		To preserve the likelihood, we rotate the orthogonal complement basis as
		\begin{eqnarray}
			\pW_{\perp} \leftarrow \pW_{\perp} \pR_{\perp}.
		\end{eqnarray}
		We then update the covariance components as
		\begin{eqnarray}
			\pC_{\perp} &=& \pW_{\perp} \pPsi \pW_{\perp}^{\top}, \\
			\pU_{\pY} &=& \pW \pU \pW^{\top} + \pC_{\perp}.
		\end{eqnarray}
		Finally, we update
		\begin{eqnarray}
			\pJ^{+} = \pW_{\perp}^{\top}.
		\end{eqnarray}

		The empirical application reported in the paper uses this general (dense) $\pPsi$ update. 

		Nothing about the collapsing or PXDA machinery requires $\pPsi$ to be dense, however. A scalar restriction remains available as a computationally cheaper alternative. Under $\pPsi=\psi\eye_{n-r}$, the PXDA step above collapses to a single scalar draw: combining the same sufficient statistic $\pS_\psi$ with the prior $\psi\sim\text{IG}\Lp\underline{a}_\psi,\underline{b}_\psi\Rp$ gives
		\begin{eqnarray}
			\psi \mid \pY_{1:T},\ldots \sim \text{IG}\Lp \underline{a}_\psi + \frac{n-r}{2}\sum_{t=1}^{T} m_t,\ \ \underline{b}_\psi + \frac{1}{2}\text{tr}\Lb\pS_\psi\Rb\Rp,
		\end{eqnarray}
		with $\pPsi=\psi\eye_{n-r}$ and $\pC_\perp,\pU_{\pY}$ updated as above. No eigendecomposition or rotation of $\pW_\perp$ is needed in this restricted case, since $\psi\eye_{n-r}$ is already invariant to any orthogonal rotation of the complement basis.

		\item Draw the matrix $\pU$. We integrate out the latent states using the Kalman filter. Let $\pV_t$ and $\pF_t$ denote the prediction errors and their covariance matrices, with $\pV_t \mid \pU \sim \text{MN}\Lp \pzero, \pF_t, \pU\Rp$. The log-likelihood can be written as
		\begin{eqnarray*}
			\log p(\pY_{1:T} \mid \pU, \ldots)
			= -\frac{1}{2} \sum_{t=1}^{T} \left[
			r\log |\pF_t| + m_t \log|\pU| + \text{tr}\Lp \pF_t^{-1} \pV_t \pU^{-1} \pV_t^{\top} \Rp
			\right].
		\end{eqnarray*}
		Only the terms involving $\pU$ matter for updating it. Using $\text{tr}\Lp \pF_t^{-1} \pV_t \pU^{-1} \pV_t^{\top} \Rp = \text{tr}\Lp \pU^{-1} \pV_t^{\top} \pF_t^{-1} \pV_t \Rp$, the likelihood contribution can be summarized by the sufficient statistic
		\begin{eqnarray}
			\pS_U = \sum_{t=1}^{T} \pV_t^{\top} \pF_t^{-1} \pV_t,
		\end{eqnarray}
		so that, up to a constant not involving $\pU$, $\log p(\pY_{1:T} \mid \pU, \ldots) = -\frac{1}{2}\Lp \sum_{t=1}^{T} m_t\Rp \log|\pU| - \frac{1}{2}\text{tr}\Lp \pU^{-1}\pS_U\Rp$. We employ a parameter expanded data augmentation (PXDA) step. Let $\widetilde{\pU}$ denote the expanded covariance matrix. Combining the likelihood with the inverse Wishart prior,
		\[
		\widetilde{\pU} \sim \text{IW}(\underline{\nu}, \underline{\pS}),
		\]
		the posterior is
		\begin{eqnarray}
			\widetilde{\pU} \mid \pY_{1:T}, \ldots \sim 
			\text{IW}(\overline{\nu}, \overline{\pS}),
		\end{eqnarray}
		where
		\begin{eqnarray}
			\overline{\nu} &=& \underline{\nu} + \sum_{t=1}^{T} m_t, \\
			\overline{\pS} &=& \underline{\pS} + \pS_U.
		\end{eqnarray}
		Let the eigenvalue decomposition of $\widetilde{\pU}$ be
		\[
		\widetilde{\pU} = \pR \pD \pR^{\top},
		\]
		where $\pD = \text{diag}(d_1,\ldots,d_r)$ with $d_1 \geq \cdots \geq d_r$.
		
		We set
		\begin{eqnarray}
			\pU = \pD,
		\end{eqnarray}
		and interpret $\pR$ as a rotation matrix. 
		
		To preserve the likelihood, we rotate
		\begin{eqnarray}
			\pW &\leftarrow& \pW \pR, \\
			\pA_t &\leftarrow& \pA_t \pR, \quad t=1,\ldots,T.
		\end{eqnarray}
		
		We impose a sign normalization on each column $j=1,\ldots,r$ using a prior anchor vector $\underline{\pM}_{W}$:
		\begin{eqnarray}
			\text{if } \pW_{\cdot j}^{\top} \underline{\pM}_{W,\cdot j} < 0,
			\quad \text{then } 
			\pW_{\cdot j} \leftarrow -\pW_{\cdot j}, \quad
			\pA_{\cdot j,t} \leftarrow -\pA_{\cdot j,t}.
		\end{eqnarray}
		To identify the scale of the covariance matrices, we impose the normalization
		\[
		\frac{1}{s} \text{tr}(\pOmega) = 1,
		\]
		matching the identifying restriction stated in the empirical application section of the main paper. Let
		\[
		c = \frac{\text{tr}(\pOmega)}{s}.
		\]
		As derived in Section 2.3 of the main paper, the model's Kronecker structure makes this a single scalar redundancy linking $\pU$, $\pOmega$ (through $\pQ_t$), the observation left scale $\pH_t$, and $\pPsi$: rescaling $\pU\to c\pU$ leaves $\pU\otimes\pQ_t$, $\pU\otimes\pH_t$, and $\pPsi\otimes\pH_t$ unchanged only if $\pQ_t$, $\pH_t$, and $\pPsi$ are rescaled in step. We therefore rescale all four jointly,
		\begin{eqnarray}
			\pOmega &\leftarrow& \pOmega / c, \\
			\pU &\leftarrow& \pU \, c, \\
			\pPsi &\leftarrow& \pPsi \, c,
		\end{eqnarray}
		and absorb the compensating $c^{-1}$ factor on $\pH_t$ into the baseline row-scale matrix underlying $\pH_t$ (the Student-$t$ mixture scales $h_{it}$ are left untouched, as they are unit-scale by construction). When $\pPsi=\psi\eye_{n-r}$ is imposed, only the scalar $\psi\leftarrow\psi\,c$ needs updating.

		Finally, we update the orthogonal complement
		\begin{eqnarray}
			\pW_{\perp} = \text{null}(\pW^{\top}),
		\end{eqnarray}
		and define
		\[
		\pJ^{*} = \pW^{\top}, \quad \pJ^{+} = \pW_{\perp}^{\top}.
		\]
				
		\item Draw the states $\pA_{1:T}$ from the full conditional distribution $p\Lp \pA_{1:T}|\pY_{1:T}^{o},\pU,\pPsi,\ldots,\Rp$ using the simulation smoothing algorithm. This step does not use the quarterly missing observations. 
		
		\item Draw the nissing quarterly observations. Conditional on the latent states and parameters, the observation equation implies
		\begin{eqnarray}
			\pY_t \mid \pA_t,\ldots \sim \text{MN}(\pM_t, \pSigma_t, \pU_{\pY}),
		\end{eqnarray}
		where
		\[
		\pM_t = \pZ_t \pA_t \pW^\top + \pD_t.
		\]
		
		Let $\pY_t = (\pY_{t}^{o}, \pY_{t}^{q})$ denote the partition into observed and missing (quarterly) rows. Then,
		\begin{eqnarray}
			\pY_{t}^{q} \mid \pY_{t}^{o}, \pA_t,\ldots \sim \text{MN}(\pM_{t}^{q|o}, \pSigma_{t}^{q|o}, \pU_{\pY}),
		\end{eqnarray}
		where
		\begin{align}
			\pM_{t}^{q|o} &= \pM_{t}^{q} + \pSigma_{qo}\pSigma_{oo}^{-1}(\pY_{t}^{o} - \pM_{t}^{o}), \\
			\pSigma_{t}^{q|o} &= \pSigma_{qq} - \pSigma_{qo}\pSigma_{oo}^{-1}\pSigma_{oq}.
		\end{align}
		
		We draw $\pY_t^{q}$ using the Cholesky decomposition of $\pSigma_{t}^{q|o}$ and $\pU_{\pY}$.
				 
	\end{itemize}
	
	\item[3.] Draw the ragged-edge missing monthly observations.
	
	For entries that are missing within partially observed rows, we work with the vectorized system
	\begin{eqnarray}
		\text{vec}(\pY_t) \mid \pA_t,\ldots \sim 
		\mathcal{N}\left( \text{vec}(\pM_t), \ \pU_{\pY} \otimes \pSigma_t \right).
	\end{eqnarray}
	
	Let $\mathbf{y}_t = \text{vec}(\pY_t)$ and partition into observed and missing elements:
	\[
	\mathbf{y}_t = (\mathbf{y}_{t}^{o}, \mathbf{y}_{t}^{m}).
	\]
	
	Then the conditional distribution is
	\begin{eqnarray}
		\mathbf{y}_{t}^{m} \mid \mathbf{y}_{t}^{o}, \pA_t,\ldots \sim 
		\mathcal{N}(\boldsymbol{\mu}_{t}^{m|o}, \boldsymbol{\Omega}_{t}^{m|o}),
	\end{eqnarray}
	where
	\begin{align}
		\boldsymbol{\mu}_{t}^{m|o} &= \mathbf{m}_{t}^{m} + \boldsymbol{\Omega}_{mo}\boldsymbol{\Omega}_{oo}^{-1}(\mathbf{y}_{t}^{o} - \mathbf{m}_{t}^{o}), \\
		\boldsymbol{\Omega}_{t}^{m|o} &= \boldsymbol{\Omega}_{mm} - \boldsymbol{\Omega}_{mo}\boldsymbol{\Omega}_{oo}^{-1}\boldsymbol{\Omega}_{om},
	\end{align}
	with $\boldsymbol{\Omega} = \pU_{\pY} \otimes \pSigma_t$.
	
	In implementation, we avoid forming the Kronecker product explicitly. Instead, linear systems involving $\boldsymbol{\Omega}^{-1}$ are solved using the identity
	\[
	(\pU_{\pY} \otimes \pSigma_t)^{-1}\text{vec}(\pX)
	= \text{vec}\left( \pSigma_t^{-1} \pX \pU_{\pY}^{-1} \right),
	\]
	which allows efficient computation of the conditional mean and draws.
						
	\item[4.] Draw the left factor loadings $\pLambda$. The joint full conditional distribution of $\pLambda$ belongs to the matrix Bingham--von Mises--Fisher family. Direct sampling from this distribution is computationally challenging. Instead, we construct a Metropolis--Hastings algorithm on the Stiefel manifold that updates one column at a time. Conditional on the remaining columns, each column lies on the unit sphere in the orthogonal complement, and its conditional density has a quadratic exponential form. The resulting Markov chain leaves the full conditional distribution invariant.
	
	Conditional on the latent factors and other parameters, the observation equation implies a Gaussian likelihood for each row $i=1,\ldots,m$. Let $\plambda_{i}$ denote the $i$-th row of $\pLambda$. Then, the full conditional distribution satisfies
	\begin{eqnarray}
		p(\pLambda \mid \ldots) \propto \prod_{i=1}^{m} \exp\left\{ 
		\plambda_{i}^{\top} \pS_{i} 
		- \frac{1}{2} \plambda_{i}^{\top} \pM_{i} \plambda_{i}
		\right\}
		\label{Lambda_kernel}
	\end{eqnarray}
	subject to the orthonormality constraint
	\[
	\pLambda^{\top}\pLambda = \eye.
	\]
	Recall $\pLambda \sim \text{vMF}(\underline{\pM}_\Lambda)$ a priori, with density proportional to $\exp\Lp\mathrm{tr}\Lp\underline{\pM}_\Lambda^\top\pLambda\Rp\Rp$ on the Stiefel manifold; row-wise, this contributes $\plambda_i^\top \underline{\pM}_{\Lambda,i}^\top$ to the kernel above, where $\underline{\pM}_{\Lambda,i}$ is row $i$ of $\underline{\pM}_\Lambda$. The sufficient statistics $\pS_{i}$ and $\pM_{i}$ are given by
	\begin{eqnarray}
		\pS_{i} &=& \underline{\pM}_{\Lambda,i} + \sum_{t=1}^{T} w_{it} \, \py_{i,t}^{\top} \pW \pU^{-1} \widetilde{\pFcal}_{t}^{\top}, \\
		\pM_{i} &=& \sum_{t=1}^{T} w_{it} \, \widetilde{\pFcal}_{t} \pU^{-1} \widetilde{\pFcal}_{t}^{\top},
	\end{eqnarray}
	where $w_{it} = \sigma_{it}^{-2}$ and
	\[
	\widetilde{\pFcal}_{t} =
	\begin{cases}
		\pFcal_{t}, & \text{for monthly observations}, \\
		\frac{1}{3}(\pFcal_{t} + \pFcal_{t-1} + \pFcal_{t-2}), & \text{for quarterly observations}.
	\end{cases}
	\]
	Equation (\ref{Lambda_kernel}) corresponds to a matrix-valued quadratic exponential kernel. Absent the orthonormality constraint, each row would follow a Gaussian distribution. The constraint $\pLambda^{\top}\pLambda = \eye$ restricts $\pLambda$ to lie on the Stiefel manifold.
	
	We draw $\pLambda$ using a Metropolis--Hastings algorithm that updates one column at a time while preserving the orthonormality constraint.
	
	Let $\plambda_{j}$ denote column $j$ of $\pLambda$ and let $\pLambda_{-j}$ denote the remaining columns. Conditional on $\pLambda_{-j}$, the feasible set for $\plambda_{j}$ is the unit sphere in the orthogonal complement:
	\[
	\plambda_{j} \in \mathcal{S}^{m-r} \cap \text{span}(\pLambda_{-j})^{\perp}.
	\]
	
	Let $\pN_{j}$ denote an orthonormal basis for this subspace. We reparameterize
	\[
	\plambda_{j} = \pN_{j} \mathbf{u}_{j}, \qquad \|\mathbf{u}_{j}\| = 1.
	\]
	
	A proposal $\mathbf{u}_{j}^{*}$ is generated using a random rotation on the unit sphere:
	\begin{eqnarray*}
		\mathbf{u}_{j}^{*} = \mathbf{u}_{j} \cos(\delta) + \boldsymbol{\eta} \sin(\delta),
	\end{eqnarray*}
	where $\boldsymbol{\eta}$ is a random unit vector orthogonal to $\mathbf{u}_{j}$ and $\delta \sim \mathcal{N}(0, s^{2})$ is a tuning parameter.
	The proposed column is $\plambda_{j}^{*} = \pN_{j} \mathbf{u}_{j}^{*}$.
	
	The Metropolis--Hastings acceptance probability is
	\begin{eqnarray*}
		\alpha = \min\left\{1, \exp\left( \log p(\plambda_{j}^{*} \mid \ldots) - \log p(\plambda_{j} \mid \ldots) \right) \right\},
	\end{eqnarray*}
	where the log-target for column $j$ is obtained from (\ref{Lambda_kernel}) as
	\begin{eqnarray}
		\log p(\plambda_{j} \mid \ldots) 
		= \sum_{i=1}^{m} \left[
		\lambda_{ij} S_{ij} - \frac{1}{2} \plambda_{i}^{\top} \pM_{i} \plambda_{i}
		\right].
	\end{eqnarray}
	We cycle through columns $j=1,\ldots,r$ for a fixed number of sweeps. After each update, the orthonormality constraint is preserved by construction. A sign normalization is imposed on each column after the full update.
			
	\item[5.] Draw the right factor loadings $\pW$.
	The joint full conditional distribution belongs to the matrix Bingham--von Mises--Fisher family. Direct sampling from this distribution is infeasible in this setting, so we instead construct a Metropolis--Hastings algorithm targeting the full conditional density. Rather than condition on the current draw of $\pPsi$, we integrate $\pPsi$ out of the complement likelihood analytically before proposing $\pW$. Conditioning on a fixed $\pPsi$ induces a strong posterior coupling between $\pW$ and $\pPsi$ -- the eigenbasis of $\pPsi$ effectively pins down the preferred orientation of $\pW_{\perp}$ -- which produces poor mixing; marginalizing $\pPsi$ removes this coupling.

	The observation covariance matrix is
	\begin{eqnarray}
		\pU_{\pY} = \pW \pU \pW^\top + \pW_{\perp} \pPsi \pW_{\perp}^\top,
	\end{eqnarray}
	where $\pW^\top \pW = \eye$ and $\pW_{\perp}$ spans the orthogonal complement. Since $\pPsi \sim \text{IW}(\underline{\nu}_\psi, \underline{\pS}_\psi)$ and $\pY_t^+ - \pD_t^+ \mid \pPsi \sim \text{MN}(\pzero, \pH_t, \pPsi)$ independently across $t$ (Section \ref{loglik derivation}), integrating $\pPsi$ out of the $\pY^+$ likelihood gives the standard matrix-variate $t$ marginal kernel. As a function of $\pW$, acting only through the complement subspace it defines,
	\begin{eqnarray}
		\log p\Lp \pY_{1:T}^+ \mid \pW \Rp = \text{const} - \frac{\overline{\nu}_\psi}{2} \log \Lv \pG_Q(\pW) + \underline{\pS}_\psi \Rv,
	\end{eqnarray}
	where $\pQ(\pW) = \text{null}(\pW^\top)$ is any orthonormal basis for the complement of $\pW$,
	\[
	\pG_{\star} = \sum_{t=1}^{T} \pY_t^\top \pH_t^{-1} \pY_t, \qquad \pG_Q(\pW) = \pQ(\pW)^\top \pG_{\star} \pQ(\pW), \qquad \overline{\nu}_\psi = \underline{\nu}_\psi + \sum_{t=1}^{T} m_t.
	\]
	Because $\underline{\pS}_\psi$ is proportional to the identity in our specification, this log-kernel does not depend on which orthonormal basis $\pQ(\pW)$ is used for the complement of $\pW$: $\log\Lv \pO^\top \pX \pO + c\eye \Rv = \log\Lv \pX + c\eye \Rv$ for any orthogonal $\pO$. The Metropolis--Hastings step below can therefore recompute $\pQ(\pW)$ afresh at every proposal with no basis-consistency issue. (This relies on the isotropic prior; a non-isotropic $\underline{\pS}_\psi$ would reintroduce exactly the basis dependence that conditioning on a fixed, non-scalar $\pPsi$ has.)

	Let $\pM_t = \pZ_t \pA_t$ and define the additional sufficient statistic
	\begin{eqnarray}
		\pF = \sum_{t=1}^{T} \pY_t^\top \pH_t^{-1} \pM_t \pU^{-1}.
	\end{eqnarray}
	The conditional density of $\pW$ can be written as
	\begin{eqnarray*}
		p(\pW \mid \cdots) \propto
		\exp\left\{
		\mathrm{tr}(\pF^\top \pW)
		- \frac{1}{2}\mathrm{tr}(\pW^\top \pG_{\star} \pW \pU^{-1})
		\right\}
		\times
		\exp\left\{ -\frac{\overline{\nu}_\psi}{2} \log \Lv \pG_Q(\pW) + \underline{\pS}_\psi \Rv \right\},
	\end{eqnarray*}
	subject to $\pW^\top \pW = \eye$.

	The log-density decomposes into two components:
	\begin{description}
	\item[(i)] a term corresponding to the factor space likelihood,
	\item[(ii)] a term corresponding to the marginal orthogonal complement likelihood, with $\pPsi$ integrated out.
	\end{description}
	Both components are required for correctness.

	We draw $\pW$ using a Metropolis--Hastings algorithm that updates one column at a time while preserving the orthonormality constraint.
	
	Let $\mathbf{w}_j$ denote column $j$ of $\pW$ and $\pW_{-j}$ the remaining columns. Conditional on $\pW_{-j}$, the feasible set is the unit sphere in the orthogonal complement:
	\[
	\mathbf{w}_j \in \text{span}(\pW_{-j})^{\perp}, \quad \|\mathbf{w}_j\| = 1.
	\]
	
	Let $\pN_j$ be an orthonormal basis for this subspace and write
	\[
	\mathbf{w}_j = \pN_j \mathbf{u}_j, \quad \|\mathbf{u}_j\| = 1.
	\]
	We generate proposals using a geodesic random walk on the sphere:
	\begin{eqnarray*}
		\mathbf{u}_j^{*} = \mathbf{u}_j \cos(\delta) + \boldsymbol{\eta} \sin(\delta),
	\end{eqnarray*}
	where $\boldsymbol{\eta}$ is a random unit vector orthogonal to $\mathbf{u}_j$ and $\delta \sim \mathcal{N}(0,s^2)$.
	The proposed column is $\mathbf{w}_j^{*} = \pN_j \mathbf{u}_j^{*}$.
	
	The acceptance probability is
	\begin{eqnarray}
		\alpha = \min\left\{1, \exp(\Delta_{\mathcal{Y}^*} + \Delta_{\mathcal{Y}^+}) \right\},
	\end{eqnarray}
	where, writing $\pQ = \pQ(\pW)$ for the complement basis of the current $\pW$ and $\pQ^{*} = \pQ(\pW^{*})$ for that of the column-$j$ proposal,
	\begin{align}
		\Delta_{\mathcal{Y}^*} &=
		\mathbf{f}_j^\top \mathbf{w}_j^{*}
		- \frac{1}{2} u_j^{-1} \mathbf{w}_j^{*\top} \pG_{\star} \mathbf{w}_j^{*}
		- \left(
		\mathbf{f}_j^\top \mathbf{w}_j
		- \frac{1}{2} u_j^{-1} \mathbf{w}_j^\top \pG_{\star} \mathbf{w}_j
		\right), \\
		\Delta_{\mathcal{Y}^+} &=
		-\frac{\overline{\nu}_\psi}{2} \left[
		\log \Lv \pQ^{*\top} \pG_{\star} \pQ^{*} + \underline{\pS}_\psi \Rv
		- \log \Lv \pQ^\top \pG_{\star} \pQ + \underline{\pS}_\psi \Rv
		\right].
	\end{align}
	Because $\underline{\pS}_\psi \propto \eye$, $\pQ$ and $\pQ^{*}$ may each be recomputed as $\text{null}(\pW^\top)$ and $\text{null}(\pW^{*\top})$ at every proposal without tracking any rotation between them; $\Delta_{\mathcal{Y}^+}$ is unaffected by which orthonormal basis `null' happens to return.

	After updating all columns, we set $\pW_{\perp} = \text{null}(\pW^\top)$.

	After updating $\pW$, we impose a sign normalization to maintain a consistent orientation of the factors. For each column $j=1,\ldots,r$, if
	\[
	\pW_{\cdot j}^{\top} \underline{\pM}_{W,\cdot j} < 0,
	\]
	we set
	\begin{eqnarray}
		\pW_{\cdot j} &\leftarrow& -\pW_{\cdot j}, \\
		\pA_{\cdot j,t} &\leftarrow& -\pA_{\cdot j,t}, \quad t=1,\ldots,T.
	\end{eqnarray}
	We then update $\pJ^{*} = \pW^\top$, $\pW_{\perp} = \text{null}(\pW^\top)$, and $\pJ^{+} = \pW_{\perp}^{\top}$.

	\textbf{Redrawing $\pPsi$ conditional on the new $\pW$.} The Metropolis--Hastings step above integrated $\pPsi$ out to update $\pW$. It did not update $\pPsi$ itself, and the value of $\pPsi$ carried in from the earlier PXDA step (used, e.g., in $\pU_{\pY}$ for any intervening draws) is conditional on the old $\pW$. Partially-collapsed-Gibbs validity requires redrawing $\pPsi$ from its exact conditional posterior given the just-updated $\pW$ before it is used again:
	\begin{eqnarray}
		\widetilde{\pPsi} \mid \pW, \pY_{1:T}, \ldots \sim \text{IW}\Lp \overline{\nu}_\psi,\ \underline{\pS}_\psi + \pQ(\pW)^\top \pG_{\star} \pQ(\pW) \Rp,
	\end{eqnarray}
	reusing $\pQ(\pW)^\top \pG_{\star} \pQ(\pW)$ already computed at the accepted final $\pW$. As in the PXDA step for $\pU$, we diagonalize this draw and rotate the complement basis into its eigenbasis for identification: with eigendecomposition $\widetilde{\pPsi} = \pR_\perp \pD_\perp \pR_\perp^\top$ ($\pD_\perp$ diagonal, descending),
	\begin{eqnarray}
		\pPsi &=& \pD_\perp, \\
		\pW_{\perp} &\leftarrow& \pW_{\perp} \pR_\perp, \\
		\pC_{\perp} &=& \pW_{\perp} \pPsi \pW_{\perp}^{\top}, \\
		\pJ^{+} &=& \pW_{\perp}^\top, \\
		\pU_{\pY} &=& \pW \pU \pW^\top + \pC_{\perp}.
	\end{eqnarray}

	We update the latent factor blocks and fitted values:
	\begin{align}
		\pFcal_t &= \text{first block of } \pA_t, \\
		\pFcal_{t-1} &= \text{second block}, \\
		\pFcal_{t-2} &= \text{third block}, \\
		\widehat{\pY}_t &= \pZ_t \pA_t \pW^\top.
	\end{align}

	Finally, we redraw $\pPsi$ (using the same PXDA step as in the discussion of $\pPsi$ above, now conditional on the updated $\pW$) and update $\pC_\perp,\pU_{\pY}$ accordingly. This second draw is required. The update of $\pW$ above was obtained by collapsing $\pPsi$ out of the conditional density, holding it fixed at its value from before $\pW$ was redrawn. Once $\pW$ changes, $\pPsi$'s conditional distribution given the new $\pW$ is in general different, so redrawing it here restores the correct partially collapsed target distribution for the sampler (the alternative -- carrying the pre-update $\pPsi$ forward unchanged -- would leave the chain targeting the wrong conditional).
		  
	\item[6.] Draw the matrix $\pOmega$. Let the state innovations be defined as
	\begin{eqnarray}
		\pE_{3t} = \pFcal_{t} - \pPhi_{1}\pFcal_{t-1} - \pPhi_{2}\pFcal_{t-2}, \qquad t=3,\ldots,T,
	\end{eqnarray}
	with the obvious modification for $t=2$. The state equation implies
	\[
	\pE_{3t} \sim \text{MN}(\pzero, \kappa_t \pOmega, \pU).
	\]
	Conditional on $\pU$, the likelihood contribution can be summarized by
	\begin{eqnarray}
		\pS_{\Omega} = \sum_{t=2}^{T} \frac{1}{\kappa_t} \pE_{3t} \pU^{-1} \pE_{3t}^{\top}.
	\end{eqnarray}
	We employ a PXDA step. Let $\widetilde{\pOmega}$ denote the expanded covariance matrix. Combining with the inverse Wishart prior
	\[
	\widetilde{\pOmega} \sim \text{IW}(\underline{\kappa}, \underline{\pGamma}),
	\]
	the posterior is
	\begin{eqnarray}
		\widetilde{\pOmega} \mid \ldots \sim \text{IW}(\overline{\kappa}, \overline{\pGamma}),
	\end{eqnarray}
	where
	\begin{eqnarray}
		\overline{\kappa} &=& \underline{\kappa} + r(T-1), \\
		\overline{\pGamma} &=& \underline{\pGamma} + \pS_{\Omega}.
	\end{eqnarray}
	Let the eigenvalue decomposition be
	\[
	\widetilde{\pOmega} = \pR_{\ell} \pD_{\ell} \pR_{\ell}^{\top},
	\]
	where $\pD_{\ell}$ is diagonal with descending entries.
	
	We set
	\begin{eqnarray}
		\pOmega = \pD_{\ell},
	\end{eqnarray}
	and interpret $\pR_{\ell}$ as a rotation matrix in the factor space.
	
	To preserve the likelihood, we rotate the parameters as
	\begin{eqnarray}
		\pLambda &\leftarrow& \pLambda \pR_{\ell}, \\
		\pPhi_{1} &\leftarrow& \pR_{\ell}^{\top} \pPhi_{1} \pR_{\ell}, \\
		\pPhi_{2} &\leftarrow& \pR_{\ell}^{\top} \pPhi_{2} \pR_{\ell}.
	\end{eqnarray}
	Let $\pL = \text{blkdiag}(\pR_{\ell}^{\top}, \pR_{\ell}^{\top}, \pR_{\ell}^{\top})$. We update
	\begin{eqnarray}
		\pA_t \leftarrow \pL \pA_t, \qquad t=1,\ldots,T.
	\end{eqnarray}
	
	We impose a sign normalization using a prior anchor $\underline{\pM}_{\Lambda}$. For each column $j$,
	\begin{eqnarray}
		\text{if } \pLambda_{\cdot j}^{\top} \underline{\pM}_{\Lambda,\cdot j} < 0,
	\end{eqnarray}
	we flip the sign of the $j$-th factor:
	\begin{eqnarray}
		\pLambda_{\cdot j} &\leftarrow& -\pLambda_{\cdot j}, \\
		\pPhi_{1,j\cdot} &\leftarrow& -\pPhi_{1,j\cdot}, \quad
		\pPhi_{1,\cdot j} \leftarrow -\pPhi_{1,\cdot j}, \\
		\pPhi_{2,j\cdot} &\leftarrow& -\pPhi_{2,j\cdot}, \quad
		\pPhi_{2,\cdot j} \leftarrow -\pPhi_{2,\cdot j}, \\
		\pA_{j,t} &\leftarrow& -\pA_{j,t}, \quad
		\pA_{s+j,t} \leftarrow -\pA_{s+j,t}, \quad
		\pA_{2s+j,t} \leftarrow -\pA_{2s+j,t}.
	\end{eqnarray}
	
	Finally, we update the factor blocks
	\begin{align*}
		\pFcal_t &= \text{first block of } \pA_t, \\
		\pFcal_{t-1} &= \text{second block}, \\
		\pFcal_{t-2} &= \text{third block}.
	\end{align*}

	\item[7.] Draw the autoregressive matrices $\pPhi = (\pPhi_1,\pPhi_2)$.

	We rewrite the state equation as
	\begin{eqnarray}
		\pFcal_t = \pPhi \pX_t + \pE_{3t},
	\end{eqnarray}
	where
	\[
	\pX_t =
	\begin{bmatrix}
		\pFcal_{t-1} \\
		\pFcal_{t-2}
	\end{bmatrix},
	\quad
	\pPhi = \begin{bmatrix} \pPhi_1 & \pPhi_2 \end{bmatrix}.
	\]
	The innovations satisfy
	\[
	\pE_{3t} \sim \text{MN}(\pzero, \kappa_t \pOmega, \pU).
	\]
	Conditional on $\pU$, the likelihood can be written in weighted form. Define
	\begin{eqnarray}
		\pS_{XX} &=& \sum_{t=2}^{T} \frac{1}{\kappa_t} \pX_t \pU^{-1} \pX_t^{\top}, \\
		\pS_{FX} &=& \sum_{t=2}^{T} \frac{1}{\kappa_t} \pFcal_t \pU^{-1} \pX_t^{\top}.
	\end{eqnarray}
	Combining with the prior
	\[
	\pPhi \sim \text{MN}(\underline{\pPhi}, \underline{\pV}, \pOmega),
	\]
	the posterior is
	\begin{eqnarray}
		\pPhi \mid \ldots \sim \text{MN}(\overline{\pPhi}, \overline{\pV}, \pOmega),
	\end{eqnarray}
	where
	\begin{eqnarray}
		\overline{\pV} &=& \left( \underline{\pV}^{-1} + \pS_{XX} \right)^{-1}, \\
		\overline{\pPhi} &=& \left( \underline{\pPhi}\,\underline{\pV}^{-1} + \pS_{FX} \right)
		\left( \underline{\pV}^{-1} + \pS_{XX} \right)^{-1}.
	\end{eqnarray}
	A draw is obtained as
	\begin{eqnarray}
		\pPhi = \overline{\pPhi}
		+ \pOmega^{1/2} \, \pXi \, \overline{\pV}^{1/2},
	\end{eqnarray}
	where $\pXi$ is a matrix of independent standard normal draws.
	We partition $\pPhi = (\pPhi_1, \pPhi_2)$ accordingly.
	Draws that violate stationarity are rejected.

	\item[8.] Draw the baseline observation variances $\{\sigma_i^2\}_{i=1}^{m}$.

	Define the residuals
	\begin{eqnarray}
		\pE_{1t} = \pY_t - \pZ_t \pA_t \pW^\top - \pD_t,
	\end{eqnarray}
	and let $\mathbf{e}_{it}$ denote the $i$-th row of $\pE_{1t}$. Conditional on the Student's $t$ scale mixture $h_{it}$, the observation equation implies
	\begin{eqnarray}
		\mathbf{e}_{it} \sim \mathcal{N}\Lp \pzero, h_{it}\sigma_i^2\pU_{\pY}\Rp.
	\end{eqnarray}
	Define the standardized sufficient statistic and observed-element count
	\begin{eqnarray}
		\pS_i &=& \sum_{t=1}^{T} \Lp \mathbf{e}_{it}/\sqrt{h_{it}}\Rp \pU_{\pY}^{-1} \Lp \mathbf{e}_{it}/\sqrt{h_{it}}\Rp^{\top}, \\
		N_i &=& \sum_{t=1}^{T} n_{it},
	\end{eqnarray}
	where $n_{it}$ is the number of observed elements of $\mathbf{e}_{it}$ at date $t$; if some elements are missing, the quadratic form uses only the observed components and the corresponding submatrix of $\pU_{\pY}$. Given the prior $\sigma_i^2\sim\text{IG}\Lp\underline{a}_i,\underline{b}_i\Rp$, the posterior is
	\begin{eqnarray}
		\sigma_i^2 \mid \ldots \sim \text{IG}\Lp a_i, b_i\Rp, \qquad a_i = \underline{a}_i+\frac{1}{2}N_i, \qquad b_i = \underline{b}_i+\frac{1}{2}\pS_i.
	\end{eqnarray}
	We draw $\sigma_i^2$ for every $i=1,\ldots,m$.

	\item[9.] Draw the Student's $t$ scale mixtures $\{h_{it}\}$.

	Conditional on $\sigma_i^2$, define
	\begin{eqnarray}
		Q_{it} = \mathbf{e}_{it}\pU_{\pY}^{-1}\mathbf{e}_{it}^{\top},
	\end{eqnarray}
	using only the observed components of $\mathbf{e}_{it}$ and the corresponding submatrix of $\pU_{\pY}$ if $n_{it}<n$. Given the prior $h_{it}\sim\text{I.G.}\Lp\nu_i/2,\nu_i/2\Rp$, the standard normal--inverse-gamma representation of the Student's $t$ distribution gives the conjugate posterior
	\begin{eqnarray}
		h_{it} \mid \ldots \sim \text{IG}\Lp \frac{\nu_i+n_{it}}{2},\ \frac{\nu_i + Q_{it}/\sigma_i^2}{2}\Rp,
	\end{eqnarray}
	drawn independently for every $i=1,\ldots,m$ and $t=1,\ldots,T$ (if $n_{it}=0$, $h_{it}$ is left at its prior mean, since there is no information to update it). We construct the time-varying variance matrix as
	\begin{eqnarray}
		\pH_t = \text{diag}\Lp h_{1t}\sigma_1^2,\ldots,h_{mt}\sigma_m^2\Rp.
	\end{eqnarray}

\end{description}

%
%
%

\bibliographystyle{jf}
\bibliography{creal}
